%% file: paper.tex
\documentclass[prd,nofootinbib,floatfix,preprintnumbers, twocolumn, superscriptaddress]{revtex4}%
\pdfoutput=1 

\usepackage{amssymb,amsmath}
\usepackage{color}
\usepackage{graphicx}
\usepackage{comment}
\usepackage{epsfig}
\usepackage{latexsym} % Provided \Box\begin{figure}[htbp]
\usepackage{multirow}

\def\simge{%  ``greater than about'' symbol
    \mathrel{\rlap{\raise 0.511ex
        \hbox{$>$}}{\lower 0.511ex \hbox{$\sim$}}}}
\def\simle{%  ``less than about'' symbol
    \mathrel{\rlap{\raise 0.511ex
        \hbox{$<$}}{\lower 0.511ex \hbox{$\sim$}}}}
\def\beq{\begin{equation}}
\def\eeq{\end{equation}}
\def\barr{\begin{eqnarray}}
\def\earr{\end{eqnarray}}
\def\bc{\begin{center}}
\def\ec{\end{center}}

\renewcommand{\>}{\rangle}

\newcommand{\half}{\frac{1}{2}}
\newcommand{\threehalf}{\frac{3}{2}}

\newcommand{\ninehalf}{\frac{9}{2}}
\newcommand{\osq}{\frac{1}{\sqrt{2}}}

\begin{document}

\title{Excited state baryon spectroscopy from lattice QCD}

\author{Robert~G.~Edwards}
\email{edwards@jlab.org}
\affiliation{Jefferson Laboratory, 12000 Jefferson Avenue,  Newport News, VA 23606, USA}

\author{Jozef~J.~Dudek}
\email{dudek@jlab.org}
\affiliation{Jefferson Laboratory, 12000 Jefferson Avenue,  Newport News, VA 23606, USA}
\affiliation{Department of Physics, Old Dominion University, Norfolk, VA 23529, USA}

\author{David G. Richards}
\email{dgr@jlab.org}
\affiliation{Jefferson Laboratory, 12000 Jefferson Avenue,  Newport News, VA 23606, USA}

\author{Stephen J.~Wallace}
\email{stevewal@umd.edu}
\affiliation{Department of Physics, University of Maryland, College Park, MD 20742, USA}

\collaboration{for the Hadron Spectrum Collaboration}
\date{June 6, 2011}

\begin{abstract}

We present a calculation of the Nucleon and Delta excited state spectrum on dynamical anisotropic clover lattices.  A method for operator construction is introduced that allows for the reliable identification of the continuum spins of baryon states, overcoming the reduced symmetry of the cubic lattice.  Using this method, we are able to determine a spectrum of single-particle states for spins up to and including $J = \tfrac{7}{2}$, of both parities, the first time this has been achieved in a lattice calculation. 
We find a spectrum of states identifiable as admixtures of $SU(6) \otimes O(3)$ representations and a counting of levels that is consistent with the non-relativistic $qqq$ constituent quark model. This dense spectrum is incompatible with quark-diquark model solutions to the ``missing resonance problem" and shows no signs of parity doubling of states.

\end{abstract}

\maketitle

\section{Introduction}\label{sec:intro}
\input{intro}

\section{Gauge fields}\label{sec:lattice}
\input{lattice}

\section{Correlator construction using distillation}\label{sec:distillation}
\input{distillation}

\section{Construction of baryon operators}\label{sec:ops}
\input{operators}

\section{Correlator analysis}\label{sec:fitting}
\input{fitting}

\section{Determining the spin of a state}\label{sec:spin}
\input{spin}

\section{Stability of spectrum extraction}\label{sec:stability}
\input{stability}

\section{Results}\label{sec:results}
\input{results}

\section{Multi-particle states}\label{sec:multi}
\input{multi}

\section{Summary}\label{sec:summary}
\input{summary}

%\vspace{1cm} 
 
\acknowledgements

We thank our colleagues within the Hadron Spectrum Collaboration.  We
also acknowledge illuminating discussions with Simon Capstick and
Winston Roberts.  {\tt Chroma}~\cite{Edwards:2004sx} and 
{\tt QUDA}~\cite{Clark:2009wm,Babich:2010mu} were used to perform this
work on clusters at Jefferson Laboratory under the USQCD Initiative
and the LQCD ARRA project.  
Gauge configurations were generated using resources awarded from the U.S. Department of Energy 
INCITE program at Oak Ridge National Lab, the NSF Teragrid at the Texas Advanced Computer Center 
and the Pittsburgh Supercomputer Center, as well as at Jefferson Lab.
SJW acknowledges support from
U.S. Department of Energy contract DE-FG02-93ER-40762.  RGE, JJD and
DGR acknowledge support from U.S. Department of Energy contract
DE-AC05-06OR23177, under which Jefferson Science Associates, LLC,
manages and operates Jefferson Laboratory.

\bibliography{bibliography} 
%\input{paper.bbl} 

%\clearpage
\newpage
\appendix

\section{Construction of flavor/spin/space symmetric states}\label{sec:symm_states}
\input{details}

\section{Quantum mechanics of continuum spin in the octahedral representation} \label{sec:octahedral_spin}
\input{qm}

\input{subduce_tables}

\end{document}

%% file: intro.tex
Explaining the excitation spectrum of baryons is core to our
understanding of QCD in the low-energy regime, and if we truly
understand QCD in the strong-coupling regime, we should be able to
confront experimental spectroscopic data with first-principles
calculations within QCD.  The experimental investigation of the
excited baryon spectrum has been a long-standing element of the
hadronic-physics program.  An important goal has been the search for
so-called ``missing resonances'', baryonic states predicted by the
quark model based on three constituent quarks but which have not yet
been observed experimentally; should such states not be found, it may
indicate that the baryon spectrum can be modeled with fewer effective degrees of
freedom, such as in quark-diquark models.  In the past decade, there
has been an extensive program to collect data on electromagnetic
production of one and two mesons at Jefferson Lab, MIT-Bates, LEGS,
MAMI, ELSA, and GRAAL.  To analyse these data, and thereby refine our
knowledge of the baryon spectrum, a variety of physics analysis models
have been developed at Bonn, George Washington University, Jefferson
Laboratory and Mainz.

The experimental efforts outlined above should be complemented by
high-quality \textit{ab initio} computations within lattice QCD.
Historically, the calculation of the masses of the lowest-lying
states, for both baryons and mesons, has been a benchmark
calculation of this discretized, finite-volume computational approach, 
where the aim is well-understood control
over the various systematic errors that enter into a calculation;
for a recent review, see~\cite{Hoelbling:2011kk}.  However, there is
now increasing effort aimed at calculating the excited states of the
theory, with several groups presenting investigations of the
low-lying excited baryon spectrum, using a variety of
discretizations, numbers of quark flavors, interpolating operators,
and fitting
methodologies~\cite{Mahbub:2010me,Mahbub:2010jz,Engel:2010my,Mathur:2003zf}.
Some aspects of these calculations remain unresolved and are the
subject of intense effort, notably the ordering of the Roper resonance
in the low-lying Nucleon spectrum.

A basis of baryon operators for states at rest, respecting the (cubic)
symmetry of the lattice, was developed in
Refs.~\cite{Basak:2005ir,Basak:2005aq}, and subsequently used in
calculations of the excited state Nucleon spectrum in both quenched
QCD\cite{Basak:2007kj}, and with two dynamical quark
flavors\cite{Bulava:2009jb}.  In parallel, we studied Clover fermions on
anisotropic lattices\cite{Edwards:2008ja,Lin:2008pr}, with a
finer temporal than spatial resolution, enabling the hadron
correlation functions to be observed at short temporal distances and
hence many energy levels to be extracted. Crucial to our determination
of the spectrum has been the use of the variational
method~\cite{Michael:1985ne,Luscher:1990ck,Blossier:2009kd} with a
large number of interpolating operators at both the source and the
sink; we developed and used the ``distillation'' method, enabling the
necessary correlation functions to be computed in an efficient manner.
A recent calculation of the Nucleon, $\Delta$ and $\Omega$
excited-state spectrum demonstrated the efficacy of the
method\cite{Bulava:2010yg}.

In this paper, we expand the above program of computations
considerably, extending to baryons the spin-identification techniques
developed for mesons in Refs.~\cite{Dudek:2009qf,Dudek:2010wm}.  We
develop a new basis of interpolating operators with good total angular
momentum, $J$, in the continuum, which are then subduced to the
various lattice irreducible representations (irreps).  We find that
the subduced operators retain a memory of their continuum antecedents
to a remarkable degree. For example, hadron correlation functions
between operators subduced from different continuum spins $J$ are
suppressed relative to those subduced from the same $J$, illustrating
an approximate realization of rotational symmetry at the scale of
hadrons.  This allows us to determine reliably the spins of most of
the (single-particle) states, which helps to delineate between the
nearly degenerate energy levels we observe in the spectrum.  We are
thereby able to determine the highly excited spectrum, including spins
up to $J= \tfrac{7}{2}$, and to resolve the masses of the low-lying states with
a statistical precision at or below 1\%.

The remainder of the paper is organized as follows.  In Section~\ref{sec:lattice}
we describe the parameters of the lattice gauge fields used in our calculation.  The ``distillation'' method, and
its application to the construction of baryon correlation functions
is outlined in Section~\ref{sec:distillation}. Section~\ref{sec:ops}
describes the procedure for constructing baryon interpolating
operators with good continuum spin.  Angular dependence transforming like orbital angular momenta is
incorporated through covariant derivatives that transform as
representations with $L=1$ and $L=2$; a detailed construction is provided in 
Appendix~\ref{sec:symm_states}.  We also develop subduction
matrices that allow the continuum operators to be subduced to irreducible 
representations of the cubic group; a derivation for half-integer spins up to
$\frac{9}{2}$ is given in Appendix~\ref{sec:octahedral_spin}, where tables of the subduction
matrices are also given. Our implementation of the variational
method is presented in Section~\ref{sec:fitting}, and our procedure
for determining the spins of the resulting lattice states is described in
Section~\ref{sec:spin}, which also shows tests of 
the approximate rotational invariance in the spectra.
The stability of the resultant spectra with respect to 
changes in the analysis method is discussed in Section~\ref{sec:stability}.
We present our results in
Section~\ref{sec:results}, beginning with a detailed discussion of the
spectrum at the heaviest of our quark masses before examining the
quark mass dependence. 
Discussion of multi-particle states is given in Section~\ref{sec:multi}. 
We summarize our findings and provide our
onclusions in Section~\ref{sec:summary}.

%% file: lattice.tex
%\section{Gauge fields}\label{sec:lattice}

\begin{table}[t]
  \begin{tabular}{cccc|cccc}
    $\substack{a_t m_\ell\\a_t m_s}$ & $\substack{m_\pi\\/\mathrm{MeV}}$ & $m_K/m_\pi$ & \multicolumn{1}{c|}{$a_t m_\Omega$} & volume   &$N_{\mathrm{cfgs}}$ & $N_{\mathrm{t_{srcs}}}$ & $N_{\mathrm{vecs}}$ \\
    \hline
     $\substack{-0.0808\\-0.0743}$ & 524 & 1.15 & $0.3200(7)$ & $16^3\times 128$ & 500 & 7 & 56 \\
     $\substack{-0.0830\\-0.0743}$ & 444 & 1.29 & $0.3040(8)$ & $16^3\times 128$ & 601 & 5 & 56 \\				
     $\substack{-0.0840\\-0.0743}$ & 396 & 1.39 & $0.2951(22)$ & $16^3\times 128$ & 479 & 8 & 56 \\
  \end{tabular}
  \caption{The lattice data sets and propagators used in this paper. The light and strange quark mass as well as the $\Omega$ baryon mass in temporal lattice units are shown. The pion mass, lattice size and number of configurations are listed, as well as the number of time-sources and the number of distillation vectors $N_{\mathrm{vecs}}$.}
\label{tab:lattices}
\end{table}

A major challenge in the extraction of the spectrum of excited states
from a lattice calculation is that the correlation functions, or more
specifically the principal correlators of the variational method that
correspond to excited states, decay increasingly rapidly with
Euclidean time as the energy of the state increases, whilst the noise
 behaves in the same manner as in the ground state.
Hence the
signal-to-noise ratio for excited-state correlators exhibits
increasingly rapid degradation with Euclidean time with increasing
energy. To ameliorate this problem we have adopted a dynamical
anisotropic lattice formulation whereby the temporal extent is
discretized with a finer lattice spacing than in the spatial
directions; this approach avoids the computational cost that would
come from reducing the spacing in all directions, and is core to our
excited-state spectroscopy program.  Improved gauge and fermion
actions are used, with two mass-degenerate light dynamical quarks and
one strange dynamical quark, of masses $m_l$ and $m_s$ respectively.
Details of the formulation of the actions as well as the techniques
used to determine the anisotropy parameters can be found in
Refs.~\cite{Edwards:2008ja,Lin:2008pr}.

The lattices have a spatial lattice spacing $a_s\sim 0.123~{\rm fm}$
with a temporal lattice spacing approximately $3.5$ times smaller, corresponding to
a temporal scale $a_t^{-1}\sim 5.6$ GeV.  In this work, results are
presented for the light quark baryon spectrum at quark mass parameters
$a_t m_l=(-0.0808, -0.0830, -0.0840)$ and $a_t m_s=-0.0743$, and
lattice sizes of $16^3\times 128$ with spatial extent $\sim 2$fm.  The bare strange quark
mass is held fixed to its tuned value of $a_t m_s = -0.0743$; some details
of the lattices are provided in Table~\ref{tab:lattices}. 
The lattice
scale, as quoted above, is determined by an extrapolation to the 
physical quark mass limit
using the $\Omega$ baryon mass (denoted by $a_t m_\Omega$). To
facilitate comparisons of the spectrum at different quark masses, the
ratio of hadron masses with the $\Omega$ baryon mass obtained on the
same ensemble is used to remove the explicit scale dependence,
following Ref.~\cite{Lin:2008pr}.

%% file: distillation.tex
%\section{Correlator construction}\label{sec:distillation}
The determination of the excited baryon spectrum proceeds through the
calculation of 
matrices of correlation functions between baryon creation and
annihilation operators at time $0$ and $t$ respectively:
\[
C_{ij}(t) = \big\langle 0 \big| {\cal O}_i(t) {\cal O}_j^\dag(0) \big| 0
\big\rangle .
\] 
Inserting a complete set of eigenstates of the Hamiltonian, we have
\begin{equation}
C_{ij}(t) = \sum_\mathfrak{n} \frac{1}{2 E_\mathfrak{n}} \big\langle 0  \big|  {\cal O}_i \big| \mathfrak{n} \big\rangle
 \big\langle \mathfrak{n} \big| {\cal O}_j^\dag \big| 0 \big\rangle\, e^{- E_\mathfrak{n} t} ,
\end{equation}
where the sum is over all states that have the same quantum numbers
as the interpolating operators $\{ {\cal O}_i \}$.  Note that in a
finite volume, this yields a discrete set of energies. 

Smearing is a well-established means of suppressing the short-distance
modes that do not contribute significantly to the low-energy part of the spectrum, and in turn,
allows for the construction of operators that couple predominantly to the low-lying
states. A widely adopted version is Jacobi smearing, which
uses the three-dimensional Laplacian,
\begin{equation}
  -\nabla^2_{xy}(t) = 6 \delta_{xy} - \sum_{j=1}^3
  \left(
  \tilde{U}_j(x,t)         \delta_{x+\hat\jmath,y}
  +\tilde{U}^\dagger_j(x-\hat\jmath,t) \delta_{x-\hat\jmath,y}
  \right), \nonumber
\end{equation}
where the gauge fields, $\tilde{U}$ may be constructed from an appropriate
covariant gauge-field-smearing algorithm \cite{Morningstar:2003gk}.
From this a simple smearing operator,
\[
J_{\sigma, n_\sigma} (t) = \left(1 + \frac{\sigma
  \nabla^2(t)}{n_\sigma} \right)^{n_\sigma} ,
\]
is subsequently applied to the quark fields $\psi$; for large
$n_\sigma$, this approaches a Gaussian form $\exp{\sigma \nabla^2(t)}$.
``Distillation''\cite{Peardon:2009gh}, the method we adopt, replaces
the smearing function by an outer product over the low-lying
eigenmodes of the Laplacian,
\begin{equation}
 \Box_{xy}(t) = \sum_{k=1}^N \xi_x^{(k)} (t) \xi_y^{(k)\dag} (t),
    \label{eq:box}
\end{equation}
where the $\xi^{(k)}(t)$ is the $k^{\rm th}$ eigenvector of
$\nabla^2_{xy}(t)$, ordered by the magnitude of the eigenvalue; the
(volume-dependent) number of modes $N$ should be sufficient to sample
the required angular structure at the hadronic
scale\cite{Peardon:2009gh,Dudek:2010wm}, but is small compared to the
number of sites on a time slice.  Thus distillation is a highly
efficient way of computing hadron correlation functions.

To illustrate how distillation is applied to the construction of the
baryon correlators, we specialize to the case of a positively charged isospin-$\tfrac{1}{2}$
baryon.  A generic annihilation operator can be written
\begin{equation}
  {\cal O}_i(t) = \epsilon^{abc} S^i_{\alpha\beta\gamma}
  (\mathbf{\Gamma}_1\Box d)^a_{\alpha}
  (\mathbf{\Gamma}_2\Box u)^b_{\beta}
  (\mathbf{\Gamma}_3\Box u)^c_{\gamma}(t),
\label{eq:baryon_op}
\end{equation}
where $u$ and $d$ are $u-$ and $d-$quark fields respectively,
$\mathbf{\Gamma}_j$ is a spatial operator, including possible displacements,
acting on quark $j$, $a,b,c$ are color indices, and $\alpha,\beta,
\gamma$ are spin indices; $S$ encodes the spin structure of the
operator, and is constructed so that the operator 
has the desired quantum numbers, as discussed in the next section.  We now construct
a baryon correlation function as
\begin{eqnarray}
\lefteqn{C_{ij}(t) =  \epsilon^{abc} \epsilon^{\bar{a}\bar{b}\bar{c}}
S^i_{\alpha\beta\gamma}\bar{S}^{\ast j}_{\bar{\alpha}\bar{\beta}\bar{\gamma}}}
 \nonumber \\
& &\quad\quad\quad \times\Big\langle \big[ (\mathbf{\Gamma}_1\Box d)^a_{\alpha}
(\mathbf{\Gamma}_2\Box u)^b_{\beta}
(\mathbf{\Gamma}_3\Box u)^c_{\gamma}(t) \big]  \nonumber \\
& & \quad\quad \quad\quad\;\;\cdot  \big[    
( \bar{d} \Box \overline{\mathbf{\Gamma}}_1  )^{\bar{a}}_{\bar{\alpha}}
(  \bar{u} \Box  \overline{\mathbf{\Gamma}}_2  )^{\bar{b}}_{\bar{\beta}}
(   \bar{u} \Box  \overline{\mathbf{\Gamma}}_3)^{\bar{c}}_{\bar{\gamma}}(0) \big]\Big\rangle, \nonumber
\end{eqnarray}
where the bar over $S$ and $\mathbf{\Gamma}$ indicate these belong to the creation operator.
Inserting the outer-product decomposition of $\Box$ from
Eq.~\ref{eq:box}, we can express the correlation function as
\begin{eqnarray}
  C_{ij}(t) =&& \!\!\!\!  \Phi^{i, (p,q,r)}_{\alpha\beta\gamma}(t)
    \Phi^{j,(\bar{p},\bar{q},\bar{r})\dag }_{\bar{\alpha}\bar{\beta}\bar{\gamma}}(0) 
  \nonumber \\
 & & \times\Big[\tau^{p\bar{p}}_{\alpha\bar{\alpha}}(t,0)\tau^{q\bar{q}}_{\beta\bar{\beta}}(t,0)\tau^{r\bar{r}}_{\gamma\bar{\gamma}}(t,0)
  \nonumber \\
  & &\quad\quad-\;\tau^{p\bar{p}}_{\alpha\bar{\alpha}}(t,0)\tau^{q\bar{r}}_{\beta\bar{\gamma}}(t,0)\tau^{r\bar{q}}_{\gamma\bar{\beta}}(t,0)\Big],
\label{eq:corrcons}
\end{eqnarray}
where 
\begin{equation}
\Phi^{i,(p,q,r)}_{\alpha\beta\gamma} = \epsilon^{abc}
S^i_{\alpha\beta\gamma} (\mathbf{\Gamma}_1 \xi^{(p)})^a (\mathbf{\Gamma}_2
\xi^{(q)})^b (\mathbf{\Gamma}_3 \xi^{(r)})^c \nonumber
\end{equation}
encodes the choice of operator and
\begin{equation}
\tau^{p\bar{p}}_{\alpha\bar{\alpha}}(t,0) = \xi^{\dagger(p)}(t)
M^{-1}_{\alpha\bar{\alpha}}(t,0) \xi^{(\bar{p})}(0) \nonumber
\end{equation}
is the operator-independent ``perambulator'', with $p,\bar{p}$ the eigenvector indices, and
$M$ the usual discretized Dirac operator.  The perambulators play the role of
the quark propagators between smeared sources and sinks.  Once the
perambulators have been computed, the factorization exhibited in
Eq.~(\ref{eq:corrcons}) enables the correlators to be computed between
any operators expressed through $\Phi^i, \Phi^j$, including those with
displaced quark fields, at both the source and the sink.  This feature
will be key to our use of the variational method and the subsequent
extraction of a spin-identified baryon spectrum.

%% file: operators.tex
%\section{Construction of baryon operators}\label{sec:ops}
The construction of a comprehensive basis of baryon interpolation
operators is critical to the successful application of the variational method.
The lowest-lying states in the spectrum can be
captured with color-singlet local baryon operators, of the form
given in Eq.~(\ref{eq:baryon_op})\footnote{With, in this case, $\mathbf{\Gamma}$ being just ordinary Dirac gamma matrices.}.
Combinations of such fields can be formed so as to have definite quantum
numbers, including definite symmetry properties on the cubic lattice.
The introduction of a rotationally symmetric smearing operation, be it
Jacobi smearing or distillation, in which the quark fields $\psi(x)$ 
are replaced by quasi-local, smeared fields $\Box \psi(x)$, does not change the
symmetry properties of the interpolating operators.  However, such
local and quasi-local operators can only provide access to states with
spins up to $\tfrac{3}{2}$.  In order to access higher spins in the spectrum,
and to have a sufficient basis of operators to effectively apply the
variational method, it is necessary to employ non-local operators. 

A basis of baryon interpolating operators incorporating angular structure,
respecting the cubic symmetry of the lattice, and able to access higher spins in the spectrum, was constructed in
refs.~\cite{Basak:2005aq,Basak:2005ir}, and these operators were
employed in our earlier determinations of the spectrum.  The
identification of the continuum spins corresponding to the calculated
energy levels remained challenging, however.  To overcome this
challenge, we adopt a different procedure for operator construction:
we first derive a basis of operators in the continuum, with
well-defined continuum spin quantum numbers, and then form the subduction
of those operators into the irreducible representations of the octahedral group of the lattice.

\subsection{Continuum baryon interpolating operators}
Baryons are color-singlet objects, and thus they involve totally anti-symmetric
combinations of the color indices of the three valence quarks. Furthermore baryon
interpolating operators have to be anti-symmetric under the exchange
of any pair of quarks, which is automatically satisfied since they are
constructed from anti-commuting Grassmann fields. Thus the remaining
quark labels, namely those of flavor, spin and spatial structure, have
to be in totally symmetric combinations.

We will construct our baryon interpolating operators from products
of three quark fields.  Before proceeding to classify our operators,
we note that three objects $\{x,y,z\}$ can exist in four definite symmetry
combinations: symmetric ($\mathsf{S}$), mixed-symmetric ($\mathsf{MS}$), mixed antisymmetric
($\mathsf{MA}$) and totally antisymmetric ($\mathsf{A}$)\footnote{Mixed symmetry combinations are of definite symmetry under exchange of the first two objects.}; projection operators that generate these
 combinations are specified in the Appendix, in Eqs.~(\ref{eq:projection}) and (\ref{eq:symmetry}).  We will write our
baryon interpolating operators by applying projection operators that
act on the flavor, spin and spatial labels of a generic three-quark operator,
$\psi_1 \psi_2 \psi_3$:
\begin{equation}
  B =\big( {\cal F}_\mathsf{\Sigma_F} \otimes {\cal S}_\mathsf{\Sigma_S} \otimes {\cal D}_\mathsf{\Sigma_D}\big) \{ \psi_1 \psi_2
  \psi_3\},  \nonumber
\end{equation}
where ${\cal F}, {\cal S}~{\rm and}~{\cal D}$ are flavor, Dirac spin and
spatial projection operators, respectively, and the subscripts $\mathsf{\Sigma_F,
\Sigma_S}$ and $\mathsf{\Sigma_D}$ specify the symmetry combinations of
flavor, Dirac spin and spatial labels.  For each operator $B$, we
must combine the symmetry projection operators such that the
resulting baryon operator is overall symmetric. The rules for combining 
symmetries of such direct products are given in Eq.~(\ref{eq:combine}).

To illustrate the construction, we specialize to the case of local or
quasi-local operators, where the spatial dependence is 
the same for each quark, and
thereby symmetric.  Thus we write this simplified interpolating
operator as
\begin{equation}
B = \big({\cal F}_\mathsf{\Sigma_F} \otimes {\cal S}_\mathsf{\Sigma_S}\big) \{\psi_1 \psi_2
\psi_3\}. \nonumber
\end{equation}
Furthermore, we will only consider the case of two-component Pauli spin
rather than four-component Dirac spin.  In the standard convention, 
Pauli spin involves a label $s$ that can take two values: $+$ and $-$.  
It is straightforward to extend the construction to Dirac spins using 
the Dirac-Pauli representation of Dirac matrices, as in 
Refs.~\cite{Basak:2005aq, Basak:2005ir}. 
Dirac spins involve direct products of 
two Pauli spin representations: one for $s$-spin and the other for 
$\rho$-spin (intrinsic parity).  The operator construction for Dirac spin is
described in Appendix~\ref{sec:symm_states}.  

The product
rules of Eq.~(\ref{eq:combine}) specify three ways to combine these flavor
and spin projectors to yield an overall symmetric projector:
\begin{eqnarray}
& {\cal F}_\mathsf{S} {\cal S}_\mathsf{S} \label{eq:SS} \\
 & {\cal F}_\mathsf{A} {\cal S}_\mathsf{A} \label{eq:AA} \\
& \frac{1}{\sqrt{ 2}} \left({\cal F}_\mathsf{MS} {\cal S}_\mathsf{MS} +
{\cal F}_\mathsf{MA} {\cal S}_\mathsf{MA}\right) \label{eq:MM}
\end{eqnarray}
The four symmetric spin combinations ${\cal S}_\mathsf{S}$ simply correspond to the four
states or operators of spin $\tfrac{3}{2}$: $\{{\small +++}\}_\mathsf{S}, \{{\small ++-}\}_\mathsf{S}, \{{\small +--}\}_\mathsf{S},
\{{\small ---}\}_\mathsf{S}$, while the two mixed symmetric and two mixed antisymmetric
combinations each correspond to states or operators of spin $\tfrac{1}{2}$; there
is no antisymmetric combination of three objects taking only two values.

For the case of
$SU(3)$ flavor, the flavor-symmetric combination ${\cal F}_\mathsf{S}$ yields
the decuplet ($\mathbf{10}$).  Hence Eq.~(\ref{eq:SS}), with ${\cal F}_\mathsf{S} \rightarrow \{uuu\}$,
specifies the operators for the spin-$\tfrac{3}{2}$ $\Delta^{++}$.  The
mixed-symmetric combinations ${\cal F}_\mathsf{MS,MA}$ specify the octet ($\mathbf{8}$), so
that $\{ u d u\}_\mathsf{MA,MS}$ correspond to the Nucleon.  Thus we see that
Eq.~(\ref{eq:MM}) specifies the operators for the spin-$\tfrac{1}{2}$ octet.
Since there is no ${\cal S}_\mathsf{A}$ spin combination, this example cannot
provide flavor-singlet ($\mathbf{1}$) interpolating operators - angular structure through non-local behavior is required.

Covariant derivatives defined as in Ref.~\cite{Basak:2005ir} are incorporated
into the three-quark operators in order to obtain suitable representations that 
transform like orbital angular momentum.  This is necessary in order to obtain operators with 
total angular momentum $J>\frac{3}{2}$.  First the covariant derivatives 
are combined in 
definite symmetries with respect to their action on the three quark fields. 
A single derivative %($d=1$)
is constructed from $\{\vec{D}\,\openone\,\openone\}_\mathsf{\Sigma_D}$. There are two relevant
symmetry combinations, 
\begin{equation}
  L=1:\;\left\{
  \begin{array}{l}
  	D_\mathsf{MS}^{[1]} =\frac{1}{\sqrt{6}}\big(2 D^{(3)} - D^{(1)} - D^{(2)}\big) \\
    D_\mathsf{MA}^{[1]} = \osq\big(D^{(1)} - D^{(2)}\big)
  \end{array}\right.
    \label{eq:deriv}
\end{equation}

where the notation $D^{(q)}$ means that the derivative acts
on the $q$-th quark. There is no totally antisymmetric construction of one derivative and the symmetric combination $D^{(1)} + D^{(2)} + D^{(3)}$ is a total derivative 
that gives zero when applied to a baryon with zero momentum - it is omitted. Each derivative additionally has a direction index (suppressed above) for which we choose a circular basis such that they transform under rotations like components of a spin-1 object:
\begin{eqnarray}
D_{m=\pm1} &=&\pm \tfrac{i}{2}\left(D_x \pm i D_y\right)\nonumber\\
D_{m=0} &=& -\tfrac{i}{\sqrt{2}} D_z\label{Dindex} \nonumber.
\end{eqnarray}

Totally symmetric baryon operators with one derivative are constructed
by applying Eq.~(\ref{eq:deriv}) to the mixed symmetry spin-flavor operators
according to the rules given in Eq.~(\ref{eq:combine}), 
\begin{equation}
  \psi_\mathsf{S}^{[1]} =  \osq \big(D_\mathsf{MS}^{[1]}\psi_\mathsf{MS}^{[0]}+D_\mathsf{MA}^{[1]}\psi_\mathsf{MA}^{[0]} \big), \nonumber
\end{equation}
where superscripts in brackets indicate the number of derivatives in the operator.

Two derivative operators %($d=2$) 
can be formed in definite three-quark symmetry combinations that transform like $L=0,1,2$ as follows,
\begin{alignat}{2}
  L=0,2:\; & D_\mathsf{S}^{[2]}  &=  \osq(+D_\mathsf{MS}^{[1]} D_\mathsf{MS}^{[1]}+D_\mathsf{MA}^{[1]} D_\mathsf{MA}^{[1]}),\label{eq:deriv_deriv_S}\\
  L=0,2:\; & D_\mathsf{MS}^{[2]} &=  \osq(-D_\mathsf{MS}^{[1]} D_\mathsf{MS}^{[1]}+D_\mathsf{MA}^{[1]} D_\mathsf{MA}^{[1]}),\\
  L=0,2:\; & D_\mathsf{MA}^{[2]} &=  \osq(+D_\mathsf{MS}^{[1]} D_\mathsf{MA}^{[1]}+D_\mathsf{MA}^{[1]} D_\mathsf{MS}^{[1]}),\\
  L=1: \;  & D_\mathsf{A}^{[2]}  &=  \osq(+D_\mathsf{MS}^{[1]} D_\mathsf{MA}^{[1]}-D_\mathsf{MA}^{[1]} D_\mathsf{MS}^{[1]}).
\label{eq:2_deriv}
\end{alignat}
The projection into $L=0,1,2$ comes from combining a pair of derivatives via their (suppressed) direction indices using an $SO(3)$ Clebsch-Gordan coefficient, i.e. as $\big\langle 1,m_1;1,m_2\big|L,M\big\rangle D_{m_1} D_{m_2}$. 

Although it is allowed, we omit the $D_\mathsf{S}^{[2]}$ combination that couples to $L=0$.  It
corresponds to the spatial Laplacian and vanishes at zero momentum. 
Several possibilities occur when totally symmetric baryon operators are formed using 
the rules of Eq.~(\ref{eq:combine}) to combine the spatial derivatives and spins,
\begin{equation}
  \psi_\mathsf{S}^{[2]} = D_\mathsf{S}^{[2]}\psi_\mathsf{S}^{[0]}, \ \ \osq \big(D_\mathsf{MS}^{[2]}\psi_\mathsf{MS}^{[0]}+D_\mathsf{MA}^{[2]}\psi_\mathsf{MA}^{[0]}\big),\ \ 
 D_\mathsf{A}^{[2]}\psi_\mathsf{A}^{[0]}\quad, \nonumber
\end{equation}
where no total derivatives are formed from these constructions. The angular momenta of spinors and derivatives are combined using the standard Clebsch-Gordan formula
of $SU(2)$ in order to obtain operators with good $J$ in the continuum, 
\begin{equation}
  \big|\, [J, M] \big\rangle  = \sum_{m_1,m_2}  \big|\, [J_1,m_1] \big\rangle \otimes \big|\, [J_2, m_2] \big\rangle
\big\langle J_1 m_1; J_2 m_2 \big| J M \big\rangle.\nonumber
\end{equation}

\begin{table}
\begin{tabular}{ccl}
$J$ & & irreps, $\Lambda(\mathrm{dim})$ \\
\hline
$\frac{1}{2}$      & & $G_1(2)$ \\
$\frac{3}{2}$ & & $H(4)$ \\
$\frac{5}{2}$  & & $H(4) \oplus G_2(2)$\\
$\frac{7}{2}$ & & $G_1(2) \oplus H(4) \oplus G_2(2)$\\
$\ninehalf$  & & $G_1(2) \oplus \,^1H(4) \oplus \,^2H(4) $
\end{tabular}  
\caption{Continuum spins subduced into lattice irreps $\Lambda(\mathrm{dim})$. There are two embeddings of $H$ in $J=\tfrac{9}{2}$.}
\label{tab:subduce}
\end{table}

\begin{table}
\begin{tabular}{cc}

\begin{tabular}{c|c|c|c}
  & $G_1$ & $H$ & $G_2$ \\
\hline
Singlet, $\mathbf{1}$  & 13 & 22 & 9 \\
Octet, $\mathbf{8}$    & 28 & 48 & 20 \\
Decuplet, $\mathbf{10}$ & 15 & 26 & 11 \\
\hline
\end{tabular}  
&\quad
\begin{tabular}{c|cc|c|c|c}
%\hline
  & $I$ & $S$ & $G_1$ & $H$ & $G_2$ \\
\hline
$N$       & $\frac{1}{2}$      & 0  & 28 & 48 & 20 \\
$\Delta$  & $\frac{3}{2}$ & 0  & 15 & 26 & 11 \\
$\Lambda$ & $0$          & 0  & 41 & 60 & 29 \\
$\Sigma$  & $1$          & -1 & 43 & 74 & 31 \\
$\Xi$     & $\frac{1}{2}$      & -2 & 43 & 74 & 31 \\
$\Omega$  & $0$          & -3 & 15 & 26 & 11 \\
\hline
\end{tabular}

\end{tabular}

\caption{(left) Number of distinct operators, categorized by $SU(3)_F$ in each lattice irrep $\Lambda_{g,u}$, using operators with up to two derivatives. 
(right) Number of distinct operators, categorized by $SU(2)_F$ isospin (I) and strangeness (S) in each lattice irrep $\Lambda_{g,u}$, using operators with up to two derivatives.
Each operator has $\mathrm{dim}(\Lambda)$ ``rows". There are equal numbers of operators in positive and negative parity.}
\label{tab:opnumbers}
\end{table}

The scheme outlined in this section provides a classification of all
baryon operators constructed from three quarks with either light
or strange masses, up to two derivatives in their spatial structure and
respecting classical continuum symmetries.

A naming convention for such operators is useful. 
A Nucleon operator, with spin and parity of the three quarks equal to 
$\frac{3}{2}^-$, two derivatives coupled into $L=2$ and total spin 
and parity $J^P=\frac{7}{2}^-$, is denoted as
\begin{equation}
\left(N_\mathsf{M} \otimes \big(\tfrac{3}{2}^-\big)^{\mathit{1}}_\mathsf{M} \otimes D^{[2]}_{L=2,\mathsf{S}} \right)^{J=\tfrac{7}{2}} ,
\label{eq:ops_notation}
\end{equation}
where subscripts show that the flavor construction is
mixed symmetry ($\mathsf{M}$), the spin construction also is mixed symmetry and 
the two derivatives are in a symmetric combination ($\mathsf{S}$) as in 
Eq.~(\ref{eq:deriv_deriv_S}).  Direct products of these flavor, spin
and space constructions yield an overall symmetric set of 
flavor, spin and spatial labels, as required.  

A spin state like $\left( \tfrac{3}{2}^-\right)$ can be constructed 
in several ways because Dirac spinors are used and they have four 
components, two upper and two lower components.  A superscript $\mathit{1}$ attached to
the spin part indicates that the operator is the first of several embeddings
with the same quantum numbers.  
Because of this, the basis set presented here is over-complete since operators 
featuring both derivatives and lower components are specified. However, 
this redundancy is intentional as we need a basis with multiple 
operators in each irrep for use with the variational method.
If only Pauli spinors were used, we would have an $SU(6)\otimes O(3)$ 
classification of operators.  That subset of our operators is referred to 
as ``non-relativistic''.

We reiterate that the operator construction described here provides a basis set for spectrum calculations.
We are not forcing the symmetries manifested in the operator basis into the spectrum results to be shown later. 
Rather, the dynamics of QCD will decide on the eigenstates, which can correspond to any linear superpositions of the operator basis which may exhibit quite different symmetries to those diagonalised by this basis.

\subsection{Subduction}

In lattice QCD calculations the theory is discretized on a
four-dimensional Euclidean grid.  The full three-dimensional
rotational symmetry that classifies energy eigenstates in the
continuum is reduced to the symmetry group of a cube (the cubic
symmetry group, or equivalently the octahedral group).  Instead of the
infinite number of irreducible representations labelled by spin and
parity, $J^P$, the double-cover cubic group for half-integer spin has
only six irreducible representations (\emph{irreps}): $G_{1g}$, $H_g$,
$G_{2g}$, $G_{1u}$, $H_u$, $G_{2u}$, where parity is denoted by the
$g$ (gerade) subscript for positive parity or the $u$ (ungerade)
subscript for negative parity.  The distribution of the various $M$
components of a spin-$J$ baryon into the lattice irreps is known as
\emph{subduction}, the result of which is shown in Table
\ref{tab:subduce}.

 Extending the analysis of Refs.~\cite{Dudek:2009qf,Dudek:2010wm} to baryons, we use ``subduction'' matrices for half-integer spins to project the continuum-based operators to irreducible 
representations of the octahedral group,
\begin{equation}
{\cal O}^{[J]}_{^n\!\Lambda,r} = \sum_M {\cal S}^{J,M}_{^n\!\Lambda,r} {\cal O}^{[J,M]}\ ,
\label{eq:OJM}
\end{equation}
where ${\cal O}^{[J,M]}$ is a lattice operator constructed as outlined in Appendix~(\ref{sec:deriv_ops}).
As seen in Table~\ref{tab:subduce}, subduction of $J = \frac{9}{2}$ yields two 
occurrences of the $H$ irrep.   Multiple occurrences of lattice irreps are the general rule for spins higher than $4$.  We denote by 
$ ^n\!\Lambda$ the $n^{th}$ occurrence of irrep $\Lambda$ in the subduction of spin $J$. 
For example, the two occurrences of octahedral irrep $H$ in the subduction of $J=\frac{9}{2}$ are
denoted as $^1\!H$ and $^2\!H$.   
 For each $J\rightarrow ^n\!\!\!\Lambda$ there is a subduction matrix ${\cal S}$ in the values of $M$ and the rows of
the irrep, $r$, that maps the continuum spin operators to irreducible representations of the octahedral group.  These subduction matrices are derived for
half-integer spins up to $\frac{9}{2}$ in Appendix~\ref{sec:octahedral_spin} and the resulting matrices are tabulated in Appendix~\ref{sec:octahedral_spin}.

Classifying operators by $SU(3)_F$ symmetry, we have the total number of operators in each lattice irrep as shown in Table~\ref{tab:opnumbers}(left).  Considered as a broken symmetry with an unbroken $SU(2)_F$ isospin symmetry, we have the number of operators shown in Table~\ref{tab:opnumbers}(right).
The operator basis used in this work is constructed using $SU(3)_F$ flavor symmetry.  
Of course, the $SU(3)_F$ symmetry is broken in QCD, most notably by the mass of the strange quark.
The $SU(3)_F$ operator basis allows the flavor symmetry breaking to be determined by the relative degree of overlap onto operators transforming in different representations of $SU(3)_F$. This is comparable to what is done in quark models and other approaches, which helps to relate the lattice results to phenomenology.

With the construction of operators and their subduction to irreps of the
cubic group outlined as above, we defer further details to the appendices.
Appendix~\ref{sec:symm_states} provides details of the operator constructions, 
Appendix~\ref{sec:octahedral_spin} derives the subduction matrices through an analysis of the quantum mechanics of continuum spin in the octahedral representation, and Appendix~\ref{sec:octahedral_spin} provides the subduction matrices for half-integer spins.

%% file: fitting.tex
%\section{Correlator analysis}\label{sec:fitting}

%%%%%%%%%%%%
\begin{figure}
 \centering
\includegraphics[width=0.5\textwidth,angle=0]{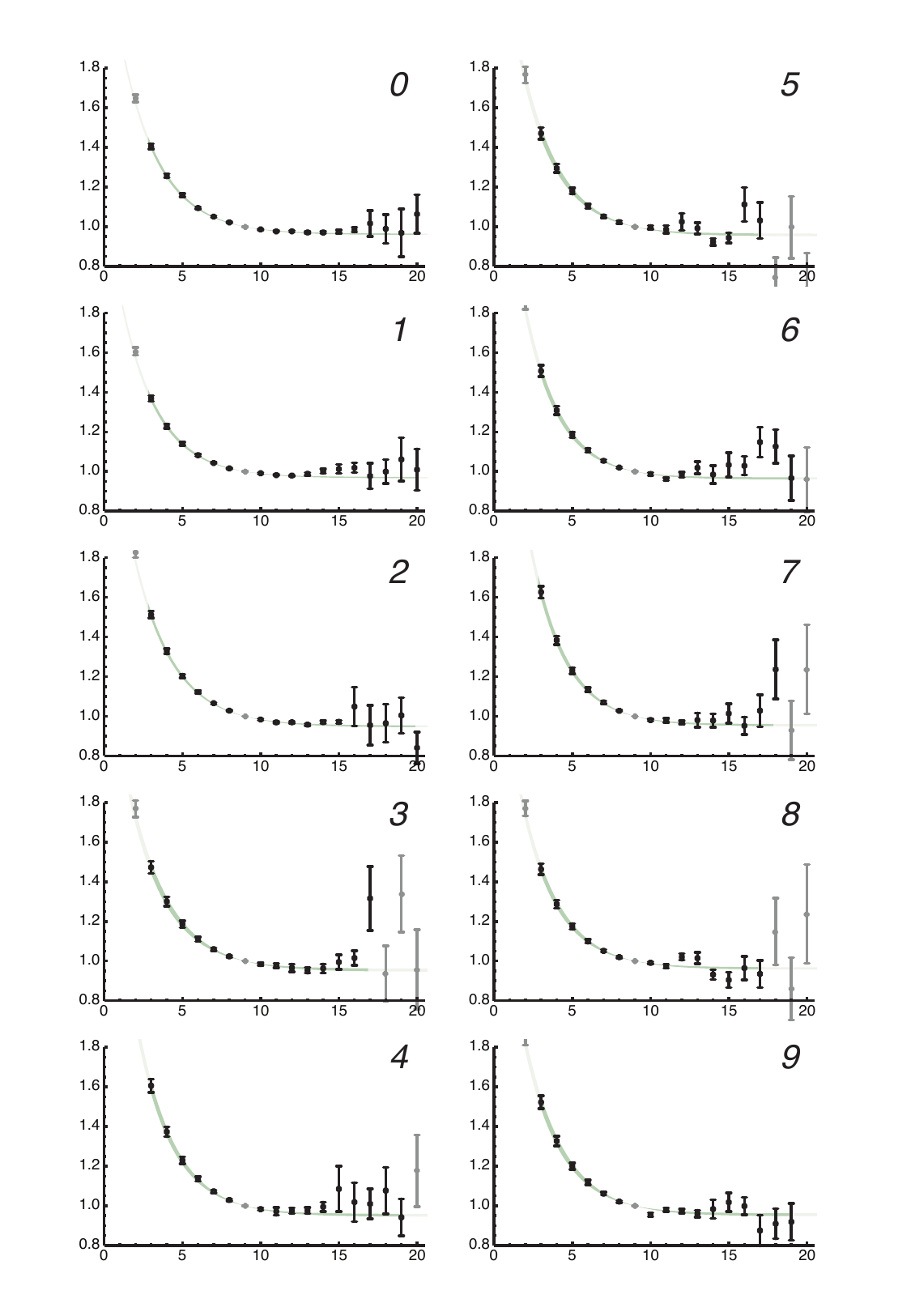}
\caption{Principal correlator fits according to Eq.~(\ref{eq:fitprincorr}).  For ten states of the Nucleon $H_u$ irrep for $m_{\pi}$ = 524 MeV, the plots show $\lambda^{\mathfrak{n}}(t)~\!\cdot~\!e^{m_\mathfrak{n}(t-t_0)}$ data and the fit for $t_0 = 9$. Data used in the fit are shown in black, while points excluded from the fit are in grey.  
\label{fig:princorrfits}}
\end{figure}

The variational method for spectral extraction
\cite{Michael:1985ne,Luscher:1990ck,Blossier:2009kd}, which takes
advantage of the multiplicity of operators within a given symmetry channel
to find the best (in a variational sense) linear combination of
operators for each state in the spectrum, is now in common 
usage\cite{Burch:2009wu,Gattringer:2008be,Burch:2006dg,Burch:2004he,Bulava:2010yg}. 
Our application of the method follows that developed in
Refs.~\cite{Dudek:2007wv,Dudek:2009qf,Dudek:2010wm}, and applied to
the analysis of the excited meson spectrum, and we summarize it here.

The starting point is the system of generalized eigenvalue equations for
the correlation matrix:
\begin{equation}
  C(t) v^\mathfrak{n}(t) = \lambda_\mathfrak{n}(t) C(t_0)  v^\mathfrak{n}(t) \label{var}
\end{equation}
where $\lambda_\mathfrak{n}(t_0) = 1$, and where there is an
orthogonality condition on the eigenvectors of different states
($\mathfrak{n}, \,\mathfrak{n}'$), $v^{\mathfrak{n}'\dag} C(t_0)
v^\mathfrak{n} = \delta_{\mathfrak{n}, \mathfrak{n}'}$. This
orthogonality condition provides eigenvectors that distinguish clearly
between nearly degenerate states, which would be difficult to distinguish by their time
dependence alone.  Equation~(\ref{var}) is solved for eigenvalues,
$\lambda_\mathfrak{n}$, and eigenvectors, $v^\mathfrak{n}$,
independently on each timeslice, $t$.  Rather than
ordering the states by the size of their eigenvalue, which can
become uncertain because of the high level of degeneracy in the baryon 
spectrum, we associate states at different timeslices using the similarity 
of their eigenvectors. We
choose a reference timeslice on which reference eigenvectors are
defined, $v^\mathfrak{n}_{\mathrm{ref}} \equiv
v^{\mathfrak{n}}(t_\mathrm{ref})$, and compare eigenvectors on other
timeslices by finding the maximum value of
$v^{\mathfrak{n}'\dag}_\mathrm{ref} C(t_0) v^{\mathfrak{n}}$ which
associates a state $\mathfrak{n}$ with a reference state
$\mathfrak{n}'$. Using this procedure we observe essentially no
``flipping" between states 
in either the principal correlators,
$\lambda_\mathfrak{n}(t)$, or the eigenvectors, $v^\mathfrak{n}(t)$, as
functions of $t$.

Any two-point correlation function on a finite spatial lattice can be 
expressed as a spectral decomposition
\begin{equation}
	C_{ij}(t) = \sum_\mathfrak{n} \frac{Z_i^{\mathfrak{n}*} Z_j^{\mathfrak{n}}}{2m_\mathfrak{n}} e^{-m_\mathfrak{n} t}
	\label{spectro_decomp}
\end{equation}
where we assume that $t \ll L_t$, the
temporal length of the box, so that the opposite-parity contributions
arising from the other time ordering on the periodic lattice can be
ignored. The ``overlap factors", $Z^\mathfrak{n}_i \equiv \langle
\mathfrak{n} | {\cal O}_i^\dag | 0 \rangle$ are related to the eigenvectors
by $Z^\mathfrak{n}_i = \sqrt{2 m_\mathfrak{n}} e^{m_\mathfrak{n}
  t_0/2}\, v^{\mathfrak{n}*}_j C_{ji}(t_0)$. 

We obtain the masses from fitting the principal correlators,
$\lambda_\mathfrak{n}(t)$, which for large times should tend to $e^{-
  m_\mathfrak{n} (t - t_0)}$. In practice we allow a second
exponential in the fit form, and our fit function is
\begin{equation}
  \lambda_\mathfrak{n}(t) = (1 - A_\mathfrak{n}) e^{-m_\mathfrak{n}(t-t_0)} + A_\mathfrak{n} e^{-m_\mathfrak{n}' (t-t_0)},
  \label{eq:fitprincorr}
\end{equation}
where the fit parameters are $m_\mathfrak{n}, m_\mathfrak{n}'$ and
$A_\mathfrak{n}$. Typical fits for a set of excited states within an
irrep are shown in Figure \ref{fig:princorrfits} where we plot the
principal correlator with the dominant time-dependence due to state
$\mathfrak{n}$ divided out. If a single exponential were to dominate
the fit, such a plot would show a constant value of unity for all
times.  For the form of Eq.~\ref{eq:fitprincorr}, the data would
approach a constant $1-A$ at large times, and this is clearly
satisfied for $t>t_0$.

Empirically we find that the size of the second exponential term
decreases rapidly as one increases $t_0$. Further we find, in
agreement with the perturbative analysis of
Ref.~\cite{Blossier:2009kd} and with our earlier meson analysis, that
for large $t_0$ values the $m_\mathfrak{n}'$ extracted are larger than
the value of $m_{\mathfrak{n}=\mathrm{dim(C)}}$, the largest ``first"
exponential mass extracted. At smaller $t_0$ values this is not
necessarily true and is indicative of an insufficient number of
states in Eq.~(\ref{spectro_decomp}) to describe $C(t_0)$ completely.  
The values of $A_\mathfrak{n}$ and
$m_\mathfrak{n}'$ are not used elsewhere in the analysis.

Our choice of $t_0$ is made using the ``reconstruction''
scheme\cite{Dudek:2007wv,Dudek:2010wm}: the masses, $m_\mathfrak{n}$, extracted
from the fits to the principal correlators, and the $Z_i^\mathfrak{n}$ extracted
from the eigenvectors at a single time slice are used to reconstruct
the correlator matrix using Eq.~(\ref{spectro_decomp}).  This
reconstructed matrix is then compared with the data for $t>t_0$, with
the degree of agreement indicating the acceptability of the spectral
reconstruction.  
Adopting too small a value of $t_0$
leads to a poor
reconstruction of the data for $t>t_0$. 
In general, the reproduction is better as $t_0$ is
increased until increased statistical noise prevents further
improvement. 
The sensitivity of extracted
spectral quantities to the value of $t_0$ used will be discussed in
detail in section \ref{sec:t0}, but in short the energies of low-lying states are rather insensitive
to $t_0$ and the reconstruction of the full correlator matrix usually is best
when $t_0 \gtrsim 7$, but not too large.

From the spectral decomposition of the correlator, equation
\ref{spectro_decomp}, it is clear that there should in fact be no time
dependence in the eigenvectors. Because of states
at higher energies than can be resolved with $\mathrm{dim}(C)$ operators, 
there generally is a contribution to energies and $Z$'s that decays more rapidly
than the lowest mass state that contributes to a principal correlator.
 As for energies, we obtain ``time-independent'' overlap factors,
$Z^\mathfrak{n}_i$ from fits of $Z^\mathfrak{n}_i(t)$, obtained
from the eigenvectors, with a constant or a constant plus an
exponential, in the spirit of the perturbative corrections outlined in
\cite{Blossier:2009kd}).

%% file: spin.tex
%\section{Determining the spin of a state}\label{sec:spin}

\subsection{Motivation and procedure}

\begin{figure}
\begin{center}
\includegraphics[width=0.45\textwidth]{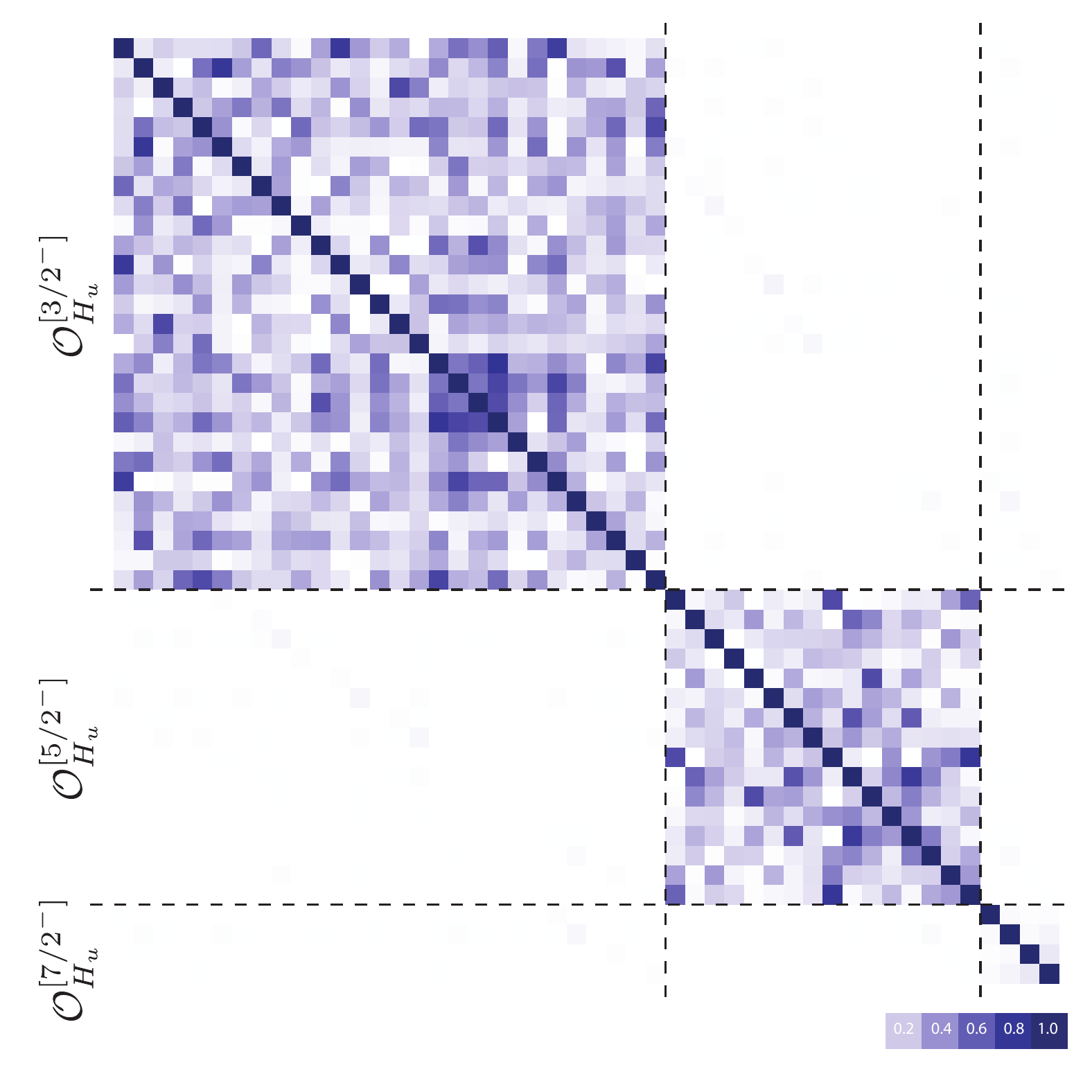}
 \caption{The magnitude of matrix elements in a matrix of correlation 
functions, $C_{ij}/\sqrt{C_{ii}C_{jj}}$, at time-slice 5 is shown according to the scale at the lower right.
The matrix is for the Nucleon $H_u$ irrep, with 28 $[J=\frac{3}{2}]$ operators, 16 $[J=\frac{5}{2}]$ operators and 4 $[J=\frac{7}{2}]$ operators.
 \label{fig:matrixplot}}
\end{center}
\end{figure}

A new method for identifying the spins of excited mesonic states 
was introduced in Refs.~\cite{Dudek:2009qf,Dudek:2010wm}. In this work, 
we extend the method to identify spins of excited baryonic states.  The
explanation for the success of the new method is that there is an 
approximate realization of rotational invariance at the scale of hadrons in
our correlation functions.  There are two reasons for this claim.
The first is that there are no dimension-five operators made of quark bilinears that respect the
symmetries of lattice actions based on the Wilson formalism and that do not also
transform trivially under the continuum group of spatial rotations. Thus, 
rotational symmetry breaking terms do not appear until ${\cal O}(a^2)$ in the action. This
argument holds even though the action used in this work describes an anisotropic
lattice. 
The second reason is 
that our baryon operators are constructed from low momentum filtered, 
smeared quark fields, where the smearing
is designed to filter out fluctuations on small scales.
Possible divergent mixing with lower dimensional operators is then expected to be suppressed.
The baryon operators then are reasonably smooth on 
a size scale typical of baryons, which is $R \approx 1$ fm. With the lattice 
spacing of $\sim 0.12$ fm, the breaking of rotational symmetry in the action or the
baryon operators can be small: 
$(a/R)^2 \approx $ 0.015.  
Of course, these qualitative arguments should be
backed up by evidence of approximate rotational invariance in explicit calculation.

The operators constructed in Section \ref{sec:ops} using subduction matrices 
transform irreducibly under the allowed cubic rotations, that is they 
faithfully respect the symmetries of the
lattice. They also carry information about the continuum angular momentum, $J$,
from which they are subduced. To the extent that approximate rotational invariance is
realized, we expect that an operator
subduced from 
spin $J$ to overlap strongly only onto states of continuum spin $J$, and have little overlap
with states of different continuum spin.
In fact this is clearly apparent even at the level of the correlator 
matrix as seen in Figure \ref{fig:matrixplot}. Here the correlator 
matrix for the Nucleon $H_u$ is observed to be approximately block diagonal 
when the operators are ordered according to the spin from which they 
were subduced.

\begin{figure*}[t]
 \centering
\includegraphics[width=0.85\textwidth,bb= 30 40 870 396]{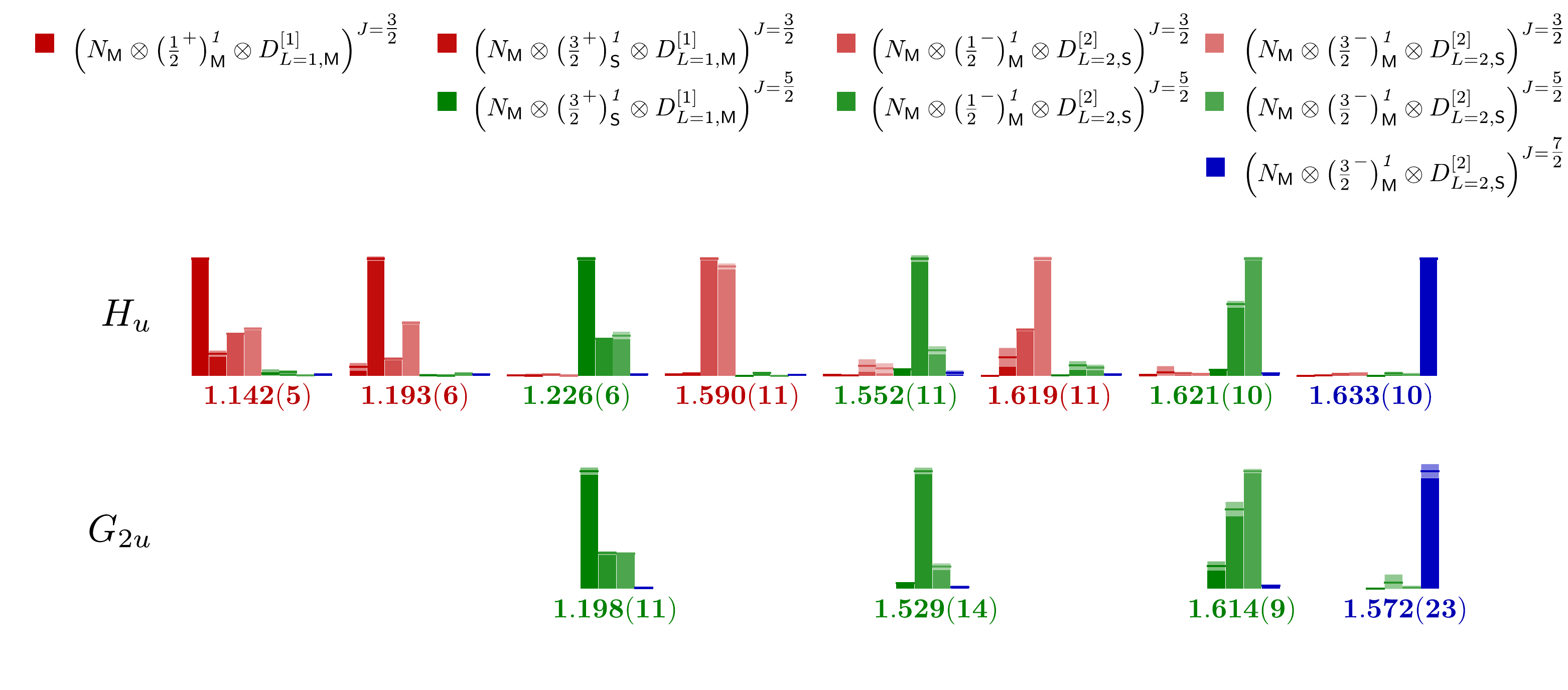}
\caption{Histograms of spectral overlaps, $Z$, are shown for a selection of eight operators 
(shown at the top using the naming convention of Eq.~(\ref{eq:ops_notation})). The 4 operators
labelled with $J=\frac{3}{2}$ (color red) have subductions only to the $H_u$ irrep.  The 3 operators 
labelled with $J=\frac{5}{2}$ (color green) and the 1 labelled with $J=\frac{7}{2}$ (color blue) have 
subductions to both $H_u$ and $G_{2u}$ irreps.  
Each histogram is labelled by the value of mass $m$ of the state (in units of $m_\Omega$) 
and has 8 vertical bars showing the relative $Z$ values for each of the operators.  
The data are from the $m_{\pi}$ = 524 MeV ensemble and $Z$'s are normalized so that the largest value across all
states, for a given operator, is equal to $1$. The lighter area at the head of each bar
represents the one sigma statistical uncertainly.  
Note that for each state only one or two operators have a large relative $Z$ value, and it is the 
same operators appearing across each of the irreps.
Note also that nearly the same energies are obtained in $H_u$ and $G_{2u}$ irreps
for state subduced from one $J$ value. 
\label{fig:histogram}}
\end{figure*}

To identify the spin of a state, we use the operator ``overlaps"  $Z^\mathfrak{n}_i = \langle \mathfrak{n} | {\cal O}_i^\dag |0\rangle$ for a given state extracted through the variational method presented in the previous section.
In Figure \ref{fig:histogram} we show the overlaps for a set of low-lying states in the Nucleon $H_u$ and $G_{2u}$ irreps of the $m_\pi = 524~{\rm MeV}$ $16^3$ calculation. The principal correlators for these states in $H_u$ are shown in Figure~\ref{fig:princorrfits}, and the corresponding mass spectrum is shown in Figure \ref{fig:nucleon_irrep_808}. The overlaps for a given state show a clear preference for overlap onto only operators of a single spin. 
While we show only a subset of the operators for clarity, the same pattern is observed for the full operator set.

\begin{figure*}[t]
 \centering
\includegraphics[width=0.95\textwidth,bb= 20 20 640 252]{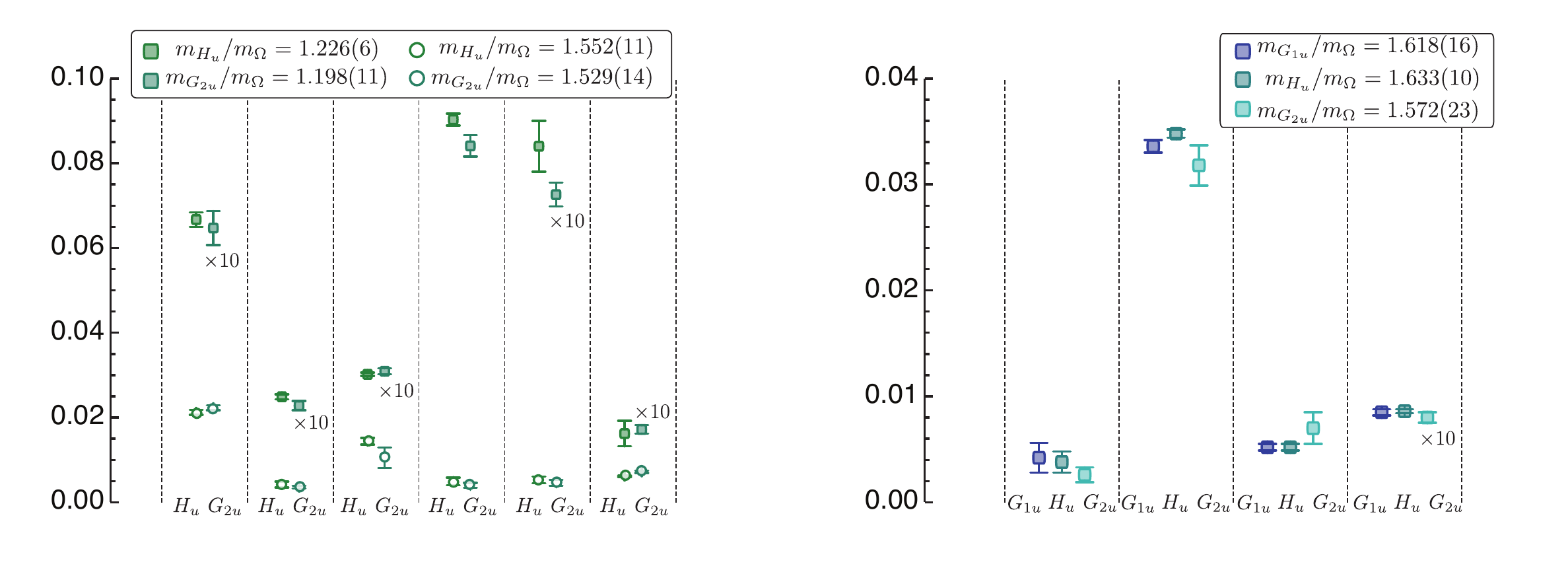}
 \caption{Selected $Z$ values across irreps $\Lambda_u$ are shown for states suspected of being
   $J=\frac{5}{2}$ (left panel) and $\frac{7}{2}$ (right panel), based on the $m_{\pi} $ = 
524 MeV ensemble. The boxes at the top show the mass for the various states. There are two states of $J=\frac{5}{2}^-$.
The operators in the left panel, all projected onto $J=\frac{5}{2}^-$, are left to right,
$N_\mathsf{M}\otimes(\tfrac{1}{2}^-)^\mathit{1}_\mathsf{M} \otimes  D^{[2]}_{L=2,\mathsf{S}}$,
$N_\mathsf{M}\otimes(\tfrac{1}{2}^-)^\mathit{1}_\mathsf{S} \otimes  D^{[2]}_{L=2,\mathsf{M}}$,
$N_\mathsf{M}\otimes(\tfrac{3}{2}^+)^\mathit{1}_\mathsf{M} \otimes  D^{[1]}_{L=1,\mathsf{M}}$,
$N_\mathsf{M}\otimes(\tfrac{3}{2}^+)^\mathit{1}_\mathsf{S} \otimes  D^{[1]}_{L=1,\mathsf{M}}$,
$N_\mathsf{M}\otimes(\tfrac{3}{2}^-)^\mathit{1}_\mathsf{M} \otimes  D^{[1]}_{L=2,\mathsf{S}}$,
$N_\mathsf{M}\otimes(\tfrac{3}{2}^-)^\mathit{1}_\mathsf{S} \otimes  D^{[1]}_{L=2,\mathsf{M}}$.
Overlaps of these operators after subduction into $H_u$ and $G_{2u}$, agree well for each of the two
states shown.
The operators in the right panel, all projected onto $J=\frac{7}{2}^-$, are left to right,
$N_\mathsf{M}\otimes(\tfrac{3}{2}^-)^\mathit{1}_\mathsf{M} \otimes  D^{[2]}_{L=2,\mathsf{M}}$,
$N_\mathsf{M}\otimes(\tfrac{3}{2}^-)^\mathit{1}_\mathsf{M} \otimes  D^{[2]}_{L=2,\mathsf{S}}$,
$N_\mathsf{M}\otimes(\tfrac{3}{2}^-)^\mathit{1}_\mathsf{S} \otimes  D^{[2]}_{L=2,\mathsf{M}}$,
$N_\mathsf{M}\otimes(\tfrac{3}{2}^-)^\mathit{2}_\mathsf{S} \otimes  D^{[2]}_{L=2,\mathsf{M}}$.
Similarly, the operator overlaps for this state agree well across $G_{1u}$, $H_u$, and $G_{2u}$.
}
 \label{fig:Zvalues}
\end{figure*}

The assignment of spin must hold for states with continuum spin $J$
subduced across multiple irreps.  In the continuum our operators are
of definite spin such that $\langle 0 | {\cal O}^{J,M\dag} |J',
M'\rangle = Z^{[J]} \delta_{J,J'} \delta_{M,M'}$, and therefore from
Eq.~\ref{eq:OJM} the overlap of the subduced operator is $\langle 0 |
{\cal O}^{[J]\dag}_{^n\!\Lambda,r} | J', M\rangle = {\cal
  S}^{J,M}_{^n\!\Lambda, r} Z^{[J]} \delta_{J,J'}$.  Only the spin $J$
states will contribute, and not any of the other spins present in the
irrep $^n\!\Lambda$.  Inserting a complete set of hadronic states
between the operators in the correlator and using the fact that the
subduction coefficients form an orthogonal
matrix~(\ref{eq:SJM_orthog}), $\sum_M {\cal S}^{J,M}_{\Lambda, r}
{\cal S}^{J,M}_{\Lambda', r'} = \delta_{\Lambda, \Lambda'} \delta_{r,
  r'}$, we thus obtain 
terms in the correlator spectral decomposition proportional to
$Z^{[J]*} Z^{[J]}$ for each $\Lambda$ we have subduced into, up to
discretization uncertainties as described above.  Hence, for example,
a $J=\frac{7}{2}$ baryon created by a $[J=\frac{7}{2}]$ operator will
have the same $Z$ value in each of the $G_1, H, G_2$ irreps. This
suggests that we compare the independently obtained $Z$-values in each
irrep.  In Figure \ref{fig:Zvalues} we show the extracted $Z$ values
for negative-parity states suspected of being spin $\frac{5}{2}$
across the $H_u$ and $G_{2u}$ irreps, and of being spin $\frac{7}{2}$ across the
$G_{1u}$, $H_u$ and $G_{2u}$ irreps.  As can be seen, there is good
agreement of $Z$ values in the different irreps that are subduced from
spin $J$, with only small deviations from exact equality.

These results demonstrate that the $Z$ values of carefully 
constructed subduced operators can be used to identify the continuum 
spin of states extracted in explicit computation for the 
lattices and operators we have used.

\begin{figure*}[t]
 \centering
\includegraphics[width=0.9\textwidth,angle=0]{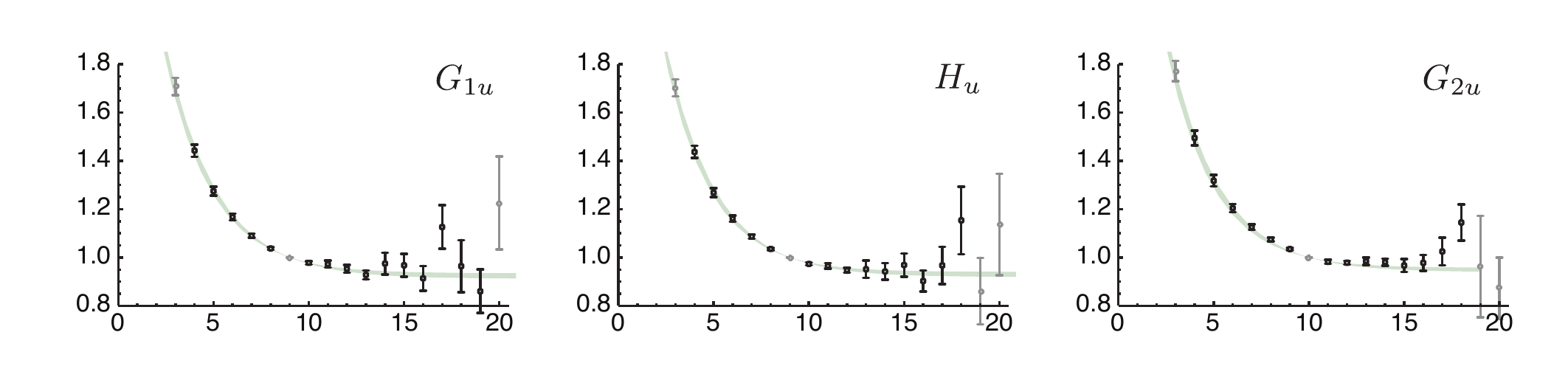}
 \caption{Fit to the three subduced principal correlators of lowest lying $\frac{7}{2}^-$ Nucleon using a common mass. 
Results are from the $m_{\pi} $ = 524 MeV ensemble. 
These levels correspond to the sixth excited $G_{1u}$, the seventh $H_u$ and the fourth $G_{2u}$. 
Plotted is $\lambda(t) \cdot e^{m(t-t_0)}$. Grey points not included in the fit.}
 \label{fig:fit_across_irreps}
\end{figure*}

We take the next step and use the identification of the 
components of the spin-$J$ baryon subduced across multiple irreps to make a 
best estimate of the mass of the state.
The mass values determined from fits to principal correlators in each irrep 
differ slightly due to what we assumed to be discretization effects and, in principle,
avoidable fitting variations (such as the fitting intervals).
We follow Ref.~\cite{Dudek:2010wm} and perform a joint fit to the principal 
correlators with the mass being common. This method provides a numerical test
that the state has been identified. We allow a differing second 
exponential in each principal correlator so that the fit parameters 
are $m_\mathfrak{n}, \{ m^{'\Lambda}_\mathfrak{n}\} $ and 
$\{A_\mathfrak{n}^\Lambda \}$. These fits are typically very successful 
with correlated $\chi^2/N_\mathrm{dof}$ close to 1, suggesting again that the
descretization effects are small. An example for 
the case of $\frac{7}{2}^-$ components identified in 
$G_{1u}, H_u, G_{2u}$ is shown in Figure \ref{fig:fit_across_irreps}. 
When we present our final, spin-assigned spectra it is the results of 
such fits that we show.

\subsection{Additional demonstration of approximate rotational invariance}

\begin{figure}
 \centering
\includegraphics[width=0.5\textwidth, bb=20 30 420 396]{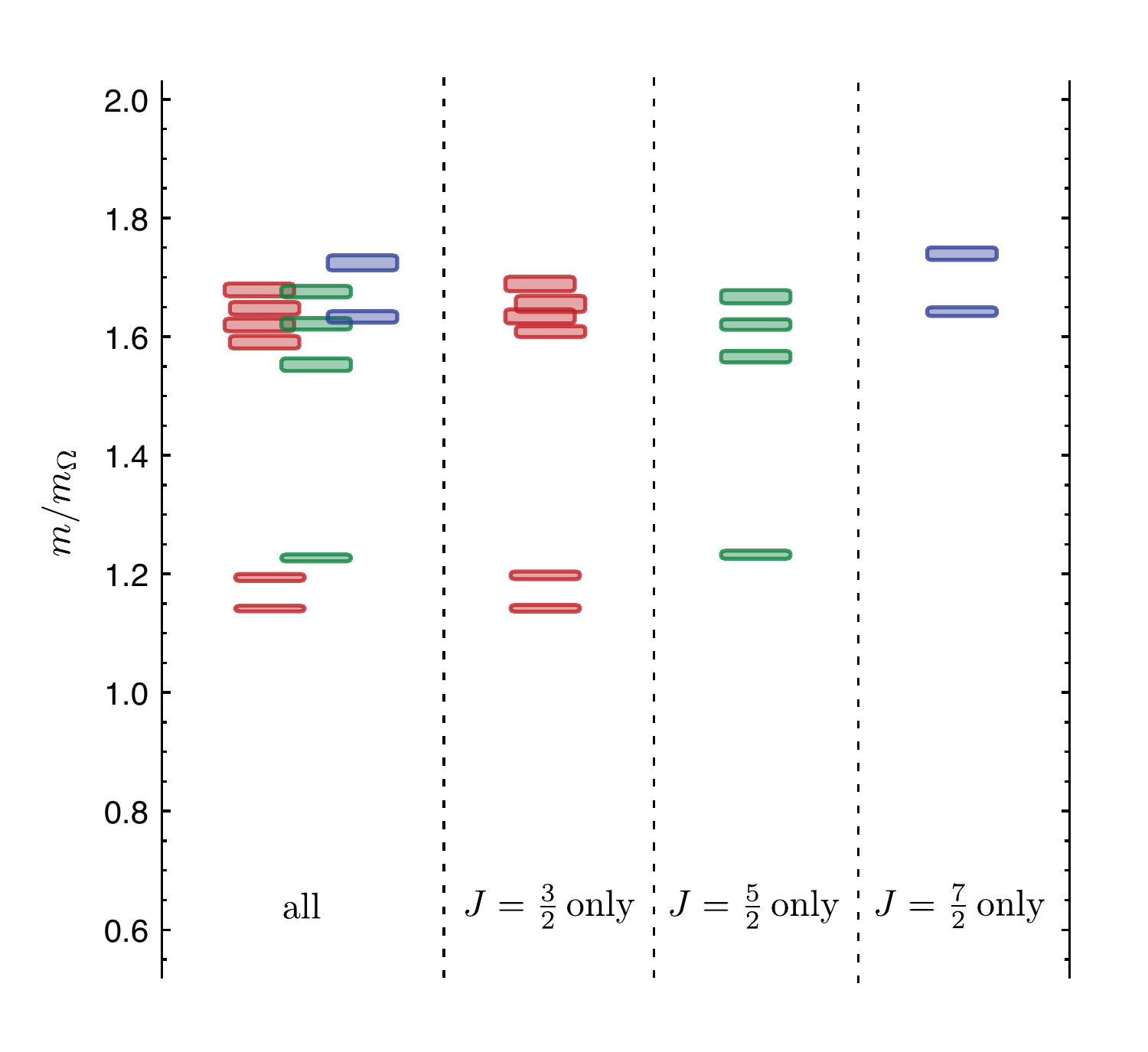}
\caption{Extracted Nucleon $H_u$ mass spectrum for various operator bases. (a) Full basis (dim = 48), (b) Only $J=\frac{3}{2}$ operators (dim = 28), (c) Only $J=\frac{5}{2}$ operators  (dim = 16), (d) Only $J=\frac{7}{2}$ operators (dim = 4).
Results are from the $m_{\pi} $ = 524 MeV ensemble. 
The dimensionality of the operator basis in each $J$ is shown in Table~\ref{tab:nuc_ops2}.
}
\label{fig:vary_ops}
\end{figure}

A further demonstration of approximate rotational invariance is based on the
spectrum of energy levels. If indeed the mixing between states of different continuum spin $J$ is
small, then the omission of such coupling should not much affect the excited state spectra. 
That proposition can be tested  by extracting energies using all operators, 
and comparing them with the energies obtained from only operators subduced 
from a single $J$ value.
If approximate rotational invariance were achieved in the spectrum, the 
energies would be nearly the same.  
As an example, we show results for the Nucleon $H_u$ irrep, in Fig.~\ref{fig:vary_ops}. The left column of
the Fig.~\ref{fig:vary_ops}, labelled ``all'', shows the lowest 12 energy levels obtained from matrices 
of correlation functions using the set of all 48 $H_u$ operators, and spin identified using the methods
previously described. The states listed in Figure~\ref{fig:histogram} correspond to a few of these 12 levels.
The second column shows the lowest 6 levels resulting from the variational method when the operator basis is restricted to only those with continuum spin $J=\frac{3}{2}$.
Similarly, the third shows the lowest 4 levels when the operator basis is restricted to only those with continuum spin $J=\frac{5}{2}$, and the last column are the lowest two levels when the basis is restricted to only $J=\frac{7}{2}$ operators.

The results are striking. 
We see that the masses of the levels in each of the restricted bases agree quite well with the results
found in the full basis. 
The agreement is quite remarkable because one expects 
operators in the $H_u$ irrep that can couple with the ground state (the lowest 
$J=\frac{3}{2}$ state), will allow for the rapid decay of correlators down to the ground state as
a function of time, $t$.  However, the 
higher-energy spin $\frac{5}{2}$ and $\frac{7}{2}$ states do not show such a 
decay; we obtain good plateaus in plots like those in Fig.~\ref{fig:princorrfits}.  
These results provide rather striking demonstration for the lack of significant rotational symmetry
breaking in the spectrum.

%% file: stability.tex
%\section{Stability of spectrum extraction}\label{sec:stability}

\begin{figure}
 \centering
\includegraphics[width=0.5\textwidth,bb=20 0 385 274]{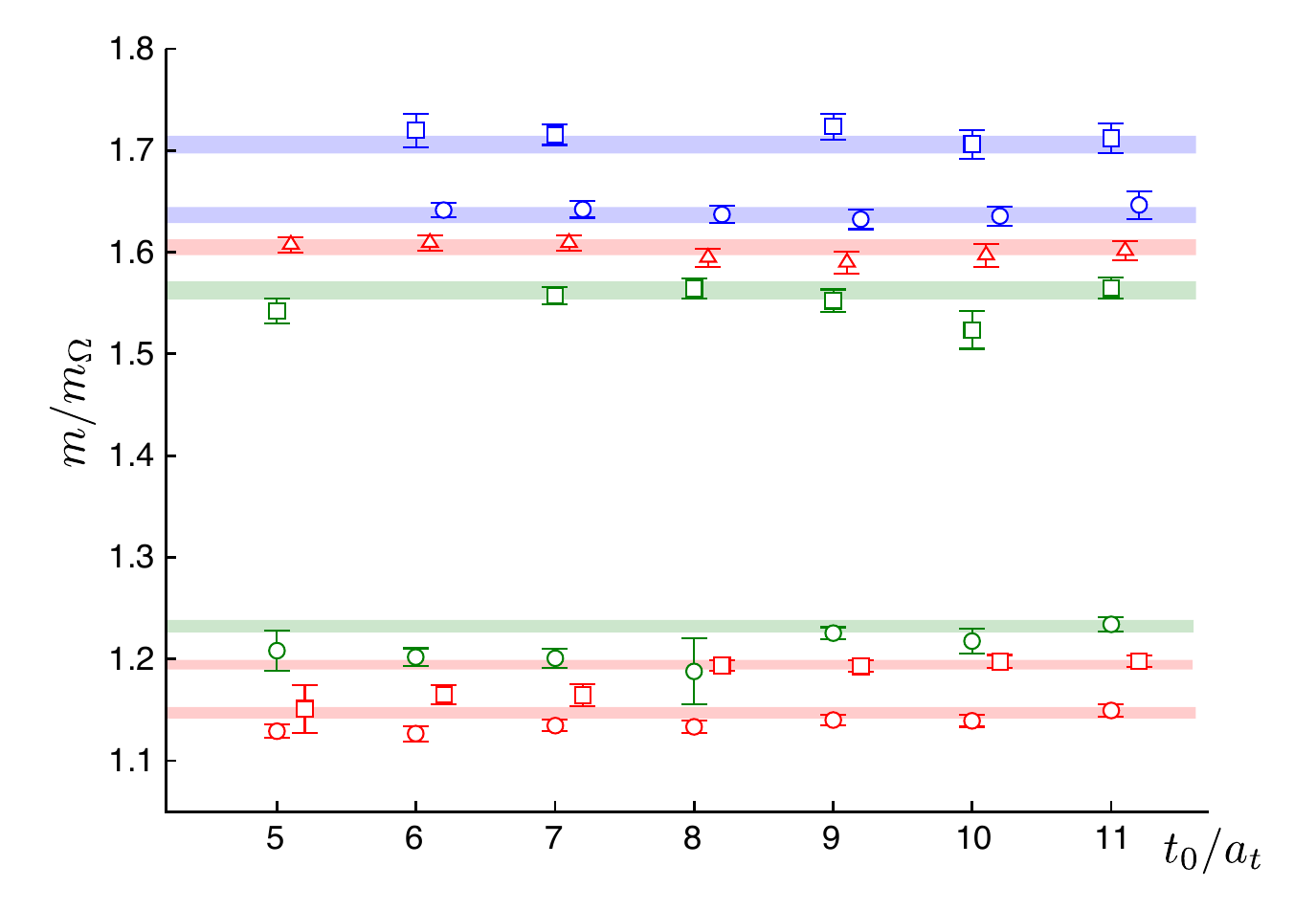}
\caption{Extracted Nucleon $H_u$ mass spectrum as a function of $t_0$. Horizontal bands to guide the eye.
For clarity of presentation, only one of the highly excited $J=\frac{3}{2}^-$ and $J=\frac{5}{2}^-$ levels are shown.
} 
\label{fig:masses_vs_t0}
\end{figure}

In this section we consider to what extent the extracted spectrum changes as we 
vary details of the calculation, such as the metric timeslice, $t_0$, used in the variational analysis, 
and the number of distillation vectors.
We will use the Nucleon $H_u$ and the Delta $H_g$ irreps in the $m_\pi = 524~{\rm MeV}$ dataset to demonstrate our findings.

\subsection{Variational analysis and $t_0$}\label{sec:t0}
Our fitting methodology was described in Section \ref{sec:fitting} where reconstruction of the correlator was used to guide us to an appropriate value of $t_0$. As seen in Figure \ref{fig:masses_vs_t0}, for $t_0 \gtrsim 7$, the low-lying mass spectrum is quite stable with
respect to variations of $t_0$. This appears to be mostly due to the inclusion of 
a second exponential term in Eq.~(\ref{eq:fitprincorr}), which is able to absorb much 
of the effect of states outside the diagonalization space.
The contribution of this second exponential typically falls rapidly with increasing $t_0$ both by having a smaller $A$ and a larger $m'$.

Overlaps, $Z^i_\mathfrak{n} = \big\langle \mathfrak{n} \big| {\cal O}_i^\dag \big| 0 \big\rangle$, 
can show more of a sensitivity to $t_0$ values being too low, as was found also 
in the analysis of mesons, Ref.~\cite{Dudek:2010wm}.

In summary it appears that variational fitting is reliable provided $t_0$ is ``large enough". Using two-exponential fits in principal correlators we observe relatively small $t_0$ dependence of masses, but more significant dependence for the $Z$ values which we require for spin-identification.

%\begin{table}[t]
%\begin{tabular}{r|l}
%black & $J=\frac{1}{2}$\\
%red   & $J=\frac{3}{2}$\\
%green & $J=\frac{5}{2}$\\
%blue  & $J=\frac{7}{2}$\\
%\hline
%orange & undetermined $J$ \\
%\end{tabular}
%\caption{Color-coding used in spectrum irrep plots.\label{tab:colors}}
%\end{table}

\begin{figure}
 \centering
\includegraphics[width=0.50\textwidth,angle=0]{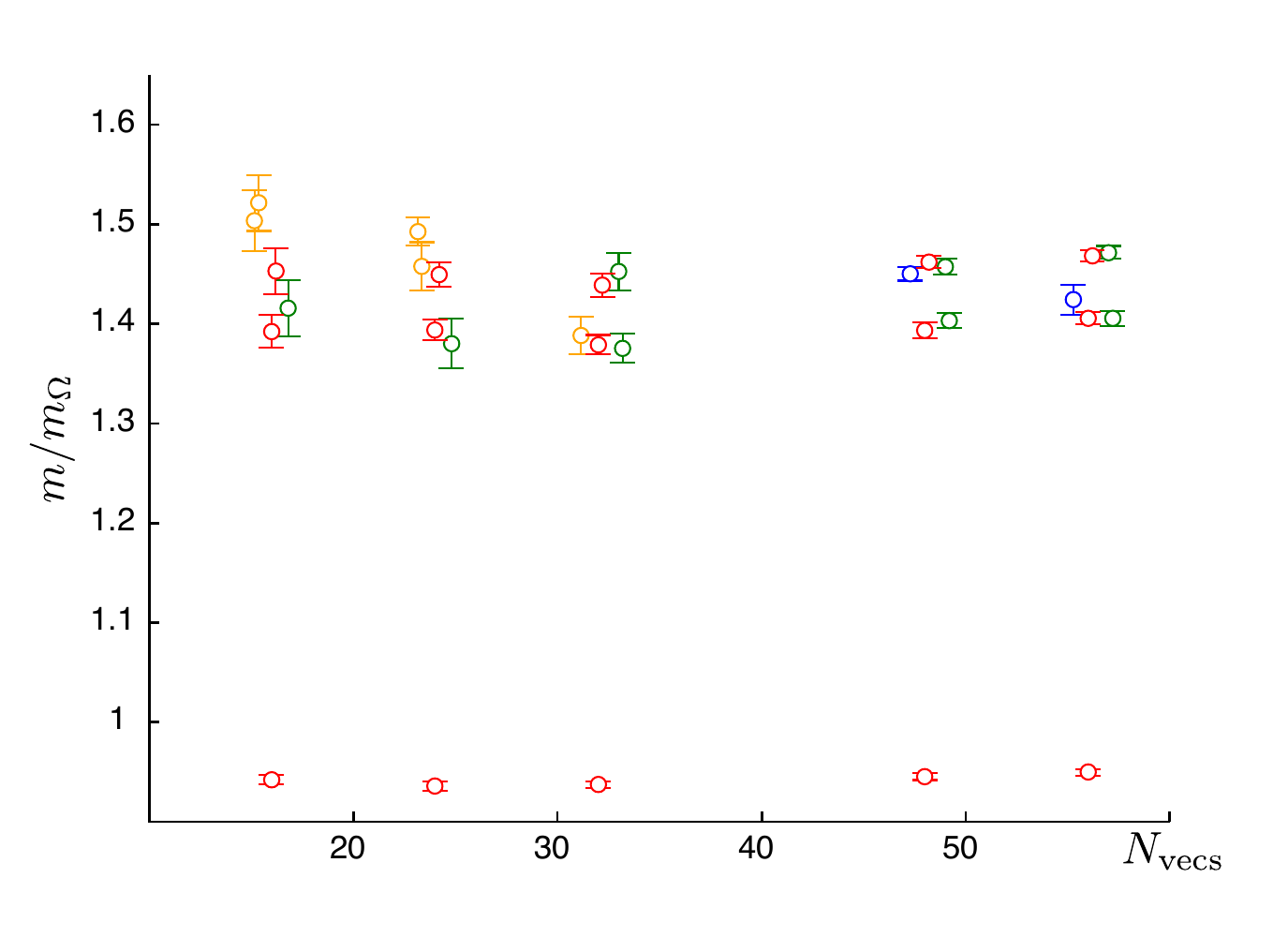}
\caption{Extracted the Delta $H_g$ mass spectrum as a function of number of distillation vectors in the
$m_\pi = 524~{\rm MeV}$ dataset. For $N\lesssim 32$, spin identification is lost for the highest $J=\frac{5}{2}^+$ and $J=\frac{7}{2}^+$ levels.}
\label{fig:masses_vs_vecs}
\end{figure}

\subsection{Number of distillation vectors}\label{sec:nvecs}
The results presented so far are based on the analysis of correlators computed on
$16^3$ lattices using 56 distillation vectors. We consider how the
determination of the spectrum varies if one reduces the number of distillation
vectors and thus reduces the computational cost of the calculation. This is
particularly important given that, as shown in \cite{Peardon:2009gh}, to get the
same smearing operator on larger volumes one must scale up the number of
distillation vectors by a factor equal to the ratio of spatial volumes. To scale
up to a $32^3$ lattice this would require $56 \times
\left(\tfrac{32}{16}\right)^3 = 448$ vectors which is not currently a realizable
number without using stochastic estimation \cite{Morningstar:2011ka}.

In figure \ref{fig:masses_vs_vecs} we show the low-lying part of the extracted Delta $H_g$ spectrum on the $m_\pi = 524~{\rm MeV}$ lattice as a function of the number of distillation vectors used in the correlator construction. It is clear that the spectrum is reasonably stable for $N \gtrsim 32$ but that the spectrum quality degrades rapidly for fewer vectors. In particular, the methods for spin identification fail for the highest $J^P=\frac{5}{2}^+$ state as well as the first $J^P=\frac{7}{2}^+$ state.

The need for a large numbers of distillation vectors has been discussed in Ref.~\cite{Dudek:2010wm} for the case of isovector mesons. The conclusion drawn is that to describe high spin hadrons having large orbital angular momentum, one needs to include sufficient vectors to sample the rapid angular dependence of the wavefunction over the typical size of a hadron. The results for baryons presented here are consistent with these observations.

In summary one is limited as to how few distillation vectors can be used if one requires reliable extraction of high-spin states. The results shown here suggest 32 distillation vectors on a $16^3$ lattice is the minimum, so at least 64 distillation vectors on a $20^3$ lattice are likely to be required.

%% file: results.tex
%\section{Results}\label{sec:results}

Our results are obtained on $16^3\times 128$ lattices with pion masses
between 396 and 524~MeV.  More complete details of the number of
configurations, time sources and distillation vectors used are given
in Table~\ref{tab:lattices}. The full basis of operators in each irrep listed in
Table~\ref{tab:opnumbers} is used for the variational method
construction.

\subsection{$m_\pi  = 524$ MeV results}
\input{phenom}

\subsection{Quark mass dependence}\label{sec:quark_mass}
\input{quark_mass}

\subsection{Comparisons}\label{sec:comparisons}
\input{comparison}

%% file: phenom.tex
\begin{figure}[h!]
\centering
\includegraphics[width=0.55\textwidth, bb=35 20 600 288]{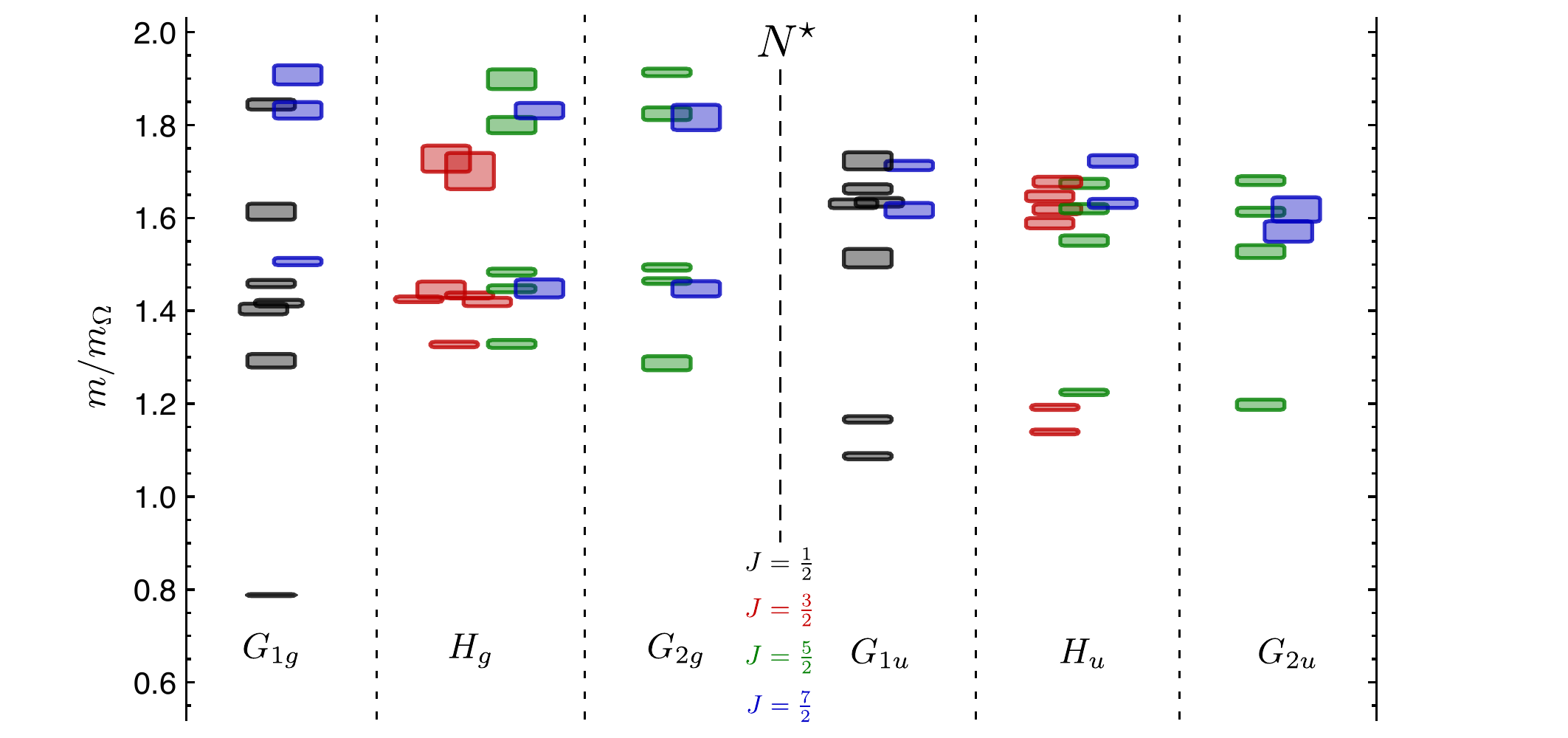}
\caption{Extracted Nucleon spectra by irrep for $m_\pi = 524~{\rm
    MeV}$. Colors are black ($J=\frac{1}{2}$), red ($\frac{3}{2}$),
  green ($\frac{5}{2}$), blue ($\frac{7}{2}$).  Masses are shown in
  ratios of the $\Omega$ baryon mass.}
\label{fig:nucleon_irrep_808}
\end{figure}
\begin{figure}[h!]
\centering
\includegraphics[width=0.55\textwidth, bb=35 20 600 288]{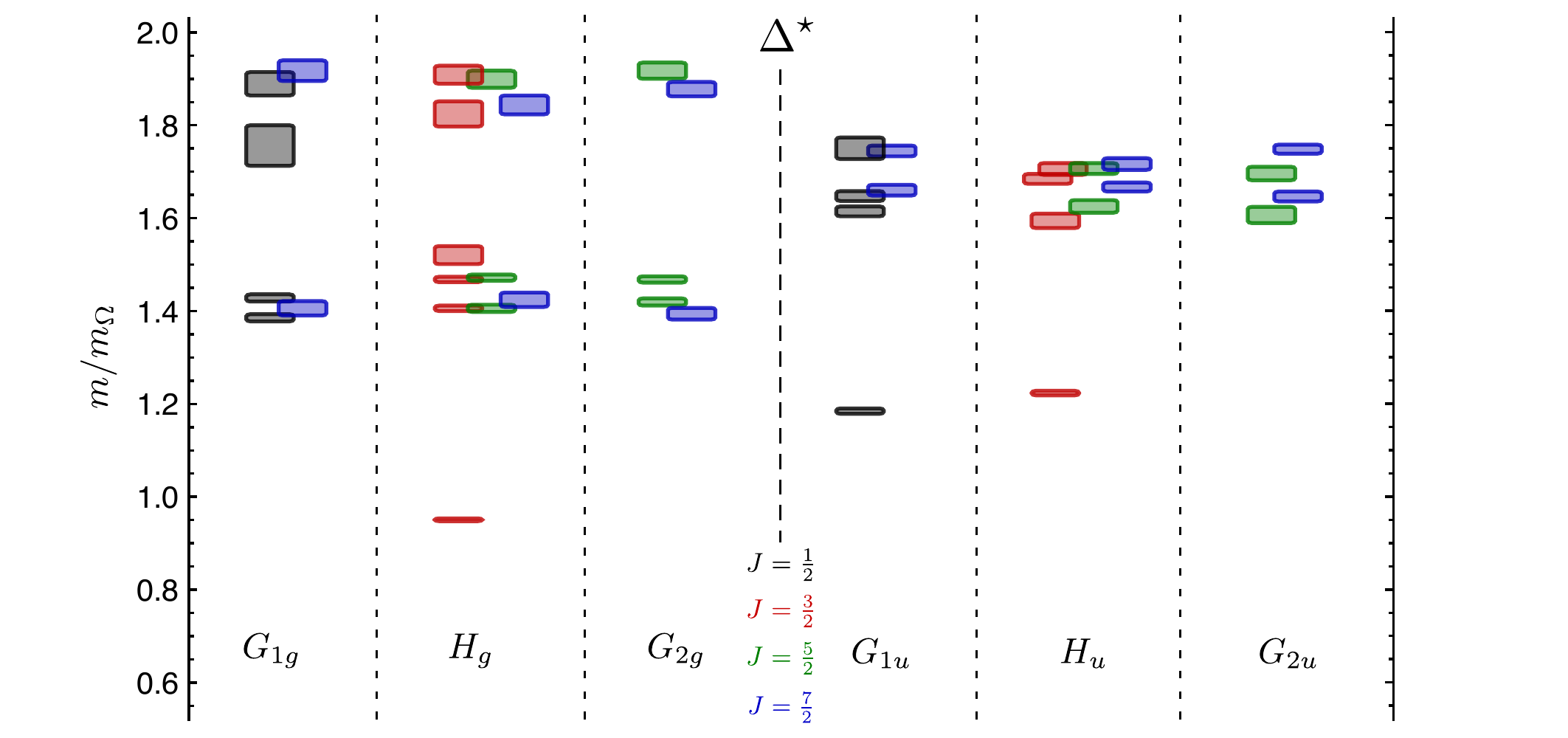}
\caption{Extracted Delta spectra by irrep for $m_\pi = 524~{\rm
    MeV}$. Colors are black ($J=\frac{1}{2}$), red ($\frac{3}{2}$),
  green ($\frac{5}{2}$), blue ($\frac{7}{2}$).  Masses are shown in
  ratios of the $\Omega$ baryon mass.}
\label{fig:delta_irrep_808}
\end{figure}
\begin{figure*}[ht]
\begin{center}
\includegraphics[width=0.8\textwidth]{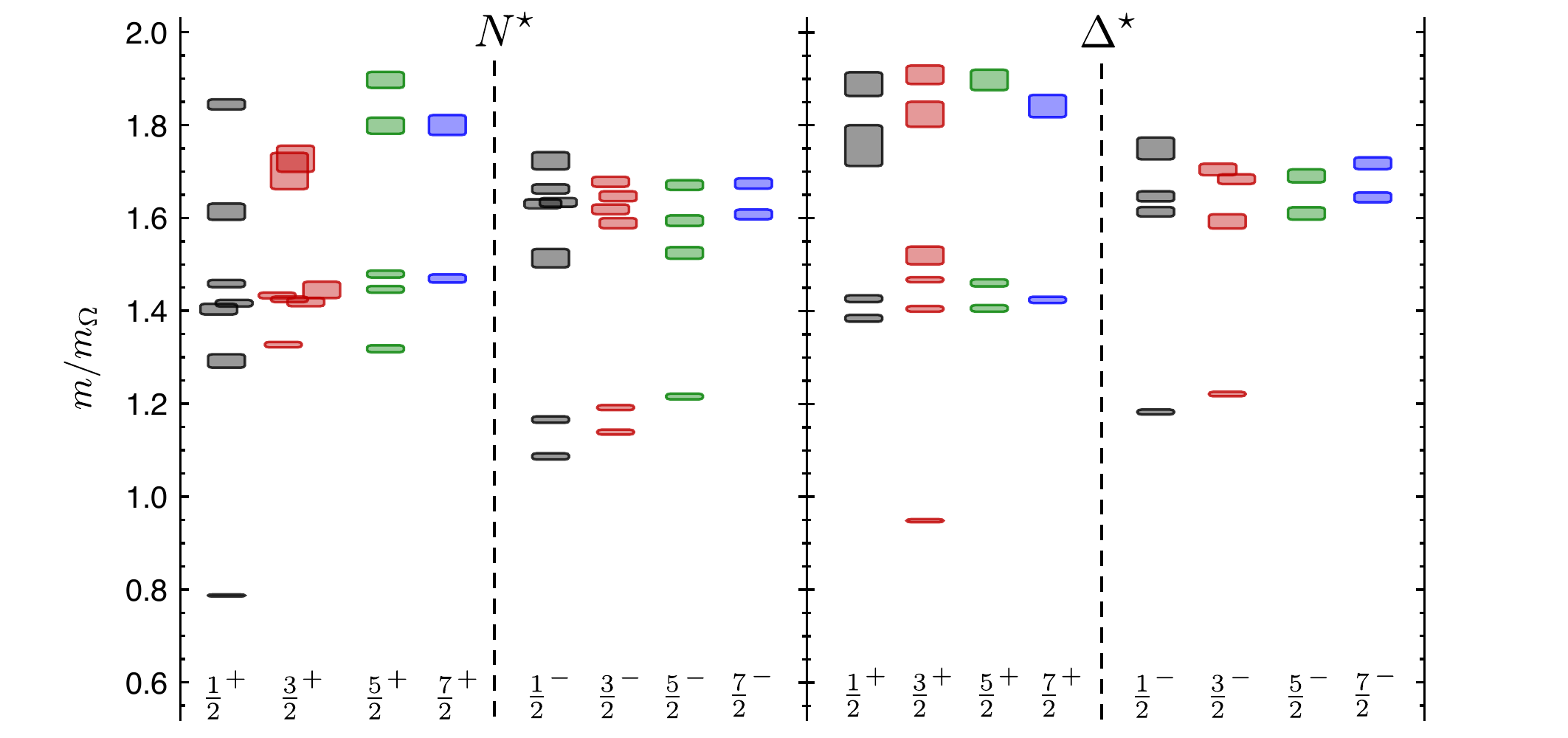}
\caption{Spin-identified spectrum of Nucleons and Deltas from the
  lattices at $m_\pi = 524~{\rm MeV}$, in units of the calculated
  $\Omega$ mass. \label{fig:spin_808}
}
\end{center}
\end{figure*}

Using the variational method outlined in Section~\ref{sec:fitting},
the spectrum of energies in each lattice irrep for the Nucleon and for
the Delta are shown in Figures~\ref{fig:nucleon_irrep_808} and
\ref{fig:delta_irrep_808}, respectively.  By applying the
spin-identification procedure described above, we can subsequently
associate continuum $J^P$ labels to each of these states, as labelled
on the figures.  For the remainder of this paper, we will therefore
label the energies by their assigned continuum spins; these are shown
for the lattices at $m_\pi = 524~{\rm MeV}$ in
Figure~\ref{fig:spin_808}, with the energies obtained from a joint fit
to the principal correlators as described in
Section~\ref{sec:fitting}.

\begin{figure*}
\begin{center}
\includegraphics[width=0.8\textwidth,angle=0]{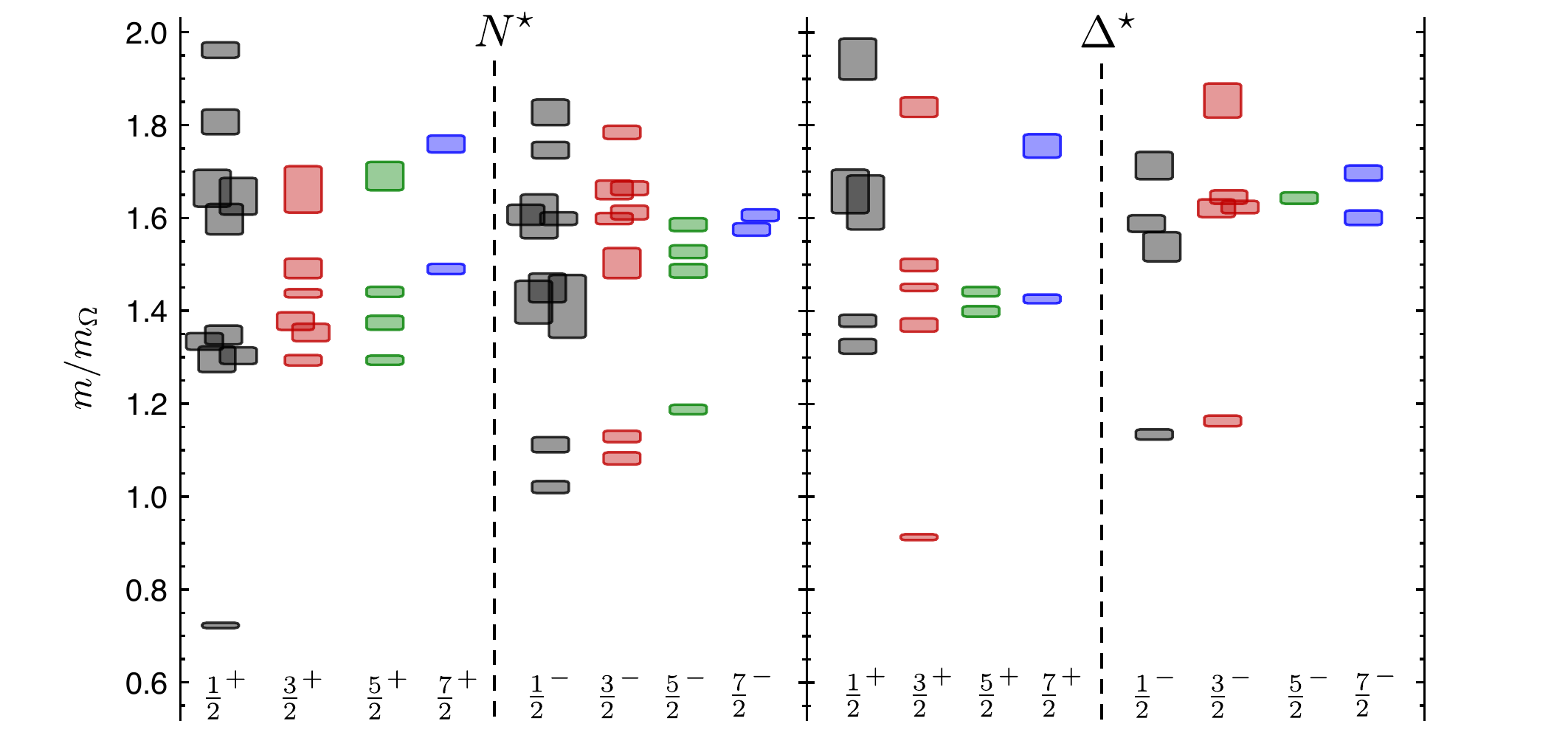}
\caption{Spin-identified spectrum of Nucleons and Deltas from the
  lattices at $m_\pi = 396~{\rm MeV}$, in units of the calculated
  $\Omega$ mass. \label{fig:spin_840}
}
\end{center}
\end{figure*}

There are several notable features in these spectra. As we will discuss,
the patterns of states have a good correspondence with single-hadron states as classified by $SU(6)\otimes O(3)$ symmetry. 
The numbers of low-lying states in each $J^P$ are similar to the numbers obtained in the non-relativistic quark model which is a particular realisation of the symmetry above (e.g., \cite{Isgur:1978xj,Isgur:1978wd}). 
For the purposes of these comparisons, it is helpful to introduce a spectroscopic notation: $X \,^{2S+1}L_\pi J^P$, where $X$ is the 
Nucleon $N$ or the Delta $\Delta$, $S$ is the Dirac spin, $L=S$, $P$, $D$,\ldots denotes the 
combined angular momentum of the derivatives, $\pi=\mathsf{S}$, $\mathsf{M}$, or $\mathsf{A}$ is the permutational symmetry of the derivative, and $J^P$ is the total angular momentum and parity. This notation also is used in Table~\ref{tab:qm}, which we discuss now.

\begin{table}
\begin{tabular}{r|ll||ll||c}
\multicolumn{3}{c||}{Nucleon ($\bf{8}$)} & \multicolumn{2}{c||}{$SU(6)\otimes O(3)$} & $n$ \\
\hline
%                    & Nucleon & Octet & $SU(6)\otimes$&$O(3)$ &~ n~ \\
\hline
$J^P=\frac{1}{2}^-$ & $N \,^2\! P_\mathsf{M} \frac{1}{2}^-$ & $N \,^4\! P_\mathsf{M} \frac{1}{2}^-$ & $[\mathbf{70},1^-]$ & $[\mathbf{70},1^-]$ & 2 \\
\hline
$J^P=\frac{3}{2}^-$ & $N \,^2\! P_\mathsf{M} \frac{3}{2}^-$ & $N \,^4\! P_\mathsf{M} \frac{3}{2}^-$ & $[\mathbf{70},1^-]$ & $[\mathbf{70},1^-]$ & 2 \\
\hline
$J^P=\frac{5}{2}^-$ &                                & $N \,^4\! P_\mathsf{M} \frac{5}{2}^-$ & & $[\mathbf{70},1^-]$ & 1 \\
\hline
\hline
$J^P=\frac{1}{2}^+$ & $N \,^2\! S_\mathsf{S} \frac{1}{2}^+$ & $N \,^4\! D_\mathsf{M} \frac{1}{2}^+$ & $[\mathbf{56},0^+]$  & $[\mathbf{70},2^+]$ & \multirow{3}{*}{4} \\
                   & $N \,^2\! S_\mathsf{M} \frac{1}{2}^+$ & & $[\mathbf{70},0^+]$ & &  \\
                   & $N \,^2\! P_\mathsf{A} \frac{1}{2}^+$ & & $[\mathbf{20},1^+]$ & &  \\
\hline
$J^P=\frac{3}{2}^+$ & $N \,^2\! D_\mathsf{S} \frac{3}{2}^+$ & $N \,^4\! S_\mathsf{M} \frac{3}{2}^+$ & $[\mathbf{56},2^+]$ & $[\mathbf{70},0^+]$ & \multirow{3}{*}{5} \\
                   & $N \,^2\! D_\mathsf{M} \frac{3}{2}^+$ & $N \,^4\! D_\mathsf{M} \frac{3}{2}^+$ & $[\mathbf{70},2^+]$ & $[\mathbf{70},2^+]$ &  \\
                   & $N \,^2\! P_\mathsf{A} \frac{3}{2}^+$ & &$[\mathbf{20},1^+]$ & &  \\
\hline
$J^P=\frac{5}{2}^+$ & $N \,^2\! D_\mathsf{S} \frac{5}{2}^+$ & $N \,^4\! D_\mathsf{M} \frac{5}{2}^+$ & $[\mathbf{56},2^+]$  & $[\mathbf{70},2^+]$ & \multirow{2}{*}{3} \\
                   & $N \,^2\! D_\mathsf{M} \frac{5}{2}^+$ & &$ [\mathbf{70},2^+]$ & & \\
\hline
$J^P=\frac{7}{2}^+$ & & $N \,^4\! D_\mathsf{M} \frac{7}{2}^+$ &$[\mathbf{70},2^+]$ & & 1 \\
\hline
\multicolumn{6}{c}{}\\

\multicolumn{3}{c||}{Delta ($\bf{10}$)} & \multicolumn{2}{c||}{$SU(6)\otimes O(3)$} & $n$ \\
\hline
%                    & Delta & Decuplet & $SU(6)\otimes$&$O(3)$ \\
\hline
$J^P=\frac{1}{2}^-$ &  $\Delta \,^2\! P_\mathsf{M} \frac{1}{2}^-$ & & $[\mathbf{70},1^-]$ & & 1 \\
\hline
$J^P=\frac{3}{2}^-$ &  $\Delta \,^2\! P_\mathsf{M} \frac{3}{2}^-$ & & $[\mathbf{70},1^-]$ & & 1 \\
\hline
$J^P=\frac{5}{2}^-$ &  &   &   &  & 0 \\
\hline
\hline
$J^P=\frac{1}{2}^+$ & $\Delta \,^2\! S_\mathsf{M} \frac{1}{2}^+$ & $\Delta \,^4\! D_\mathsf{S} \frac{1}{2}^+$ & $[\mathbf{70},0^+]$ & $[\mathbf{56},2^+]$ & 2  \\
\hline
$J^P=\frac{3}{2}^+$ & $\Delta \,^2\! D_\mathsf{M} \frac{3}{2}^+$ & $\Delta \,^4\! S_\mathsf{S} \frac{3}{2}^+$ & $[\mathbf{70},2^+]$ & $[\mathbf{56},0^+]$ & \multirow{2}{*}{3} \\
                   &  &  $\Delta \,^4\! D_\mathsf{S} \frac{3}{2}^+$  & & $[\mathbf{56},2^+]$ & \\
\hline
$J^P=\frac{5}{2}^+$ &  $\Delta \,^2\! D_\mathsf{M} \frac{5}{2}^+$ & $\Delta \,^4\! D_\mathsf{S} \frac{5}{2}^+$ &  $[\mathbf{70},2^+]$ & $[\mathbf{56},2^+]$ & 2 \\
\hline
$J^P=\frac{7}{2}^+$ & & $\Delta \,^4\! D_\mathsf{S} \frac{7}{2}^+$ &  & $[\mathbf{56},2^+]$ & 1  \\
\hline
\end{tabular}  

\caption{Subset of the operator basis classified by $SU(6)\otimes O(3)$ multiplets and total spin and parity $J^P$. The entries are the orbital angular momentum structures outlined in Section~\ref{sec:ops} and Appendix~\ref{sec:symm_states} that contribute within each $J^P$.
The operators listed here are all from the first embedding in Dirac spin in Table~\ref{tab:dirac_spin}, and correspond to only upper components (referred to as ``non-relativistic'').
A spectroscopic notation of 
 $X \,^{2S+1}L_{\mathsf{\pi}} J^P$ is used, where $X=N$ or $\Delta$, $S$ is the Dirac spin, $L=S$, $P$, $D$,\ldots is the combined angular momentum of the derivatives, $\pi=\mathsf{S}$, $\mathsf{M}$, or $\mathsf{A}$ is the permutational symmetry of the derivatives, and $J^P$ is the total angular momentum and parity.
In Sections~\ref{sec:ops}-\ref{sec:spin}, an operator notation like $\left(N_\mathsf{M}\otimes \big(\frac{3}{2}^+\big)_\mathsf{S}^{\mathit{1}}\otimes D^{[1]}_{L=1,\mathsf{M}}\right)^{J=\tfrac{5}{2}}$ was used which in this spectroscopic notation would be $N \,^4\! P_\mathsf{M}\frac{5}{2}^-$.
Dimensions and parities of $SU(6)\otimes O(3)$ representations are listed in column 4 for the 
doublet spin states of column 2, and in column 5 for the quartet spin states of column 3.
The number, $n$, of operators for each $J^P$ is listed in the rightmost column.  
This same spectroscopic notation and classification of spatial structure is also used for comparisons with models where $L$ represents the orbital angular momentum.
\label{tab:qm}
}
\end{table}

In the negative parity $N^*$ spectrum, there is a pattern of five low-lying levels, consisting of two $N\tfrac{1}{2}^-$ levels, 
two $N\tfrac{3}{2}^-$ levels, and one $N\tfrac{5}{2}^-$ level. The triplet of higher levels in this group
of five is nearly degenerate with a pair of $\Delta\tfrac{1}{2}^-$ and $\Delta\tfrac{3}{2}^-$ levels. This pattern of 
Nucleon and Delta levels is consistent with an $L=1^{-}$ $P$-wave spatial structure with mixed symmetry, $P_\mathsf{M}$. 
As shown in Table~\ref{tab:qm}, the same numbers of states are obtained in the $SU(6)\otimes O(3)$ classification 
for the negative-parity Nucleon and Delta states constructed from the ``non-relativistic'' Pauli spinors  
as we find in the lattice spectra.
The lowest two $N^{*-}$ states are dominated by operators constructed in the notation of Eq.~\ref{eq:ops_notation} as 
$N_\mathsf{M}\otimes (S=\tfrac{1}{2}^+)_\mathsf{M} \otimes (L=1^-)_\mathsf{M}\rightarrow J^P= \tfrac{1}{2}^-$ and $\tfrac{3}{2}^-$, 
while the three higher $N^{*-}$ levels are dominated by operators constructed according to 
$N_\mathsf{M}\otimes (S=\tfrac{3}{2}^+)_\mathsf{S}\otimes (L=1^-)_\mathsf{M}$ with $J^P=\tfrac{1}{2}^-$, $\tfrac{3}{2}^-$ and $\tfrac{5}{2}^-$. Similarly, the low-lying Delta levels are consistent with a $\Delta\tfrac{1}{2}^-$ and $\Delta\tfrac{3}{2}^-$ assignment. 
There are no low-lying negative-parity $S=\tfrac{3}{2}$ Delta states since a totally symmetric state (up to antisymmetry in color) cannot be formed. 
Consequently, there is no low-lying $\Delta\tfrac{5}{2}^-$, which agrees with the lattice spectrum. 
In the non-relativistic quark model~\cite{Isgur:1978xj}, a hyperfine contact term is introduced to split the doublet and quartet states up and down, respectively, compared to 
unperturbed levels and the tensor part of the interaction provides some additional splitting. 
The result is that the doublet Delta states are nearly 
degenerate with the quartet Nucleon states as is observed in the lattice spectra, Fig.~\ref{fig:spin_808}.
In the language of $SU(6)\otimes O(3)$, these low-lying $N$ and $\Delta$ states constitute the strangeness zero part of a $[\mathbf{70},1^-]$ multiplet, as indicated
in Table~\ref{tab:qm}.

\begin{figure}
\begin{center}
\includegraphics[width=0.45\textwidth]{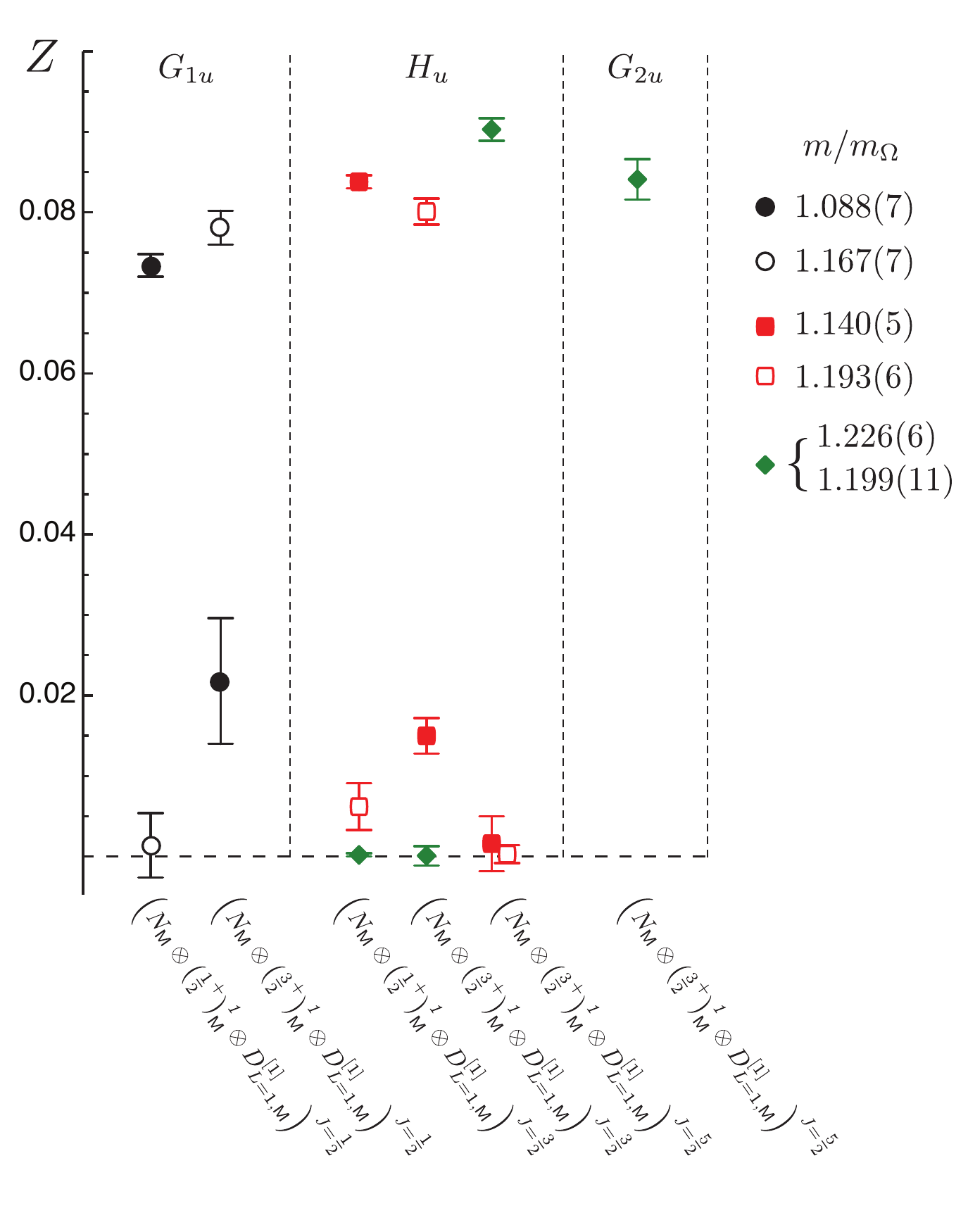}
\caption{Spectral overlaps for the negative parity Nucleon operators
  in Table~\ref{tab:qm} within the lowest lying states for the $m_\pi
  = 524$ MeV ensemble. The operators are subduced into the $G_{1u}$, $H_u$ and $G_{2u}$ irreps, and their spectral overlaps are shown for the lowest lying states with the masses listed in units of the $\Omega$ baryon mass. In the spectroscopic notation, these are the $N \,^2\! P_\mathsf{M}\frac{1}{2}^-$ and $N \,^4\! P_\mathsf{M}\frac{1}{2}^-$ subduced into $G_{1u}$, the $N\,^2\! P_\mathsf{M}\frac{3}{2}^-$ and $N\,^4\! P_\mathsf{M}\frac{5}{2}^-$ subduced into $H_u$, and $N\,^4\! P_\mathsf{M}\frac{5}{2}^-$ subduced into both $H_u$ and $G_{2u}$. As can be seen, for each level one operator is dominant. \label{fig:nonrel}
}
\end{center}
\end{figure}

In the positive parity sector, there also are interpretable patterns of lattice states in the range $m/m_\Omega\sim 1.3 - 1.5$. There 
are four $N\tfrac{1}{2}^+$ levels, five $N\tfrac{3}{2}^+$ levels, three $N\tfrac{5}{2}^+$ levels and one $N\tfrac{7}{2}^+$ level, which 
are the same numbers of levels for each $J^P$ as in Table~\ref{tab:qm}. 
The lattice spectra also have two $\Delta\tfrac{1}{2}^+$ levels, three $\Delta\tfrac{3}{2}^+$ levels, two $\Delta\tfrac{5}{2}^+$ levels, and 
one $\Delta\tfrac{7}{2}^+$ level, which are the same numbers of levels for each $J^P$ as in Table~\ref{tab:qm}.
In this case we are considering the multiplets $[\mathbf{70},0^+]$, $[\mathbf{56},2^+]$, $[\mathbf{70},2^+]$,  $[\mathbf{20},1^+]$ and a radially excited $[\mathbf{56},0^+]$ and within non-relativistic $qqq$ constituent quark models~\cite{Isgur:1978wd}, the mass eigenstates are admixtures of these basis states.

In general for both the Nucleon and Delta spectrum, there are reasonably well-separated bands of levels across the range of $J$ values, 
alternating in parity, with each band higher in energy than the previous one. 
We remark that there are no obvious patterns of degenerate levels with opposite parities for the same total spin, $J$ as in Ref.~\cite{Glozman:1999tk}.

The discussion up to now has focused on observables; namely, the level energies. 
More information about the internal structure of the lattice states can be obtained by analyzing the spectral overlaps. 
We remind the reader that the full basis of operators listed in Table~\ref{tab:opnumbers} is used within the variational method.
We find that only a few operators have large overlaps in the lowest negative-parity 
Nucleon levels of the $G_{1u}$, $H_u$ and $G_{2u}$ irreps, and these are the subduced versions of ``non-relativistic'' operators. From the 
construction of operators in Appendix~\ref{sec:symm_states}, those featuring only upper components in spin are in the 
first embedding of Dirac spin in Table~\ref{tab:dirac_spin}. There is a one-to-one correspondence between these operators, 
and those listed in Table~\ref{tab:qm}, and we will adopt the spectroscopic names as a shorthand.
This spectroscopic notation is used in Table~\ref{tab:qm} to identify operators, but is also applicable for identification of states.
(Details of the derivative construction for the operators can be found in Appendix~\ref{sec:deriv_ops}.)
The Nucleon (and Delta) $J=\tfrac{1}{2}^-$, $\tfrac{3}{2}^-$ and $\tfrac{5}{2}^-$ operators all feature one derivative in 
$P_\mathsf{M}$ coupled to either a spin $S=\tfrac{1}{2}$ (doublet) or $S=\tfrac{3}{2}$ (quartet). Spectral overlaps ($Z$)
for these operators can be directly compared since they all have a consistent normalization.
An $S=\tfrac{3}{2}$ spin coupled to one derivative can be projected to either $J=\tfrac{1}{2}$, $J=\tfrac{3}{2}$ or 
$J=\tfrac{5}{2}$. The first and second constructions are subduced into the $G_{1u}$ and $H_u$ irreps, while 
the $J=\tfrac{5}{2}$ construction is subduced into the $H_u$ and $G_{2u}$ irreps.

In Figure~\ref{fig:nonrel}, we compare these non-relativistic operator overlaps for the lowest lying states and across irreps. 
We see that one operator is dominant for each state, with magnitudes that are roughly consistent across all irreps. In 
particular, the constructions coupled to different $J$ have similar magnitudes. These results suggest that these 
low-lying levels form a multiplet with little mixing among the states. In the language of $SU(6)\otimes O(3)$ 
(spin-flavor and space), these states, and their negative parity Delta partners, are part of a $[\mathbf{70},1^-]$ multiplet. 

Similarly, we can examine the first group of excited positive-parity levels in the Nucleon channel that cluster around 
$m/m_\Omega\sim 1.4$. Again, we find that the positive parity Nucleon operators from Table~\ref{tab:qm} are dominant 
in the spectral overlaps. One of them has a 
quasi-local structure (indicated as $S_\mathsf{S}$ for no derivatives) with $J=\tfrac{1}{2}$, while the others 
involve two derivatives coupled either to $L=0$, $1$, or $2$, and are labeled as $S_\mathsf{M}$, $P_\mathsf{A}$ or 
$D_\mathsf{S}$ and $D_\mathsf{M}$. We find that there is not a unique mapping of each state to one particular operator, rather each operator contributes in varying magnitude to each state, indicating significant mixing in this basis.

These results, along with the observation that the numbers of states are consistent with the numbers of non-relativistic 
operators in each $J^+$, suggest that this band of positive-parity states belongs to more than one $SU(6)\otimes O(3)$ multiplet,
with now mixing among the multiplets mentioned before; namely, the  $[\mathbf{70},0^+]$, $[\mathbf{56},2^+]$, $[\mathbf{70},2^+]$ 
$[\mathbf{20},1^+]$ and a radially excited $[\mathbf{56},0^+]$. It is notable that there is overlap with all the allowed $L_\pi J^P$ multiplets with $L\le 2$, and in particular, 
there is mixing with the $[\mathbf{20},1^+]$ multiplet. 
There does not appear to be any 
``freezing" of degrees of freedom as suggested in some diquark models (for some reviews see Refs.~\cite{Anselmino:1992vg,Capstick:2000qj}).  
We will return to this point in the summary.

As we move up to the second excited negative-parity band in the Nucleon and Delta channels, we find the ``non-relativistic'' $P_\mathsf{M}$ operators discussed previously do not feature prominently in the spectral overlaps in these higher lying excited states. Instead, operators using two derivatives together with one quark having a lower component ($\rho = -$) Dirac structure appear, some of which appear in Figures~\ref{fig:histogram} and \ref{fig:Zvalues}.  
The lower component spinor contributes a factor $-1$ to the parity. 
Similarly, those operators that featured prominently in the first excited positive parity band do not appear significantly in the second excited positive band in the Nucleon and Delta channels. Instead, operators involving two derivatives and two lower-component Dirac spinors appear.
The lower components effectively bring in angular momentum, but are not equivalent in spatial structure.
To adequately resolve the internal structure of these higher lying states will require the introduction of operators featuring three and four derivatives.

%% file: quark_mass.tex
%\subsection{Quark mass dependence}\label{sec:quark_mass}

At the two lighter pion masses, $m_\pi = 444$ and $396$ MeV, we find
the spectra for Nucleons and Deltas, classified by irreps, to be
qualitatively similar to the heaviest pion mass case.  The spin
identification techniques described above are successful, and the
resulting spectrum, identified by spin is shown in
Fig.~\ref{fig:spin_840} for the ensemble at $m_\pi = 396~{\rm MeV}$,
normalized in units of the measured $\Omega$ baryon mass determined at
this quark mass.

The Nucleon and Delta spectrum is qualitatively similar to that
determined at the heaviest pion mass - the $m_\pi = 524~{\rm MeV}$ lattices -
albeit typically at smaller mass in units of the $\Omega$ baryon
mass. There are the same number and pattern of low lying negative
parity Nucleon and Delta states, as well as in the first excited band
of positive parity states. Again, the low lying negative parity Delta
states are slightly higher than the corresponding Nucleon states. 
There is a slightly
enhanced splitting for the first excited band of positive parity Delta
states across $J^+$, as well as more mixing among the positive parity Nucleon
states around $m/m_\Omega \sim 1.3$.

\begin{figure}[t]
 \centering   
\includegraphics[width=0.50\textwidth, bb=25 20 455 535]{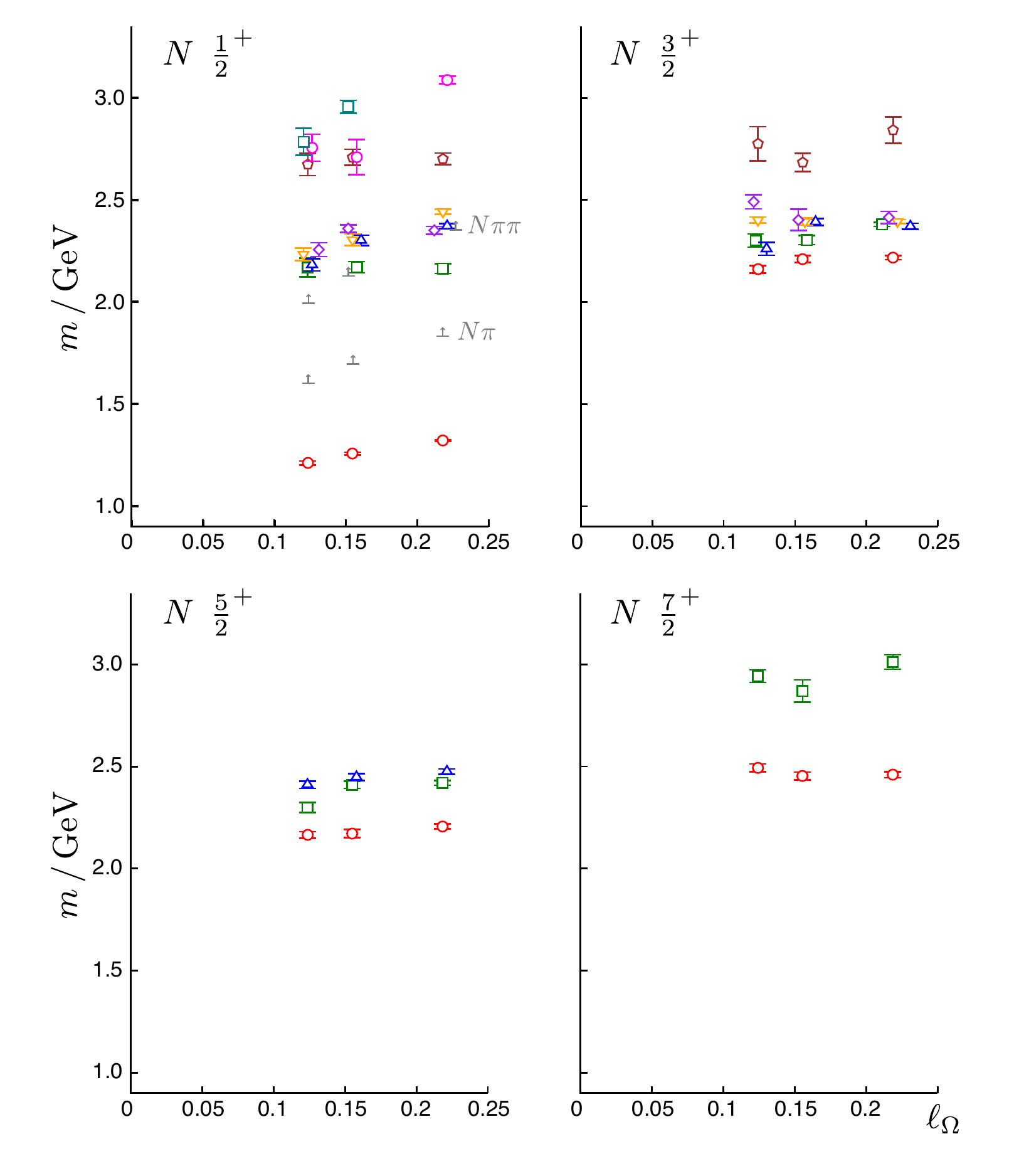}
\caption{Lightest few Nucleon $J=\tfrac{1}{2}^+$, $\tfrac{3}{2}^+$, $\tfrac{5}{2}^+$, and $\tfrac{7}{2}^+$ states.
Also shown in $N\tfrac{1}{2}^+$ is the threshold for $N\pi$ and $N\pi\pi$. 
The influence in the spectrum from these thresholds is complicated by the use of a finite spatial cubic box. Further
discussion is in Section~\ref{sec:multi}.
}
\label{fig:nucleon_pos}
\end{figure}

\begin{figure}[t]
 \centering   
\includegraphics[width=0.50\textwidth, bb=25 20 455 535]{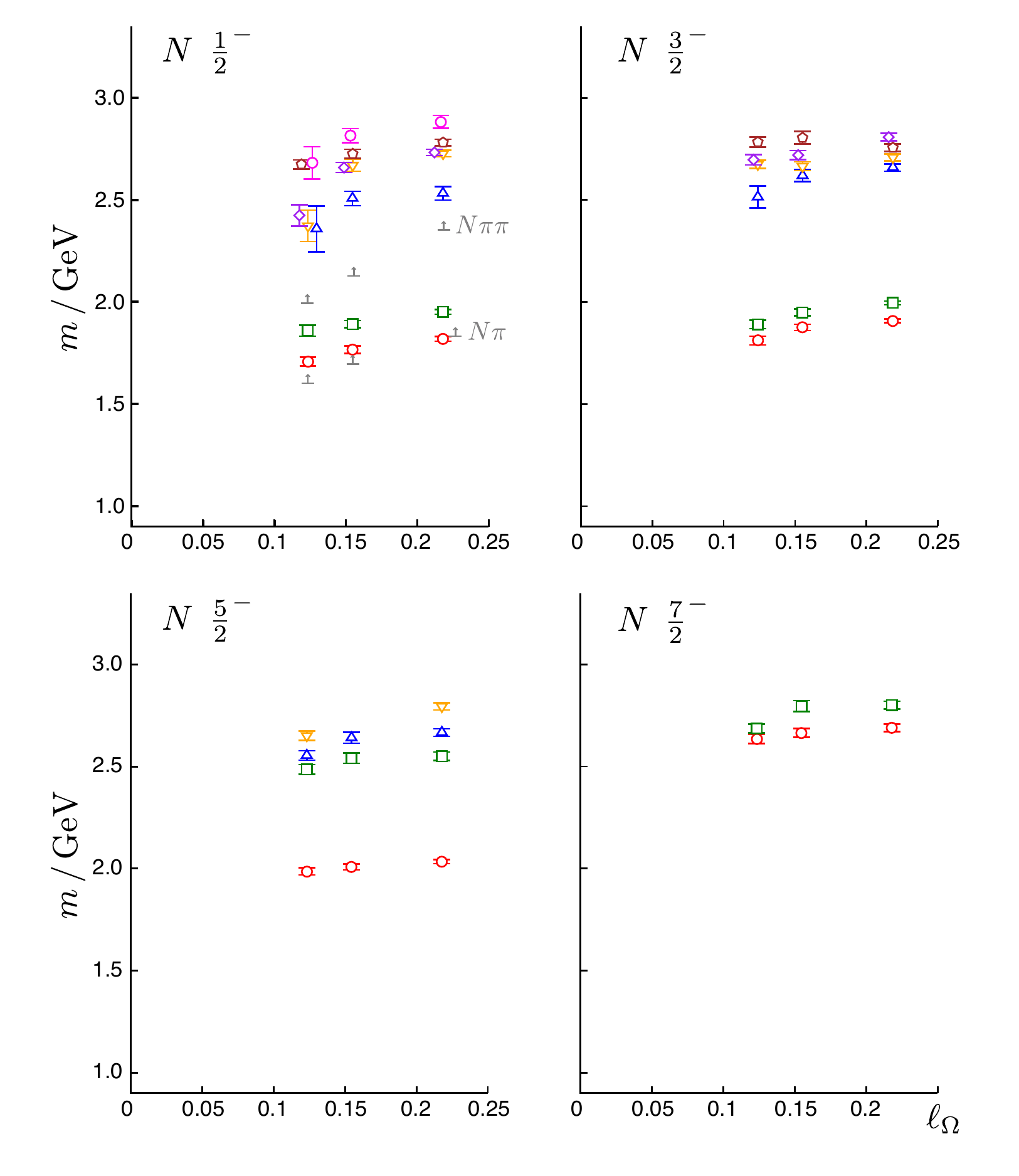}
\caption{Lightest few Nucleon $J=\tfrac{1}{2}^-$, $\tfrac{3}{2}^-$, $\tfrac{5}{2}^-$, and $\tfrac{7}{2}^-$ states.
Also shown in $N\tfrac{1}{2}^-$ is the threshold for $N\pi$ and $N\pi\pi$.
The influence in the spectrum from these thresholds is complicated by the use of a finite spatial cubic box. Further
discussion is in Section~\ref{sec:multi}.
}
\label{fig:nucleon_neg}
\end{figure}

\begin{figure}[t]
 \centering   
\includegraphics[width=0.50\textwidth, bb=25 20 455 535]{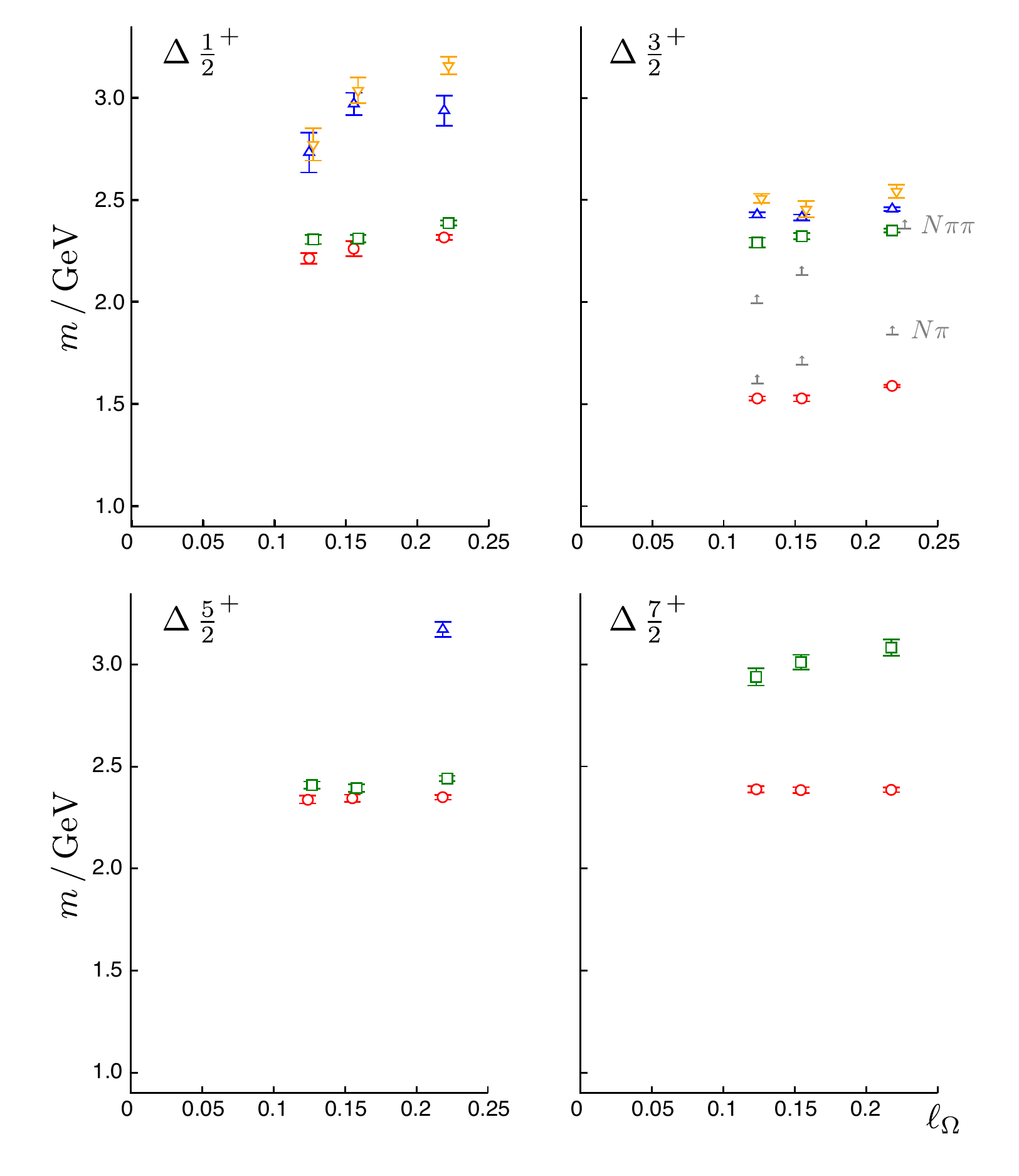}
\caption{Lightest few Delta $J=\tfrac{1}{2}^+$, $\tfrac{3}{2}^+$, $\tfrac{5}{2}^+$, and $\tfrac{7}{2}^+$ states.
Also shown in $\Delta\tfrac{3}{2}^+$ is the threshold for $N\pi$ and $N\pi\pi$.
The influence in the spectrum from these thresholds is complicated by the use of a finite spatial cubic box. Further
discussion is in Section~\ref{sec:multi}.
}
\label{fig:delta_pos}
\end{figure}

\begin{figure}[t]
 \centering   
\includegraphics[width=0.50\textwidth, bb=25 20 455 535]{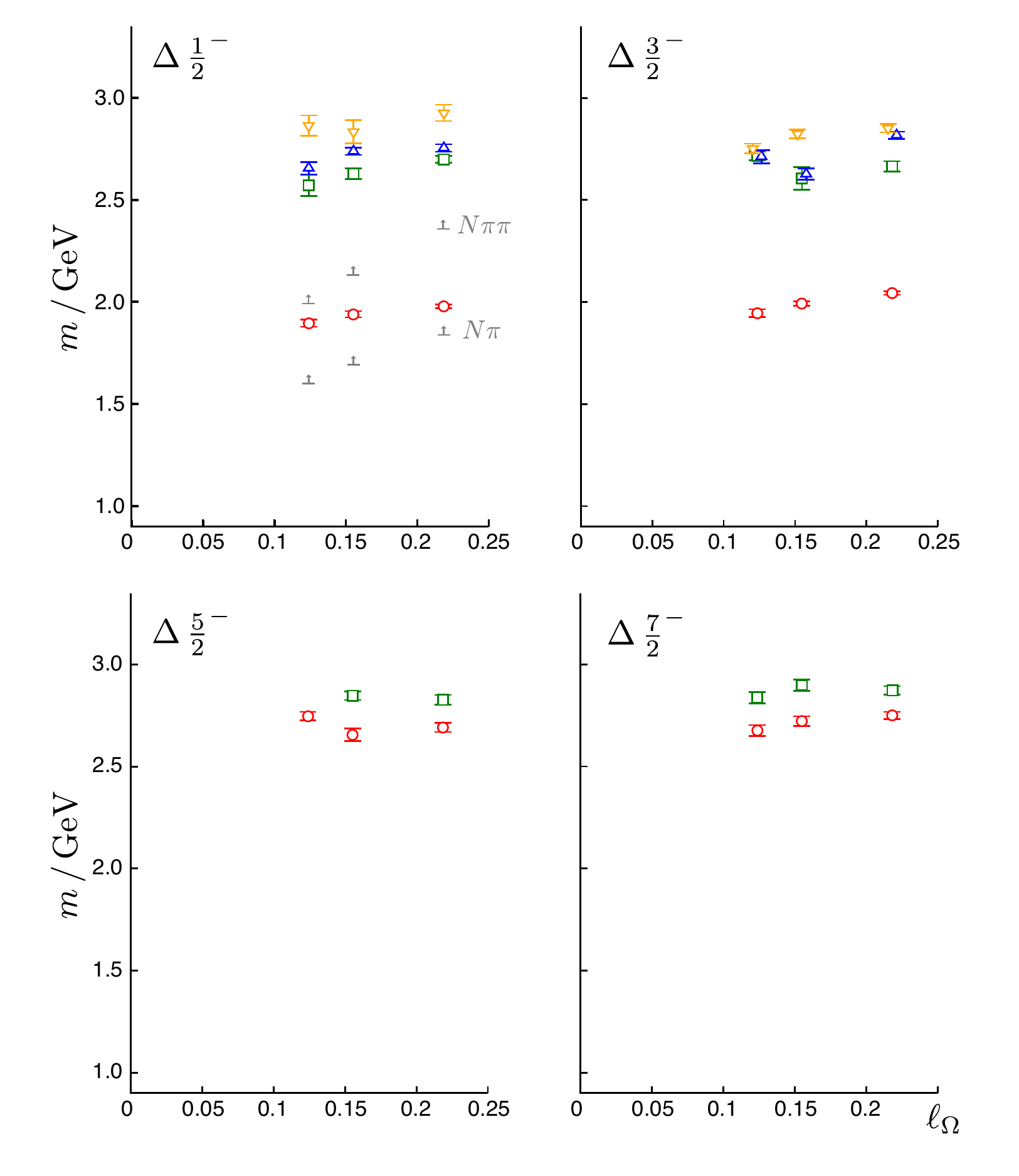}
\caption{Lightest few Delta $J=\tfrac{1}{2}^-$, $\tfrac{3}{2}^-$, $\tfrac{5}{2}^-$, and $\tfrac{7}{2}^-$ states.
Also shown in $\Delta\tfrac{1}{2}^-$ is the threshold for $N\pi$ and $N\pi\pi$.
The influence in the spectrum from these thresholds is complicated by the use of a finite spatial cubic box. Further
discussion is in Section~\ref{sec:multi}.
}
\label{fig:delta_neg}
\end{figure}

In Figures~\ref{fig:nucleon_pos} - \ref{fig:delta_neg}, we show
extracted state masses as a function of $\ell_\Omega \equiv
\tfrac{9}{4} \tfrac{(a_t m_\pi)^2}{(a_t m_\Omega)^2}$ which we use as
a proxy for the quark mass \cite{Lin:2008pr} for the three quark
masses listed in Table~\ref{tab:lattices}. The state masses are
presented via $\tfrac{ a_t m_H }{a_t m_\Omega}
m_\Omega^\mathrm{phys.}$.  The ratio of the state mass ($m_H$) to the
$\Omega$-baryon mass computed on the same lattice removes the explicit
scale dependence and multiplying by the physical $\Omega$-baryon mass
conveniently expresses the result in MeV units. This is clearly not a
unique scale-setting prescription, but it serves to display the data
in a relatively straightforward way. We remind the reader that the
data between different quark masses are uncorrelated since they follow
from computations on independently generated dynamical gauge-fields.

In Figure~\ref{fig:nucleon_pos} we show the mass of some of the lowest
identified positive parity levels. In Figure~\ref{fig:nucleon_neg} we show the
lowest few negative parity Nucleon levels. To help resolve near
degeneracies, we slightly shift symbols at the same pion mass
horizontally in $l_\Omega$. In some cases, for comparison, we plot the
mass of the lightest $N\pi$ and $N\pi\pi$ thresholds -- the mass
follows from the simple sum of the extracted masses on these
lattices. In general, the lattice levels decrease with the quark mass.
There is no observed dramatic behavior, for example, from crossing of thresholds.

Notable in these plots is the clustering of bands of levels as seen in
Figures~\ref{fig:spin_808} and \ref{fig:spin_840} and described
above. In the first excited positive parity Nucleon band, there is a
tendency for the levels across $J^+$ to cluster, and is most readily
apparent in the $J=\tfrac{1}{2}^+$ to cluster around $m/m_\Omega\sim
1.3$ as the pion mass decreases.  Another notable feature in these
figures is the absence of an excited Nucleon $J=\tfrac{1}{2}^+$ state
comparable, or slightly below, the lowest lying $\tfrac{1}{2}^-$ level
as reported in the PDG~\cite{Nakamura:2010zzi}. Similarly, there is no low lying
Delta $J=\tfrac{3}{2}^+$ state comparable in mass to the
$\tfrac{3}{2}^-$.  The ordering of the low-lying states in the Nucleon
spectrum, and in particular the spectrum of the $N\tfrac{1}{2}^+$ channel,
has been the subject of much effort in lattice QCD.

%% file: comparison.tex
%\subsection{Comparisons}\label{sec:comparisons}

\begin{figure*}
\begin{center}
\includegraphics[width=0.65\textwidth]{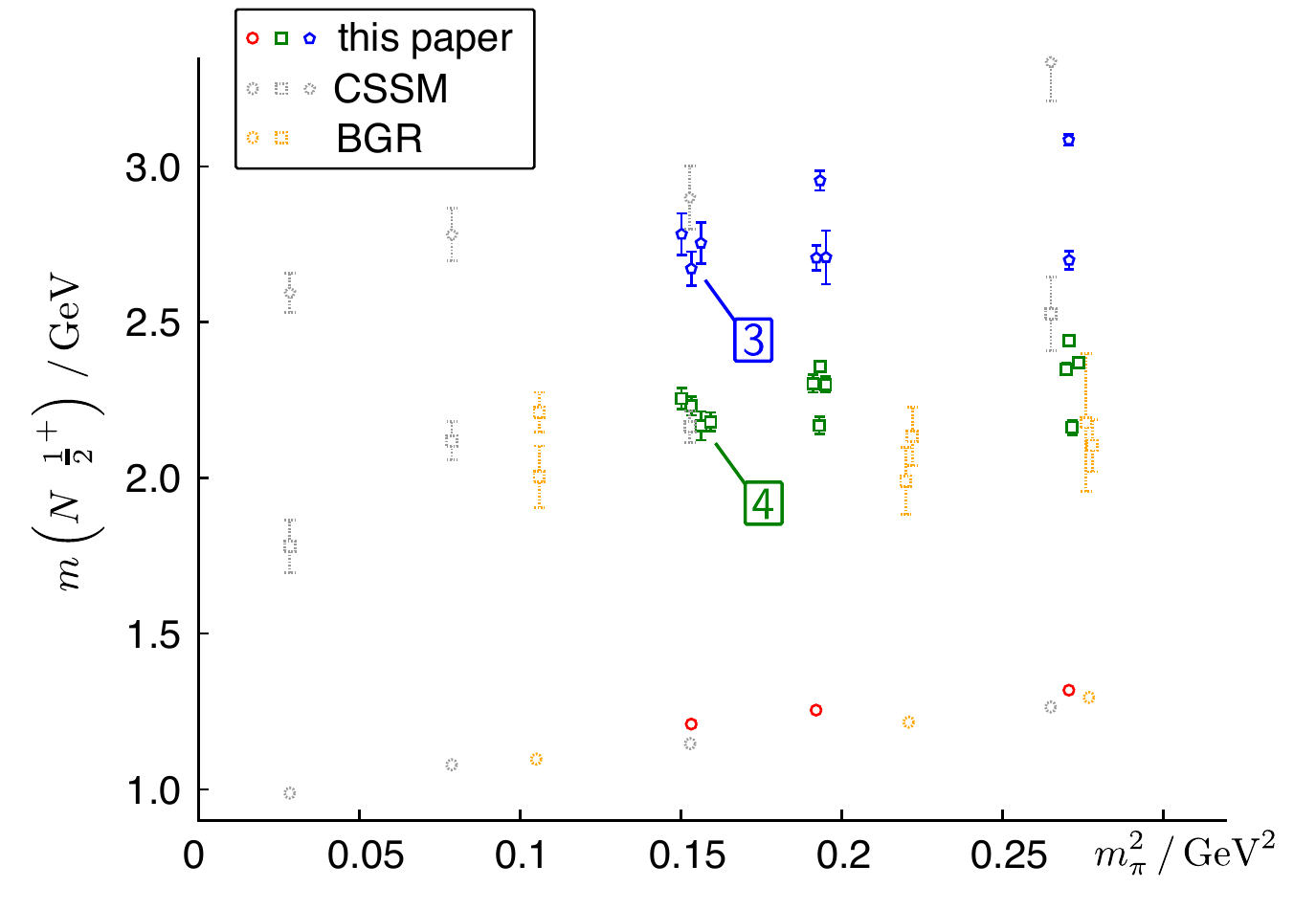}
\caption{Comparison of results for the Nucleon $J=\frac{1}{2}^+$
channel. The results shown in grey are from Ref.~\cite{Mahbub:2010rm},
while those in orange are from Ref.~\cite{Engel:2010my}. Note that
data are plotted using the scale-setting scheme in the respective
papers. Results from this paper are shown in red (the ground state), green and blue. At the lightest pion mass,
there is a clustering of four states as indicated near 2 GeV, while there are three nearly degenerate states 2.7 GeV. 
Operators featuring the derivative constructions discussed in this paper feature prominently in these excited states.
\label{fig:world}}
\end{center}
\end{figure*}
In Figure~\ref{fig:world}, we show a comparison of our result for the
Nucleon $J=\frac{1}{2}^+$ spectrum with other calculations in full
QCD. Those shown in grey are from Ref.~\cite{Mahbub:2010rm}
using $2 \oplus 1$ dynamical quark flavors, while those in
orange are from Ref.~\cite{Engel:2010my} using two dynamical
light-quark flavors. Each of these calculations employs a different
means to set the lattice scale, and we have made no attempt to resolve
these calculations to a common scale-setting scheme.  Rather our aim
in this discussion is to compare the pattern of states that has
emerged in each calculation, and to provide a possible explanation
for the differences.  

References~\cite{Mahbub:2010rm} and~\cite{Engel:2010my} find only two
excited levels below 2.8 GeV, notably isolated from one another in the
case of Ref.~\cite{Mahbub:2010rm}, at pion masses comparable to those
in this study.  At the lightest quark mass reported in this work,
there are \emph{four} nearly degenerate excited states found at approximately
2.2 GeV, and three nearly degenerate states near 2.8 GeV. A possible
explanation for the discrepancy in the number of levels is the
operator constructions used.  Refs.~\cite{Mahbub:2010rm,Engel:2010my}
use a basis of local or quasi-local operators, without, for example,
derivatives, but with multiple smearing radii.  These operators,
which have an $S$-wave spatial structure concentrated at the origin, can
have overlap with radial excitations of $S$-waves, but will have limited
overlap with higher orbital waves.  The results presented here suggest
the observed excited $J=\frac{1}{2}^+$ states are admixtures of radial
excitations as well as $D$-wave and anti-symmetric $P$-waves structures,
and the inclusion of operators featuring such structures is essential
to resolve the degeneracy of states.  The impact of an incomplete
basis of operators will be addressed more in
Section~\ref{sec:multi}.

The authors of Ref.~\cite{Mahbub:2010rm} note the large drop in the
energies of the first and second excited states at their lightest pion
mass, and ascribe this to the emergence of a light ``Roper'' in their
calculation. While we do not reach correspondingly light pion masses
and large volumes in this study, the work presented here clearly
shows the need for a sufficiently complete basis of operators before
a faithful description of the spectrum can emerge, 
and the identification of the Roper resonance be warranted.  
Indeed, even our
work remains incomplete since we have entered a regime of open decay
channels, discussed next, requiring the addition of multi-particle
operators to the basis.

%% file: multi.tex
%\section{Multi-particle states}\label{sec:multi}

In the previous section we presented the extracted spectra from calculations with three different light quark masses. In each case we were able, using the operator overlaps, to match states across irreps that we believe are subduced from the same continuum spin state. This suggests an interpretation of the spectrum in terms of single-hadron states, while in principle our correlators should receive contributions from all eigenstates of finite-volume QCD having the appropriate quantum numbers. This includes baryon-meson states which in finite volume have a discrete spectrum. 
Where are these multi-particle states?

This issue was investigated in the meson sector~\cite{Dudek:2010wm} where the spectrum was compared between multiple volumes and on multiple mass data sets. In particular, under change of volume, the extracted spectrum did not resemble the changing pattern of levels one would expect from two-meson states, but rather was largely volume-independent. As such, the interpretation of the observed levels was that of a single particle spectrum. In this work, some initial investigations were made with a $20^3\times 128$ lattice at the lightest pion mass available, and similar observations are made; namely, the spectrum between different volumes does not change substantially. We are thus led to interpret the spectrum is terms of single hadron states. The subsequent observations we make are quite similar to those made for mesons~\cite{Dudek:2010wm}.

The overlap of a localized three-quark operator onto a baryon-meson state will be suppressed by $1/\sqrt{V}$, where $V$ is the lattice volume, if the operator creates a resonance with a finite width in the infinite volume limit.  This fall-off is matched by a growth in the density of states with the volume and the resonant state thus maintains a finite width as the mixing with each discrete state falls.  The simulations in this study are carried out in cubic volumes with side-lengths $\sim 2\, \mathrm{fm}$, which might be sufficiently large that the mixing between one of the low-lying two-particle states and a resonance is suppressed sufficiently for it to be undetectable with the three-quark operator basis.  

Even if the mixing between localized single-hadron states and
baryon-meson states to form resonance-like finite volume eigenstates
is not small, there still remains a practical difficulty associated
with using only three-quark operators. In this case the state can be
produced at the source time-slice through its localized single-hadron
component, while the correlator time dependence obtained from
$e^{-Ht}$ will indicate the mass of the resonant eigenstate. Consider
a hypothetical situation in which a single baryon-meson state, denoted
by $|2\rangle$, mixes arbitrarily strongly with a single localized
single-hadron state, $|1\rangle$, with all other states being
sufficiently distant in energy as to be negligible. There will be two
eigenstates
\begin{align}
	\big| \mathfrak{a} \big\rangle &= \cos \theta \big|1\big\rangle + \sin \theta \big|2\big\rangle  \nonumber\\
	\big| \mathfrak{b} \big\rangle &= -\sin \theta \big|1\big\rangle + \cos \theta \big|2\big\rangle, \nonumber
\end{align}
with masses $m_\mathfrak{a}, m_\mathfrak{b}$. At the source (and sink) only the 
localized single-hadron component of each state overlaps with the operators in our basis and hence the overlaps, $Z^{\mathfrak{a},\mathfrak{b}}_i \equiv \langle \mathfrak{a},\mathfrak{b} | {\cal O}_i^\dag | 0 \rangle$, will differ only by an overall multiplicative constant, $Z^{\mathfrak{a}}_i = \cos \theta Z^{|1\rangle}_i, \;Z^{\mathfrak{b}}_i = -\sin \theta Z^{|1\rangle}_i$. As such the eigenvectors $v^{\mathfrak{a}}, v^{\mathfrak{b}}$ point in the same direction and cannot be made orthogonal. Thus the time dependence of both states will appear in the \emph{same} principal correlator as
\begin{equation}
	\lambda(t) \sim A_\mathfrak{a} e^{-m_\mathfrak{a}(t-t_0)} + A_\mathfrak{b} e^{-m_\mathfrak{b}(t-t_0)} + \ldots   \nonumber
\end{equation}

Since $m_\mathfrak{a}$ and $m_\mathfrak{b}$ most likely do not differ
significantly (on the scale of $a_t^{-1}$) it will prove very
difficult to extract a clear signal of two-exponential behavior from
the principal correlator. This is precisely why the variational
method's orthogonality condition on near degenerate states is so
useful, but we see that it cannot work here and we are left trying to
extract two nearby states from a $\chi^2$ fit to
time-dependence. Typically this is not possible and reasonable looking
fits to data are obtained with just one low-mass exponential. This is
analogous to the interpretation made in the comparison between our
computed $N\frac{1}{2}^+$ spectrum, and those computed using only
rotationally-symmetric smeared (quasi-)local sources\cite{Mahbub:2010rm,Engel:2010my}, shown
in Figure~\ref{fig:world}; our results showed a cluster of four
near-degenerate states, while the other analyses showed one or perhaps
two since all four states would only couple through their $S$-wave
components.

Thus, in some portions of the extracted spectrum, we might be
observing admixtures of ``single-particle'' and baryon-meson states. A
conservative interpretation then of our spectrum is that the mass
values are only accurate up to the hadronic width of the state
extracted, since this width is correlated with mixing with
baryon-meson states via a scattering phase-shift.  

In order to truly compare with the experimental situation, we would like to explicitly observe resonant behavior, thus to obtain
a significant overlap with multi-hadron states, we should include
operators with a larger number of fermion fields into our basis, and
in particular, multi-particle operators.  The construction of
single-meson and single baryon operators of definite continuum
helicity and subduced into the `in flight' little-group irreps can be
done using the tables in \cite{Moore:2005dw, Moore:2006ng}. Spin
identification is possible and will be reported in future work.  These
in-flight operators can be used in two-particle constructions, and
along with single particle operators, provide a far more complete
determination of the excited levels in an irrep.

As shown by L\"uscher\cite{Luscher:1991cf}, one can map these discrete
energy levels onto the the continuum energy dependent phase shift
within a partial wave expansion, including the phase shift for higher
partial waves. Such a technique was recently used to determine the $L=0$ and
$L=2$ phase shift for non-resonant $I=2$ $\pi\pi$
scattering~\cite{Dudek:2010ew}. The mapping of the phase shift is
both volume and irrep dependent. This is the origin of the cautionary remarks in the
captions of Figures~\ref{fig:nucleon_pos} -
\ref{fig:delta_neg}. Namely, the location of the threshold energies
are in fact irrep dependent and not solely determined by the energy of
the continuum states.

With suitable understanding of the discrete energy spectrum of the
system, the L\"uscher formalism can be used to extract the energy
dependent phase shift for a resonant system, such as has been
performed for the $I=1$
$\rho$ system~\cite{Feng:2010es}. The energy of the resonant state is
determined from the energy dependence of the phase shift. It is this
resonant energy that is suitable for chiral extrapolations.

Annihilation dynamics feature prominently in resonant systems, and
these dynamics arise from quark disconnected diagrams in
multi-particle constructions. Utilizing the techniques developed
recently for the study of isoscalar systems\cite{Dudek:2011tt},
distillation, and stochastic variants~\cite{Morningstar:2011ka}, can be used
for the efficient numerical evaluation of multi-particle systems.

%% file: summary.tex
%\section{Summary}\label{sec:summary}

We have described in detail our method for extracting a large number
of Nucleon and Delta excited states using the variational method on
dynamical anisotropic lattices.  Key to the success of the method has
been the use of a large basis of carefully constructed operators,
namely 
all
three quark baryon operators consistent with
classical continuum symmetries, and with up to two derivatives, 
that are subsequently projected onto the irreducible
representations of the cubic group.  We have exploited the observed
approximate realization of rotational symmetry to devise a method of
spin identification based on operator overlaps, enabling us to
confidently assign continuum spin quantum numbers to many states.  
We have demonstrated the importance of having a suitable operator basis with overlap
to all the continuum 
spin states that contribute to the spectrum.
We have then demonstrated the stability of the spectra with respect to changing
the number of
distillation vectors and the details of the variational analysis.
We have successfully applied these techniques at one lattice volume with three different light quark masses. We are able to reliably extract a large number
of excited states with $J^P$ ranging from $J=\frac{1}{2}$ up through
$J=\frac{7}{2}$ in both positive and negative parity.  These are the
first lattice calculations to achieve such a resolution of states in
the baryon sector with spin assignments and $J\ge\frac{5}{2}$.

We find a high multiplicity of levels spanning across $J^P$ which is
consistent with $SU(6) \otimes O(3)$ multiplet counting, and hence with that of
the non-relativistic $qqq$ constituent quark model. In particular, the counting of
levels in the low lying negative parity sectors are consistent with
the non-relativistic quark model and with the observed experimental
states~\cite{Nakamura:2010zzi}. The spectrum observed in the first excited positive
parity sector is also consistent in counting with the quark model, but
the comparison with experiment is less clear with the quark model
predicting more states than are observed experimentally, spurring
phenomenological investigations to explain the discrepancies (e.g., see
Refs.~\cite{Nakamura:2010zzi,Isgur:1978wd,Capstick:1986bm,Anselmino:1992vg,Glozman:1995fu,Capstick:2000qj,Goity:2003ab}).

We find that each of the operators in our basis features prominently
in some energy level, and there is significant mixing among each of the allowed multiplets,
including the $\mathbf{20}$-plet that is present in the non-relativistic $qqq$ quark
model, but does not appear in quark-diquark
models\cite{Anselmino:1992vg}, and in particular Ref.~\cite{Lichtenberg:1967zz}. 
This adds further credence to the
assertion that there is no ``freezing" of degrees of freedom with respect
to those of the non-relativistic quark model.  These qualitative
features of the calculated spectrum extend across all three of our
quark-mass ensembles.  Furthermore, we see no evidence for the
emergence of parity-doubling in the spectrum\cite{Glozman:1999tk}.

We have argued that the extracted spectrum can be interpreted in
terms of single-hadron states, and based on investigations in the
meson sector\cite{Dudek:2010wm} and initial investigations of the
baryon sector at a larger volume, we find little evidence for
multi-hadron states.  To study multi-particle states, and hence the resonant nature of excited states, we need to
construct operators with a larger number of fermion fields. Such
constructions are in progress, and we believe that the addition of
these operators will lead to a denser spectrum of states which can be
interpreted in terms of resonances via techniques like L\"uscher's and
its inelastic extensions\cite{Lage:2009zv}. 

The extraction and identification of a highly excited, spin identified
single-hadron spectrum, represents an important step
towards a determination of the excited baryon spectrum. 
The calculation of the single-baryon
spectrum including strange quarks is ongoing.
Combining the
methods developed in this paper with finite volume techniques for the extraction
of phase shifts, future
work will focus on the determination of hadronic resonances within QCD.

%% file: details.tex
%\section{Construction of flavor/spin/space symmetric states}\label{sec:symm_states}

Consider the construction of sets of definite symmetry for
%succinctly described in the appendix of Ref.~\cite{Feynman:1971wr}.
 three objects that can be labeled by $x$, $y$,
and $z$, where the first object is labeled by $x$, the second is 
labeled by $y$ and the third is labeled by $z$.
There are in general, four definite symmetry combinations:
symmetric, mixed-symmetric, mixed anti-symmetric, and totally
antisymmetric,  denoted by $\mathsf{S,\ MS,\ MA,\ A}$.  

Let symmetry projection operator $S_\mathsf{\Sigma}\begin{pmatrix} x & y & z \\ x^{\prime} & y^{\prime} & z^{\prime} \end{pmatrix}$ 
be defined so that its action on a generic object with labels $x^{\prime}$, $y^{\prime}$ and $z^{\prime}$
is to create a superposition of objects, denoted by $\{xyz\}_\mathsf{\Sigma}$,  with symmetry $\mathsf{\Sigma}$ of their labels as follows,
\begin{equation}
\{xyz\}_\mathsf{\Sigma} = \sum_{x^{\prime}  y^{\prime} z^{\prime}} 
S_\mathsf{\Sigma}\begin{pmatrix} x & y & z \\ x^{\prime} & y^{\prime} & z^{\prime} \end{pmatrix} \{ x^{\prime} y^{\prime} z^{\prime}\} .
\label{eq:projection}
\end{equation}

Permutation operators $S_\mathsf{\Sigma}$ can be inferred from their action on an 
 object with three labels to produce the four allowed symmetry combinations as follows,
\begin{align}
  \{xyz\}_\mathsf{S}  &= N_\mathsf{S}\big[ \{xyz\}+\{yxz\}+\{zyx\}+\{yzx\} 
    \nonumber \\ &\quad \quad \quad+\{xzy\} + \{zxy\}\big], 
    \nonumber \\
  \{xyz\}_\mathsf{MS} &= N_\mathsf{MS}\big[\{xyz\}+\{yxz\}+\{zyx\}+\{yzx\}
    \nonumber \\
    &\quad\quad\quad\quad- 2\{xzy\}-2\{zxy\}\big], \nonumber\\
  \{xyz\}_\mathsf{MA} &= N_\mathsf{MA}\big[\{xyz\}-\{yxz\}+\{zyx\}-\{yzx\}\big], \nonumber\\
  \{xyz\}_\mathsf{A}  &= N_\mathsf{A}\big[-\{xyz\}+\{xzy\}-\{yzx\}+\{yxz\}
    \nonumber \\
    &\quad\quad\quad\quad- \{zxy\}+\{zyx\}\big]\label{eq:symmetry}.
\end{align}
 If two or more 
labels are the same, then equivalent terms must be combined and the normalization constants 
 adjusted to give an appropriate normalization, e.g., for orthonormal quantum
states 
  $N_\mathsf{S} = \frac{1}{\sqrt{6}}$ when $x$, $y$ and
$z$ all are different. By convention, mixed symmetries are $\mathsf{MS}$ or $\mathsf{MA}$ according to whether
 the first two labels are symmetric or antisymmetric. 
 
 The use of projection operators allows the same constructions to be applied to the labels of operators
 as are applied to the labels of the states created by the operators.  They also provide the basis for a
 straightforward computational algorithm that yields the desired superpositions.  

Baryon operators have sets of labels for flavor, spin, and spatial arguments that transform 
independently, therefore as direct products.  Each of these sets of labels can be 
arranged according to Eq.~(\ref{eq:symmetry}).  Then the symmetries of the sets of
flavor labels must be
combined with the symmetries of the sets of spin and spatial labels in order to make
an overall symmetric object as discussed in the text.
The general rules for combining direct products of objects with independent labels sets $1$ and $2$ to
make overall symmetries of the combined sets of all labels, denoted by $1$,$2$, are as follows,
\begin{alignat}{2}
  \{1\}_\mathsf{S}\{2\}_\mathsf{S}  &= \{1,2\}_\mathsf{S},  & \quad\{1\}_\mathsf{S}\{2\}_\mathsf{MS} &= \{1,2\}_\mathsf{MS},\nonumber\\
  \{1\}_\mathsf{S}\{2\}_\mathsf{MA} &= \{1,2\}_\mathsf{MA}, & \quad\{1\}_\mathsf{S}\{2\}_\mathsf{A} &= \{1,2\}_\mathsf{A}, \nonumber\\
  \{1\}_\mathsf{A}\{2\}_\mathsf{S}  &= \{1,2\}_\mathsf{A}, & \quad\{1\}_\mathsf{A}\{2\}_\mathsf{MS} &= \{1,2\}_\mathsf{MA}, \nonumber\\
  -\{1\}_\mathsf{A}\{2\}_\mathsf{MA}  &= \{1,2\}_\mathsf{MS}, & \quad\{1\}_\mathsf{A}\{2\}_\mathsf{A} &= \{1,2\}_\mathsf{S}, \nonumber
\end{alignat}
\begin{eqnarray}
  \osq\big(+\{1\}_\mathsf{MS}\{2\}_\mathsf{MS} + \{1\}_\mathsf{MA}\{2\}_\mathsf{MA}\big) &=& \{1,2\}_\mathsf{S}, \nonumber\\
  \osq\big(-\{1\}_\mathsf{MS}\{2\}_\mathsf{MS} + \{1\}_\mathsf{MA}\{2\}_\mathsf{MA}\big) &=& \{1,2\}_\mathsf{MS}, \nonumber\\
  \osq\big(+\{1\}_\mathsf{MS}\{2\}_\mathsf{MA} + \{1\}_\mathsf{MA}\{2\}_\mathsf{MS}\big) &=& \{1,2\}_\mathsf{MA}, \nonumber\\
  \osq\big(-\{1\}_\mathsf{MS}\{2\}_\mathsf{MA} + \{1\}_\mathsf{MA}\{2\}_\mathsf{MS}\big) &=& \{1,2\}_\mathsf{A} \nonumber\\
  \label{eq:combine}
\end{eqnarray}

\subsection{Dirac spin}

The construction of Dirac spin representations follows Table 14 in Appendix B of 
Ref.~\cite{Basak:2005aq}.  
A Dirac spinor with an index that takes four values is formed from direct 
products of two spinors that have indices that take two values:
one for ordinary spin ($s$-spin) and the other for intrinsic parity ($\rho$-spin).
The $s$-spin indices are $s=+$ for spin up and $s=-$ for spin down while the 
$\rho$-spin indices are $\rho = +$ for positive intrinsic parity and $\rho = -$ for negative intrinsic parity.
These $\rho$- and $s$-spin indices determine the  
Dirac spin index (in the rest frame) as shown in Table~\ref{tab:dirac_map}, based on the Dirac-Pauli representation of Dirac matices.

There are eight $s$-spin states for a three-quark baryon and they are obtained by projecting an arbitrary state to combinations with good symmetry.  The resulting states follow from
Eq.~(\ref{eq:symmetry}) with $x$, $y$ and $z$ taking two values, $+$ and $-$, which produces four symmetric ($\mathsf{S}$) states with total spin $\frac{3}{2}$,
\begin{eqnarray}
  \left|\tfrac{3}{2},+\tfrac{3}{2} \right\rangle_\mathsf{S} &=& \big|+++\big\rangle_\mathsf{S},\nonumber\\
  \left|\tfrac{3}{2},+\tfrac{1}{2}\right\rangle_\mathsf{S} &=& \big|++-\big\rangle_\mathsf{S},\nonumber\\
  \left|\tfrac{3}{2},-\tfrac{1}{2}\right\rangle_\mathsf{S} &=& \big|+--\big\rangle_\mathsf{S},\nonumber\\
  \left|\tfrac{3}{2},-\tfrac{3}{2}\right\rangle_\mathsf{S} &=& \big|---\big\rangle_\mathsf{S},
  \label{eq:S_spin}
\end{eqnarray}
two mixed-symmetric ($\mathsf{MS}$) states with total spin $\half$,
\begin{eqnarray}
  \left|\tfrac{1}{2},+\tfrac{1}{2}\right\rangle_\mathsf{MS} &=& +\big|+-+\big\rangle_\mathsf{MS},\nonumber\\
  \left|\tfrac{1}{2},-\tfrac{1}{2}\right\rangle_\mathsf{MS} &=& -\big|-+-\big\rangle_\mathsf{MS},
  \label{eq:MS_spin}
\end{eqnarray}
and two mixed-antisymmetric ($\mathsf{MA}$) states with total spin $\half$,
\begin{eqnarray}
  \left|\tfrac{1}{2},+\tfrac{1}{2}\right\rangle_\mathsf{MA} &=& +\big|+-+\big\rangle_\mathsf{MA},\nonumber\\
  \left|\tfrac{1}{2},-\tfrac{1}{2}\right\rangle_\mathsf{MA}  &=& -\big|-+-\big\rangle_\mathsf{MA}.
  \label{eq:MA_spin}
\end{eqnarray}
The states on the right side are normalized states of definite symmetry. 
The eight $\rho$-spin states take exactly the same form as the s-spin states.  

\begin{table}
  \begin{center}
    \begin{tabular}{|c|cc|}
      \hline
      Dirac&$\rho$&$s$\\
      \hline
      1 & $+$ & $+$\\
      2 & $+$ & $-$\\
      3 & $-$ & $+$\\
      4 & $-$ & $-$\\
      \hline
    \end{tabular}
  \end{center}
  \caption{Mapping of Dirac spin indices to $\rho$ and $s$ labels.}
  \label{tab:dirac_map}
\end{table}

\begin{table*}
  \begin{center}
    \begin{tabular}{|lrr|rc|rc|}
      \hline
      Dirac&IR&Emb&\multicolumn{4}{|c|}{$\rho\otimes s$}\\
      \hline
      $\mathsf{S}$  & $\frac{1}{2}$ & $\mathit{1}$ & \multicolumn{4}{|c|}{$\osq \big(+|\rho\>_\mathsf{MS}|s\>_\mathsf{MS}+|\rho\>_\mathsf{MA}|s\>_\mathsf{MA}\big)$}\\
      & $\frac{3}{2}$ & $\mathit{1,2}$ & \multicolumn{4}{|c|}{$|\rho\>_\mathsf{S}|s\>_\mathsf{S}$}\\
      \hline
      $\mathsf{M}$ & $\frac{1}{2}$ & $\mathit{1,2}$ & $\mathsf{MS}$ & $|\rho\>_\mathsf{S}|s\>_\mathsf{MS}$& $\mathsf{MA}$ & $|\rho\>_\mathsf{S}|s\>_\mathsf{MA}$\\
      &      & $\mathit{3}$ &  & $\osq\big(-|\rho\>_\mathsf{MS}|s\>_\mathsf{MS}+|\rho\>_\mathsf{MA}|s\>_\mathsf{MA}\big)$& & $\osq\big(+|\rho\>_\mathsf{MS}|s\>_\mathsf{MA}+|\rho\>_\mathsf{MA}|s\>_\mathsf{MS}\big)$\\
      & $\frac{3}{2}$   & $\mathit{1}$ && $|\rho\>_\mathsf{MS}|s\>_\mathsf{S}$& &$|\rho\>_\mathsf{MA}|s\>_\mathsf{S}$\\
      \hline
      $\mathsf{A}$  & $\frac{1}{2}$ & $\mathit{1}$ & \multicolumn{4}{|c|}{$\osq\big(-|\rho\>_\mathsf{MS}|s\>_\mathsf{MA}+|\rho\>_\mathsf{MA}|s\>_\mathsf{MS}\big)$}\\
      \hline
    \end{tabular}
  \end{center}
  \caption{Symmetries of Dirac spin states based on direct products of $\rho$-spin and $s$-spin states
    for three quarks.
    States labeled as $|\rho\>_\mathsf{\Sigma}$ refer to a $\rho$-spin state with symmetry $\mathsf{\Sigma}$ from Eqs.~(\ref{eq:S_spin}-\ref{eq:MA_spin}). Similarly, states labeled as $|s\>_\mathsf{\Sigma}$ refer to a $s$-spin state
    with symmetry $\mathsf{\Sigma}$ from Eqs.~(\ref{eq:S_spin}-\ref{eq:MA_spin}).  Direct products of 
    the $\rho$-spin and $s$-spin states yield sums of three-quark terms in which each 
    quark has a $\rho$ and $s$ label. That determines each quark's Dirac index according to Table~\ref{tab:dirac_map}.
  }
  \label{tab:dirac_spin}
\end{table*}

The 64 Dirac spin labels of three quarks are obtained from direct products of $\rho$-spin 
and $s$-spin states of the quarks.  The possible symmetries of the Dirac spinors are obtained from the 
multiplication rules in Eq.~\ref{eq:combine}, together with
Table~\ref{tab:dirac_map}, which shows how each quark's Dirac index is determined by
its $s$-spin and $\rho$-spin indices.
Examples of the construction are given in Ref.~\cite{Basak:2005aq}.  Equal numbers of positive-
and negative-parity states are always produced. The octahedral irrep of the product 
is $G_1$ for $s$-spin $\frac{1}{2}$ and $H$ for $s$-spin $\frac{3}{2}$ representations of $SU(2)$.
Table~\ref{tab:dirac_spin} shows the irreps of 
$SU(2)$ that are produced. The number of embeddings of each irrep also 
is listed.  Parities of the states are determined by products of the $\rho$-spins of the three quarks.

\subsection{Flavor}

\begin{table*}
  \begin{tabular}{ccc}
    \begin{tabular}{|c|cr|r|c|c|}
      \hline
      \multicolumn{6}{|c|}{Octet, $\mathbf{8}$}\\
      \hline
        &$I$&$I_z$&$S$&$\phi_\mathsf{MS}$&$\phi_\mathsf{MA}$\\
      \hline
      $p$ & $\half$ & $+\half$ & $0$ & $|udu\>_\mathsf{MS}$ & $|udu\>_\mathsf{MA}$\\
      $n$ & $\half$ & $-\half$ & $0$ & $-|dud\>_\mathsf{MS}$ & $-|dud\>_\mathsf{MA}$\\
      $\Lambda_\mathbf{8}$ & $0$ & $0$ & $-1$ & $\osq(|sud\>_\mathsf{MS}-|uds\>_\mathsf{MS})$ & $\osq(|sud\>_\mathsf{MA}-|uds\>_\mathsf{MA})$\\
      $\Sigma_\mathbf{8}^+$ & $1$ & $+1$ & $-1$ & $|usu\>_\mathsf{MS}$ & $|usu\>_\mathsf{MA}$\\
      $\Sigma_\mathbf{8}^0$ & $1$ & $0$ & $-1$ & $|usd\>_\mathsf{MS}$ & $|usd\>_\mathsf{MA}$\\
      $\Sigma_\mathbf{8}^-$ & $1$ & $-1$ & $-1$ & $|dsd\>_\mathsf{MS}$ & $|dsd\>_\mathsf{MA}$\\
      $\Xi_\mathbf{8}^0$ & $\half$ & $+\half$ & $-2$ & $-|sus\>_\mathsf{MS}$ & $-|sus\>_\mathsf{MA}$\\
      $\Xi_\mathbf{8}^-$ & $\half$ & $-\half$ & $-2$ & $-|sds\>_\mathsf{MS}$ & $-|sds\>_\mathsf{MA}$\\
      \hline
    \end{tabular}
    
    \begin{tabular}{|c|cr|r|c|}
      \hline
      \multicolumn{5}{|c|}{Decuplet, $\mathbf{10}$ }\\
      \hline
      &$I$&$I_z$&$S$&$\phi_\mathsf{S}$\\
      \hline
      $\Delta^{++}$ & $\threehalf$ & $+\threehalf$ & $0$ & $|uuu\>_\mathsf{S}$\\
      $\Delta^{+}$ & $\threehalf$ & $+\half$ & $0$ & $|uud\>_\mathsf{S}$\\
      $\Delta^{0}$ & $\threehalf$ & $-\half$ & $0$ & $|udd\>_\mathsf{S}$\\
      $\Delta^{-}$ & $\threehalf$ & $-\threehalf$ & $0$ & $|ddd\>_\mathsf{S}$\\
      $\Sigma_\mathbf{10}^{+}$ & $1$ & $+1$ & $-1$ & $|uus\>_\mathsf{S}$\\
      $\Sigma_\mathbf{10}^{0}$ & $1$ & $0$ & $-1$ & $|uds\>_\mathsf{S}$\\
      $\Sigma_\mathbf{10}^{-}$ & $1$ & $-1$ & $-1$ & $|dds\>_\mathsf{S}$\\
      $\Xi_\mathbf{10}^{0}$ & $\half$ & $+\half$ & $-2$ & $|sus\>_\mathsf{S}$\\
      $\Xi_\mathbf{10}^{-}$ & $\half$ & $-\half$ & $-2$ & $|sds\>_\mathsf{S}$\\
      $\Omega^{-}$ & $0$ & $0$ & $-3$ & $|sss\>_\mathsf{S}$\\
      \hline
    \end{tabular}

 \begin{tabular}{|c|cr|r|c|}
      \hline
      \multicolumn{5}{|c|}{Singlet, $\mathbf{1}$}\\
      \hline
      &$I$&$I_z$&$S$&$\phi_{A}$\\
      \hline
      $\Lambda_\mathbf{1}^{0}$ & 0 & 0 & $-1$ & $|dus\>_\mathsf{A}$\\
      \hline
    \end{tabular}

  \end{tabular}
  \caption{
    Flavor octet, decuplet and singlet constructions.
  }
  \label{tab:octet}
\end{table*}

%%%%%%%%%%%%%%%%
\begin{table*}[!t]
  \begin{center}
    \begin{tabular}{|lrrr|rc|rc|}
      \hline
      $\mathsf{\Sigma}$ & $\substack{SU(3)\\SU(2)}$ & $N_\mathrm{nonrel}$ & $N_\mathrm{rel}$ & \multicolumn{4}{|c|}{} \\
      \hline
      $\mathsf{S}$ & $\bf (10,4)$ & 1 & 1 & \multicolumn{4}{|c|}{$\phi_\mathsf{S}\chi_\mathsf{S}$} \\
      & $\bf (10,2)$ & & 1 & \multicolumn{4}{|c|}{$\phi_\mathsf{S}\chi_\mathsf{S}$}  \\
      & $\bf  (8,4)$ & & 1 & \multicolumn{4}{|c|}{$\osq(\phi_\mathsf{MS}\chi_\mathsf{MS}+\phi_\mathsf{MA}\chi_\mathsf{MA})$}  \\
      & $\bf  (8,2)$ & 1 & 2 & \multicolumn{4}{|c|}{$\osq(\phi_\mathsf{MS}\chi_\mathsf{MS}+\phi_\mathsf{MA}\chi_\mathsf{MA})$}  \\
      & $\bf  (1,2)$ & & 1 & \multicolumn{4}{|c|}{$\phi_\mathsf{A}\chi_\mathsf{A}$}  \\
      \hline
      $\mathsf{M}$  & $\bf (10,4)$ & & 1 & $\mathsf{MS}$ & $\phi_\mathsf{S}\chi_\mathsf{MS}$ & $\mathsf{MA}$ & $\phi_\mathsf{S}\chi_\mathsf{MA}$  \\
      & $\bf (10,2)$ & 1 & 2 & & $\phi_\mathsf{S}\chi_\mathsf{MS}$ &      & $\phi_\mathsf{S}\chi_\mathsf{MA}$   \\
      & $\bf  (8,4)$ & 1 & 1 & & $\phi_\mathsf{MS}\chi_\mathsf{S}$ &      & $\phi_\mathsf{MA}\chi_\mathsf{S}$   \\
      & $\bf  (8,4)$ &   & 1 & & $\osq(-\phi_\mathsf{MS}\chi_\mathsf{MS}+\phi_\mathsf{MA}\chi_\mathsf{MA})$ & & $\osq(\phi_{MS}\chi_\mathsf{MA}+\phi_\mathsf{MA}\chi_\mathsf{MS})$ \\
      & $\bf  (8,2)$ &   & 1 & & $\phi_\mathsf{MS}\chi_\mathsf{S}$ &      & $\phi_\mathsf{MA}\chi_\mathsf{S}$   \\
      & $\bf  (8,2)$ & 1 & 2 & & $\osq(-\phi_\mathsf{MS}\chi_\mathsf{MS}+\phi_\mathsf{MA}\chi_\mathsf{MA})$ & & $\osq(\phi_\mathsf{MS}\chi_\mathsf{MA}+\phi_\mathsf{MA}\chi_\mathsf{MS})$ \\
      & $\bf  (8,2)$ &   & 1 & & $\phi_\mathsf{MA}\chi_\mathsf{A}$ & & $\phi_\mathsf{MS}\chi_\mathsf{A}$  \\
      & $\bf  (1,4)$ &   & 1 & & $\phi_\mathsf{A}\chi_\mathsf{MA}$ & & $\phi_\mathsf{A}\chi_\mathsf{MS}$  \\
      & $\bf  (1,2)$ & 1 & 2 & & $\phi_\mathsf{A}\chi_\mathsf{MA}$ & & $\phi_\mathsf{A}\chi_\mathsf{MS}$  \\
      \hline
      $\mathsf{A}$    & $\bf (10,2)$ & & 1 & \multicolumn{4}{|c|}{$\phi_\mathsf{S}\chi_\mathsf{A}$}  \\
      & $\bf  (8,4)$ & & 1 & \multicolumn{4}{|c|}{$\osq(\phi_\mathsf{MS}\chi_\mathsf{MA}-\phi_\mathsf{MA}\chi_\mathsf{MS})$}  \\
      & $\bf  (8,2)$ & 1 & 2 & \multicolumn{4}{|c|}{$\osq(\phi_\mathsf{MS}\chi_\mathsf{MA}-\phi_\mathsf{MA}\chi_\mathsf{MS})$}  \\
      & $\bf  (1,4)$ & 1 & 1 & \multicolumn{4}{|c|}{$\phi_\mathsf{A}\chi_\mathsf{S}$}  \\
      & $\bf  (1,2)$ & & 1 & \multicolumn{4}{|c|}{$\phi_\mathsf{A}\chi_\mathsf{S}$}  \\
      \hline
    \end{tabular}
  \end{center}
  \caption{
    Local operators classified according to symmetry of flavor and Dirac
    spin. The dimensionality of the $SU(3)$ representation is shown. The number
    of Dirac spin embeddings (number of operators) in a non-relativistic ($\rho = +$) 
    construction are shown in column $N_\mathrm{nonrel}$, and the number of 
    constructions featuring some number of lower components is shown in column $N_\mathrm{rel}$. The total number of embeddings is the sum of $N_\mathrm{nonrel}+N_\mathrm{rel}$. 
    The multiplicity of operators in the
    non-relativistic case is ${\bf 56}$ ($\mathsf{S}$), ${\bf 70}$ ($\mathsf{MS}$),
    ${\bf 70}$ ($\mathsf{MA}$) and ${\bf 20}$ ($\mathsf{A}$), and corresponds to
    the conventional non-relativistic $SU(6)\otimes O(3)$ construction. The
    relativistic construction, which involves both positive and negative
    parity operators, corresponds to the reduction of $SU(12)$. Note that the
    flavor singlet operators are distinct from the octet and decuplet
    operators. In the $SU(3)$ flavor limit, the flavor singlet states do
    not mix with the octet or decuplet states.}
  \label{tab:op_sym}
\end{table*}

The flavor states with well-defined symmetries also are constructed using
Eq.~(\ref{eq:symmetry}). For the purposes of discussion, we will
consider the flavor symmetric limit, i.e., degenerate quark
masses. There are three separate representations -- the symmetric ($\mathsf{S}$) 
states are the $SU(3)$ decuplet, $\big|\mathbf{10}\big\rangle_\mathsf{S}$, the mixed-symmetric ($\mathsf{MS, MA}$) states
are the $SU(3)$ octets, $\big|\mathbf{8}\big\rangle_\mathsf{MS,MA}$, and the antisymmetric ($\mathsf{A}$) state is the $SU(3)$ 
flavor-singlet state, $\big|\mathbf{1}\big\rangle_\mathsf{A}$. The $\Delta^{+}$ is $|udu\>_\mathsf{S}$, while the proton
is $|udu\>_\mathsf{MA,MS}$. The $\Sigma^{0}$ is $|uds\>_\mathsf{MA,MS}$. The octet, singlet
and decuplet constructions are shown in Table~\ref{tab:octet}.%, and the octet constructions are shown in Table~\ref{tab:octet}.

Within the $SU(3)$ flavor representations, we also have $SU(2)$ isospin states.
In the construction that follows, it is straightforward to generalize to the case of broken
$SU(3)$. The combinations of symmetry states remain valid, however, there are
some new states, such as $\Sigma$ in a $\mathsf{S}$ flavor state.

\subsection{Orbital angular momentum based on covariant derivatives}\label{sec:deriv_ops}

Smearing of quark fields is based on the distillation method of Ref.~\cite{Peardon:2009gh}.
It is used in order to filter out the effects of small scale fluctuations of the gauge fields
and it provides a spherically symmetric distribution of each quark field that carries
no orbital angular momentum.  In order to obtain higher spins, it is necessary to
add spatial structure using covariant derivative operators, as described in the text, which are combined in 
definite symmetries that correspond to orbital angular momenta in the continuum.  
For a single derivative, ($D^{[1]}$), the two symmetry combinations are given in
Eq.~(\ref{eq:deriv}), while for two derivativesi, ($D^{[2]}$), the combinations are given in
Eq.~(\ref{eq:2_deriv}).  

%%%%%%%%%%%%%%%%
\begin{table}[h]
  \begin{center}
    \begin{tabular}{|c|ccc|c|}
      \hline
      \multicolumn{5}{|c|}{Singlet, $\mathbf{1}$}\\
      \hline
      IR & $d=0$ & $d=1$ & $d=2$ & Total\\
      \hline
      $G_1$ & 1 & 4 & 9 & 14\\
      $H$ & 0 & 5 & 17 & 22\\
      $G_2$ & 0 & 1 & 8 & 9\\
      \hline
      \hline
      \multicolumn{5}{|c|}{Octet, $\mathbf{8}$}\\
      \hline
      IR & $d=0$ & $d=1$ & $d=2$ & Total\\
      \hline
      $G_1$ & 3 & 8 & 17 & 28\\
      $H$ & 1 & 11 & 36 & 48\\
      $G_2$ & 0 & 3 & 17 & 20\\
      \hline
      \hline
      \multicolumn{5}{|c|}{Decuplet, $\mathbf{10}$}\\
      \hline
      IR & $d=0$ & $d=1$ & $d=2$ & Total\\
      \hline
      $G_1$ & 1 & 4 & 10 & 15\\
      $H$ & 2 & 5 & 19 & 26\\
      $G_2$ & 0 & 1 & 10 & 11\\
      \hline
    \end{tabular}
    \caption{Numbers of singlet, octet and decuplet operators for each parity according to the irrep and 
      the number of derivatives, $d$. Total derivative constructions have been removed.
      \label{tab:num_ops}}
  \end{center}
\end{table}

%\begin{comment}
  \begin{table}[h]
    \begin{center}
      \begin{tabular}{|c|cccc|c|}
        \hline
        \multicolumn{6}{|c|}{Singlet, $\mathbf{1}$}\\
        \hline
        Rep & $J=\half$ & $J=\frac{3}{2}$ & $J=\frac{5}{2}$ & $J=\frac{7}{2}$ & Total\\
        \hline
        $G_1$ & 13 &  &  & 1 & 14\\
        $H$ &  & 13 & 8 & 1 & 22\\
        $G_2$ &  &  & 8 & 1 & 9\\
        \hline
      \end{tabular}
      \hspace{1cm}
      \begin{tabular}{|c|cccc|c|}
        \hline
        \multicolumn{6}{|c|}{Octet, $\mathbf{8}$}\\
        \hline
        Rep & $J=\half$ & $J=\frac{3}{2}$ & $J=\frac{5}{2}$ & $J=\frac{7}{2}$ & Total\\
        \hline
        $G_1$ & 24 &  &  & 4 & 28\\
        $H$ &  & 28 & 16 & 4 & 48\\
        $G_2$ &  &  & 16 & 4 & 20\\
        \hline
      \end{tabular}
      \hspace{1cm}
      \begin{tabular}{|c|cccc|c|}
        \hline
        \multicolumn{6}{|c|}{Decuplet, $\mathbf{10}$}\\
        \hline
        Rep & $J=\half$ & $J=\frac{3}{2}$ & $J=\frac{5}{2}$ & $J=\frac{7}{2}$ & Total\\
        \hline
        $G_1$ & 12 &  &  & 3 & 15\\
        $H$ &  & 15 & 8 & 3 & 26\\
        $G_2$ &  &  & 8 & 3 & 11\\
        \hline
      \end{tabular}
      \caption{Number of singlet, octet and decuplet operators according to continuum spin and subduced irrep.
        With two derivatives, at most $J=\frac{7}{2}$ can be reached.}
    \end{center}
    \label{tab:nuc_ops2}
  \end{table}
%\end{comment}

Operators that have 
good spin in the continuum are built by applying some number of derivatives to the Dirac spinors. 
Using the $SU(2)$ Clebsch-Gordan coefficients to combine orbital and spin
angular momenta, the one-derivative operators are,
\begin{equation}
  \left(D^{[1]} \Psi^{[S]}\right)^{[J,M]} = \sum_{m,s} \langle 1,m; S,s| J,M\rangle \vec{D}^{[1]}_{L=1,m} \Psi^{S,s},
\end{equation}
where $S=\half$ or $\frac{3}{2}$ are the possible spin states of three quarks in the absence
of derivatives.
Reference~\cite{Basak:2005aq} developed single
derivative operators for baryons and it provides some examples of 
the incorporation of combinations of covariant derivatives into three-quark operators.

Additional derivatives together with $SU(2)$ Clebsch-Gordan coefficients are used to obtain 
higher $J$ states.  For example, the two-derivative operators are first combined to get $L=2$ ,
\begin{equation}
  D^{[2]}_{L=2,M} = \sum_{m_1,m_2} \langle 1,m_1;1,m_2 | 2,M\rangle D^{[1]}_{m_1} D^{[1]}_{m_2}.
\end{equation}
This $L=2$ derivative operator is then applied to a spinor $\Psi^{S,s}$  as follows,
\begin{equation}
  \left(D^{[2]}_{L=2} \Psi^{[S]}\right)^{[J,M]} = \sum_{m,s} \langle 2,m; S,s| J,M\rangle D^{[2]}_{L=2,M} \Psi^{S,s}.
  \label{eq:cont_op}
\end{equation}
Derivative operator constructions for singlet, octet and decuplet follow
from Table~\ref{tab:op_sym}.
The single-site operators are symmetric in space, flavor and Dirac
indices. With one derivative, a mixed symmetry flavor and spin
construction is combined with mixed symmetry derivative operators to
make the overall symmetric combinations.  Similarly, with two
derivatives, various spin-flavor symmetry states are combined with the
derivative operators to make overall symmetric operators.

As noted above, three quarks with no derivatives can form at most spin $\half$ and $\frac{3}{2}$
states. The corresponding lattice irreps $G_1$ and $H$ are faithful
representations of these continuum spins, and hence form the basis of
the constructions of higher spins. The numbers of operators with up to two derivatives are shown in
Table~\ref{tab:num_ops}.
The number of operators classified according to total spin $J$ and irrep are shown in Table~\ref{tab:nuc_ops2}.
A general feature of the operator construction
is that there is always an equal number of positive and negative parity
operators. For example, for every operator in $G_{1g}$, there is a corresponding
operator in $G_{1u}$ and similarly for $H_{g}$ and $H_u$. There are no single-site
operators in $G_2$.

These constructions provide operators that have good total angular momentum up to $J=\frac{7}{2}$
in the continuum limit.  However, they are reducible with respect to 
the octahedral group that represents the symmetry of a cubic lattice.

%% file: qm.tex
%\section{Quantum mechanics of continuum spin in the octahedral representation} \label{sec:octahedral_spin}

In this appendix we develop the subduction of $SU(2)$ quantum states to irreducible octahedral 
states in the continuum.   The lowest spins are trivial as suitable subductions for
spins $J=\frac{1}{2}$ and $J=\frac{3}{2}$ are provided by the
elementary $G_1$ and $H$ octahedral irrep states, i.e.,
\begin{eqnarray}
  \left| \left[ \frac{1}{2}, m \right] \right\rangle &=& \left| G_1, r ,\left[\frac{1}{2}\right] \right\rangle , ~~~ r=\frac{3}{2} -m, 
  \nonumber \\
  \left| \left[ \frac{3}{2}, m \right] \right\rangle &=& \left| H, r , \left[\frac{3}{2}\right] \right\rangle,  ~~~  r=\frac{5}{2} -m.
  \label{eq:G1_H}
\end{eqnarray}
A suitable subduction for spin 1 also is trivial in terms of octahedral irrep $T_1$,
\begin{equation}
  \left| \left[ 1, m \right] \right\rangle = \left| T_1, r, \left[ 1 \right] \right\rangle, ~~~r=2-m. \nonumber
\end{equation}
Here and in the following we label octahedral states that carry continuum quantum numbers 
$J, M$ by placing the 
quantum numbers in brackets, i.e.,  $\left| \left[ J,M\right] \right\rangle$.  The angular momentum basis states are orthogonal in the continuum and consequently the octahedral states labelled 
as $\left| \left[ J, M \right] \right\rangle$ are orthogonal in the continuum limit as follows, 
\begin{eqnarray}
  \left\langle \left[ J,M\right] \Big| \left[ J^{\prime},M^{\prime}\right] \right\rangle = \delta_{J, J^{\prime}} \delta_{M, M^{\prime}} .
  \label{eq:su2_orthog}
\end{eqnarray}
However, they are reducible.  

Irreducible octahedral states,  $\left| \Lambda, r, \left[J\right] \right\rangle$, that are subduced from a single spin $J$ are labelled by the octahedral irrep, $\Lambda$, row $r$, and spin $[J]$ in brackets. Examples appear on the right side of Eq.~(\ref{eq:G1_H}).  Owing to the orthogonality of different octahedral 
irreps and rows, these states form an orthonormal set obeying,
\begin{equation}
  \left\langle \Lambda, r, \left[ J \right] \Big| \Lambda^{\prime}, r^{\prime},\left[ J\right] \right\rangle = \delta_{\Lambda, \Lambda^{\prime}} \delta_{r, r^{\prime}}.
  \label{eq:oct_orthog_1}
\end{equation}
A general octahedral irrep can contain an infinite number of continuum spins.  States transforming as the same octahedral irrep and row but subduced from different spins, such as $\left| H, r, \left[\frac{3}{2}\right] \right\rangle$,
$\left| H, r, \left[\frac{5}{2}\right] \right\rangle$ and $\left| H, r, \left[\frac{7}{2}\right] \right\rangle$, are distinguished by
their $\left[J\right]$ labels. These states are orthogonal to one another as shown in Eq.~(\ref{eq:oct_orthog_2}). 
The lowest spins have a one-to-one relation between the $\left| \left[ J,M\right] \right\rangle$ and $\left| \Lambda, r, \left[J\right] \right\rangle$ labellings
as in Eq.~(\ref{eq:G1_H}) but higher spins do not.

Higher-spin states can be constructed from direct products of lower-spin states by use of  
the $SU(2)$ Clebsch-Gordan formula for direct products of 
states of spins $J_1$ and $J_2$ as follows,
\begin{eqnarray}
  \big|\, [J, M]\, \big\rangle  = \sum_{m_1,m_2}  \big|\, [J_1,m_1] \big\rangle \otimes \big|\, [J_2, m_2] \big\rangle \nonumber \\
  \times \big\langle J_1 m_1; J_2 m_2 \big| J M \big\rangle,
  \label{eq:SU2-Clebsch}
\end{eqnarray}
where $\big\langle J_1 m_1; J_2 m_2 \big| J M \big\rangle $ is a $SU(2)$ Clebsch-Gordan coefficient. 
Equation~(\ref{eq:SU2-Clebsch}) provides a block-diagonal unitary transformation between the basis of 
$(2J_1+1)(2J_2+1)$ product states on the right side and the equal number of states in the basis of total angular momentum on the left side for $J$ in the range 
$ \big|J_1-J_2\big| \leq J \leq J_1+J_2$.    

Each of the octahedral states should be expanded in terms of a set of $(2J_1+1)(2J_2+1)$ linearly 
independent states transforming as irreducible representations of the 
octahedral group and subduced from a single $J$ value.  Tables~\ref{table:5/2occurrences} 
and \ref{table:7/2occurrences} show the relevant states based on using $J_1 = 1$ and $J_2 = \frac{3}{2}$ or 
$\frac{5}{2}$ to construct $J = \frac{5}{2}$ or $J = \frac{7}{2}$. 

\begin{table*}
  \begin{center}
    \begin{tabular}{|c|c|c|c|}
      \hline
      $J$  & $\frac{1}{2}$ & $\frac{3}{2}$ & $\frac{5}{2}$ \\
      \hline
      IR states & $\left|G_1, r, \left[\frac{1}{2}\right] \right\rangle$  & $\left|H, r, \left[\frac{3}{2}\right] \right\rangle$ &
      $\left|H, r, \left[\frac{5}{2}\right] \right\rangle$ , $\left|G_2, r, \left[\frac{5}{2}\right] \right\rangle$ \\
      \hline
    \end{tabular} 
  \end{center}
  \caption{Occurrences of octahedral irrep states with spins $J=\frac{1}{2}$, $\frac{3}{2}$ and $\frac{5}{2}$ based on using $J_1 = 1$ and $J_2 = \frac{3}{2}$. }
  \label{table:5/2occurrences}
\end{table*}

\begin{table*}
  \begin{center}
    \begin{tabular}{|c|c|c|c|}
      \hline
      $J$  & $\frac{3}{2}$ & $\frac{5}{2}$ & $\frac{7}{2}$ \\
      \hline
      IR states & $\big|H, r, \left[ \frac{3}{2}\right] \big\rangle$  & $\big|H, r, \left[ \frac{5}{2}\right] \big\rangle$ , $\big|G_2, r, \left[ \frac{5}{2}\right] \big\rangle$ & $\big|G_1, r, \left[ \frac{7}{2}\right] \big\rangle$ , $\big|H, r, \left[ \frac{7}{2}\right] \big\rangle$ , $\big|G_2, r, \left[ \frac{7}{2}\right] \big\rangle$ \\
      \hline
    \end{tabular}
  \end{center}
  \caption{Occurrences of octahedral irrep states with spins $J=\frac{3}{2}$, $\frac{5}{2}$ and $\frac{7}{2}$ based on $J_1 = 1$ and $J_2 = \frac{5}{2}$. }
  \label{table:7/2occurrences}
\end{table*}

The limited dimensions of the octahedral irreps require 
multiple occurrences of some irreps in the subduction of high spins.  
In the continuum, the different occurrences of the same irrep and row 
provide linearly independent 
states, as we will show by construction. Representations with multiple occurrences are denoted as $\left| ^n\!\Lambda, r, \left[J\right] \right\rangle$, where 
$^n\!\Lambda$ denotes the $n^{th}$ occurrence of irrep $\Lambda$, $r$ 
denotes the row of the irrep and $\left[J\right]$ shows the spin from which the state is subduced.
When there is a single occurrence, the left superscript is omitted. 
\begin{table*}
  \begin{center}
    \begin{tabular}{|c|c|c|c|}
      \hline
      $J$  & $\frac{5}{2}$ & $\frac{7}{2}$ & $\frac{9}{2}$ \\
      \hline
      IR states & $\left|H, r, \left[ \frac{5}{2}\right] \right\rangle$ , $\left|G_2, r, \left[ \frac{5}{2}\right] \right\rangle$ 
      & $\left| G_1, r, \left[ \frac{7}{2}\right] \right\rangle$ , $\left| H, r, \left[ \frac{7}{2}\right] \right\rangle$ , $\left| G_2, r, \left[ \frac{7}{2}\right] \right\rangle$ &
      $\left|G_1, r, \left[ \frac{9}{2}\right] \right\rangle$ , $\left| ^1H, r, \left[ \frac{9}{2}\right] \right\rangle$ , $\left|^2H , r, \left[ \frac{9}{2}\right] \right\rangle$  \\
      \hline
    \end{tabular}
  \end{center}
  \caption{Occurrences  of octahedral irrep states in the subduction of spin $\frac{9}{2}$ based on $J_1 = 1$ and $J_2 = \frac{7}{2}$ }
  \label{table:9/2occurrences}
\end{table*}

Expanding in terms of a complete set of irreducible octahedral states gives,    
\begin{eqnarray}
  \big|\left[J,M \right] \big\rangle &=& \sum_{^n\!\Lambda,  r } \big| \,  ^n\!\Lambda,  r, \left[J\right] 
  \big\rangle \, \big\langle ^n\!\Lambda,  r, \left[J\right] \big|\, \left[\, J,M\, \right]  \big\rangle,
  \nonumber \\
  &=&  \sum_{^n\!\Lambda, r }  \big|^n\!\Lambda, r, \left[J\right] \big\rangle    {\cal S}^{J,M}_{^n\!\Lambda,r}, 
  \nonumber \\
  {\cal S}^{J,M}_{^n\!\Lambda,r} &=& \big\langle ^n\!\Lambda, r, \left[ J \right] \big| 
  \left[ J,M \right]  \big\rangle ,
  \label{eq:SJM}
\end{eqnarray} 
where we sum over all irreps, including multiple occurrences of the same irrep, that are linearly independent.  The subduction matrix, ${\cal S}^{J,M}_{^n\!\Lambda, r}$, is defined by the
overlap of an irreducible octahedral state subduced from a single $J$ and the spin 
state $\big| \left[ J,M \right] \big\rangle$. 
Orthogonality properties of the subduction matrices follow from substituting Eq.~(\ref{eq:SJM}) into 
Eq.~(\ref{eq:su2_orthog}) with $J=J^{\prime}$ as follows,
\begin{eqnarray}
  \sum_{\Lambda, r} \sum_{\Lambda^{\prime} , r^{\prime}} 
  {\cal S}^{J,M}_{\Lambda, r}  {\cal S}^{J,M^{\prime}}_{\Lambda^{\prime}, r^{\prime}}
  \big\langle \Lambda, r, \left[ J \right] \big| \Lambda^{\prime}, r^{\prime},\left[ J \right] \big\rangle
  = \delta_{M, M^{\prime}}. 
  \nonumber \\
\end{eqnarray}
Because of Eq.~(\ref{eq:oct_orthog_1}), this reduces to
\begin{eqnarray}
  % \sum_{\Lambda, r} \sum_{\Lambda^{\prime} , r^{\prime}} 
  % S^{J,M}_{\Lambda, r}  S^{J^{\prime},M^{\prime}}_{\Lambda^{\prime}, r^{\prime}}
  % \delta_{\Lambda, \Lambda^{\prime}} \delta_{r, r^{\prime}} \delta_{J, J^{\prime}} &=& 
  % \delta_{J, J^{\prime}} \delta_{M, M^{\prime}}
  % \nonumber \\
  \sum_{\Lambda, r} 
  {\cal S}^{J,M}_{\Lambda, r}  {\cal S}^{J,M^{\prime}}_{\Lambda, r}  &=& \delta_{M, M^{\prime}}.  
  \label{eq:SJM_orthog}
\end{eqnarray} 
Summation of the squares of ${\cal S}^{J,M}_{\Lambda, r} $ over $\Lambda$ and $r$  gives the
normalization condition,
\begin{equation}
  \sum_{\Lambda, r}{\cal S}^{J,M}_{\Lambda, r}  {\cal S}^{J,M}_{\Lambda, r}  = 1.
  \label{eq:SJM_norm}
\end{equation}

Because the $\big| \left[ J,M \right] \big\rangle $ states are a complete set over the subspace of
spin $J$, there is a sum rule
\begin{eqnarray} 
  \sum_M  \big\langle \Lambda, r, \left[ J \right] \big| \left[ J,M \right] \big\rangle
  &&\!\!\!\!\!\! \big\langle \left[ J, M\right] \big| \Lambda^{\prime}, r^{\prime},\left[ J\right] \big\rangle
  \nonumber \\ 
  &&\!\!\!\!\!\!\!\!\!  =  \big\langle \Lambda, r, \left[ J \right]  \big| \Lambda^{\prime}, r^{\prime},\left[ J\right] \big\rangle .\nonumber
\end{eqnarray}
where $\sum_M |[J,M]\rangle \langle [J,M] | = 1$ was used to obtain the right side.  
The left side involves a sum over products of subduction matrices and the right side involves the
$\delta$-functions of Eq.~(\ref{eq:oct_orthog_1}), yielding
\begin{eqnarray}
  \sum_M {\cal S}^{J, M}_{\Lambda, r} {\cal S}^{J, M}_{\Lambda^{\prime}, r^{\prime}} = 
  \delta_{\Lambda, \Lambda^{\prime}} \delta_{r, r^{\prime}}.
  \label{eq:sum_M_SS}
\end{eqnarray}
Multiplying a subduction matrix times both sides of  Eq.~(\ref{eq:SJM}) and summing over $M$   
yields
\begin{eqnarray}
  \sum_M {\cal S}^{J,M}_{\Lambda, r} \big| \left[ J, M\right] \big\rangle &=&
  \sum_M {\cal S}^{J,M}_{\Lambda, r} \sum_{\Lambda^{\prime}, r^{\prime}} {\cal S}^{J, M}_{\Lambda^{\prime}, r^{\prime}} \big| \Lambda^{\prime}, r^{\prime}, \left[ J \right] \big\rangle
  \nonumber \\
  &=&  \big|  \Lambda, r, \left[ J\right] \big\rangle ,
  \label{eq:SJM_JM}
\end{eqnarray}
where Eq.~(\ref{eq:sum_M_SS}) was used to evaluate the summations on the right side.
This last equation shows how the subduction matrix is used.  Once one has a realization of 
$\left| \left[ J, M \right] \right\rangle$ states, the subduction matrix is applied to obtain the
octahedral irrep states that are subduced from a single $J$ value.   Using Eq.~(\ref{eq:SJM_JM})
leads to an important extension of Eq.~(\ref{eq:oct_orthog_1}), 
\begin{eqnarray}
  \big\langle \Lambda, r, \left[ J \right] \big| \Lambda^{\prime}, r^{\prime}, \left[ J^{\prime} \right] \big\rangle
  &&\!\!\!\!\!\!\!= \!\!\! \sum_{M,M^{\prime}}  {\cal S}^{J,M}_{\Lambda, r}  \big\langle \left[ J,M \right]\big|  
  \left[ J^{\prime}, M^{\prime} \right] \big\rangle {\cal S}^{J^{\prime},M^{\prime}}_{\Lambda^{\prime}, r^{\prime}},
  \nonumber \\
  &&\!\!\!\!\!\!\!\!\!\!\!\! = \!\!\! \sum_{M,M^{\prime}}  {\cal S}^{J,M}_{\Lambda, r}\  \delta_{J, J^{\prime}} \delta_{M,M^{\prime}} 
  \ {\cal S}^{J^{\prime},M^{\prime}}_{\Lambda^{\prime}, r^{\prime}} ,
  \nonumber \\
  &&\!\!\!\!\!\!\!\!\!\!\!\! = \!\!\! \sum_{M}  {\cal S}^{J,M}_{\Lambda, r} {\cal S}^{J,M}_{\Lambda^{\prime}, r^{\prime}} \delta_{J, J^{\prime}} ,
  \nonumber \\
  &&\!\!\!\!\!\!\!\!\!\!\!\! = \! \delta_{\Lambda, \Lambda^{\prime}} \delta_{r, r^{\prime}} \delta_{J, J^{\prime}},
  \label{eq:oct_orthog_2}
\end{eqnarray}
where Eq.~(\ref{eq:sum_M_SS}) was used in the last step.  The octahedral states 
subduced from single $J$ values are orthonormal with respect to $J$ as well as with respect to
octahedral irrep, $\Lambda$, and row, $r$.  

Substituting the
expansion in terms of orthonormal irrep states for each octahedral state in the Clebsch-Gordan formula of Eq.~(\ref{eq:SU2-Clebsch}) gives
\begin{align} 
  \sum_{^n\!\Lambda,  r} \big| ^n\!\Lambda, r, [ J ] \big\rangle \, &{\cal S}^{J, M}_{^n\!\Lambda, r} = \nonumber \\
  &\sum_{\substack{m_1, m_2\\ \Lambda_1, r_1\\ \Lambda_2, r_2}}
  \big|\Lambda_1, r_1, [J_1] \big\rangle
  %\nonumber \\
   \otimes \big|\Lambda_2, r_2, [J_2]\big\rangle 
  \,  \nonumber \\
  &  \quad\quad\quad \times  {\cal S}^{J_1, m_1}_{\Lambda_1, r_1} {\cal S}^{J_2, m_2}_{\Lambda_2, r_2}
  \big\langle J_1 m_1 ; J_2 m_2 \big| J M \big\rangle
  \nonumber \\
  \label{eq:sub-1}
\end{align}
Here the octahedral irrep states on the right side are assumed not to involve multiple occurrences so  left
superscripts are omitted for $\Lambda_1$ and $\Lambda_2$.   

The rules for combining direct products of octahedral group irreps are similar to those for continuum spins. 
The direct product $T_1\otimes H$ corresponds to a direct product of $J=1$ and $J=\frac{3}{2}$
continuum irreps.  It yields a $G_1$ irrep that corresponds to spin-$\frac{1}{2}$, a $H$ irrep 
that corresponds to spin-$\frac{3}{2}$ and a pair of irreps, $H^{\prime}$ and $G_2$, that taken 
together correspond to spin-$[\frac{5}{2}]$.  The $H$ and $H^{\prime}$ irrep states are orthogonal in
the continuum limit
because they are subduced from different spins, i.e.,  $J = \frac{3}{2}$ and $\frac{5}{2}$.
%Here the two occurrences are denoted as $^1H$ and $^2H$. 

Direct products of octahedral states are equal to sums of irreducible states according to the Clebsch-Gordan formula for the octahedral group,
\begin{eqnarray}
  \big|\Lambda_1, r_1, \left[J_1\right] \big\rangle &&\!\!\!\!\!\! \otimes \big|\Lambda_2, r_2, \left[J_2\right] \big\rangle = 
    \nonumber \\
    &&\!\!\!\!\!\!\!\!\! \sum_{\Lambda_{\Lambda_1\otimes\Lambda_2} , r} \big| \Lambda_{\Lambda_1\otimes \Lambda_2}, r \big\rangle \cdot    C \begin{pmatrix} \Lambda_1 & \Lambda_2 & \Lambda_{\Lambda_1\otimes \Lambda_2} \\  r_1 & r_2 & r \end{pmatrix},
    \nonumber \\
    \label{eq:Clebsch_O}
\end{eqnarray}
where the states produced on the right side are labeled by the irreps involved in the direct products.
The $[J_1]$ and  $[J_2]$ labels of the states on the left side are passive and do not
affect the direct products.  This gives the following expansion,
\begin{eqnarray}
  %   \left|\left[J,M \right]\, \right\rangle 
  \sum_{^n\!\Lambda,  r}  \big| ^n\!\Lambda, r, [ J ] \big\rangle \, {\cal S}^{J, M}_{^n\!\Lambda, r}
  &=&  \sum_{\Lambda_{\Lambda_1\otimes\Lambda_2},  r } \big| \Lambda_{\Lambda_1\otimes\Lambda_2}, r \big\rangle \, {\cal R}^{J, M}_{\Lambda_{\Lambda_1\otimes\Lambda_2}, r},
  \nonumber \\
  \label{eq:Rmat}
\end{eqnarray}
where matrix ${\cal R}^{J, M}_{\Lambda_{\Lambda_1\otimes\Lambda_2}, r} $ is defined by
\begin{align}
  %  R^{J, M}_{\Lambda_{\Lambda_1\otimes\Lambda_2}, r} &=& \sum_{\begin{tabular}{l}\scriptsize $m_1 m_2 \Lambda_1$\\ \scriptsize $r_1 r_2 \Lambda_2$\end{tabular} } 
  {\cal R}^{J, M}_{\Lambda_{\Lambda_1\otimes\Lambda_2}, r} &= \sum_{\substack{m_1,m_2\\ \Lambda_1,r_1 \\\Lambda_2,r_2}} 
  C \begin{pmatrix} \Lambda_1 & \Lambda_2 & \Lambda_{\Lambda_1\otimes \Lambda_2} \\  r_1 & r_2 & r \end{pmatrix} 
  \nonumber \\ 
  &\quad\quad\quad\times {\cal S}^{J_1, m_1}_{\Lambda_1, r_1} 
  {\cal S}^{J_2, m_2}_{\Lambda_2, r_2} \nonumber\\
  &\quad\quad\quad\times  \big\langle J_1 m_1; J_2 m_2 \big| J M \big\rangle. 
  \label{eq:expansion_1}
\end{align}

The notation used here is based on the fact that the states produced by the Clebsch-Gordan 
expansion are general octahedral irrep states,  not states subduced from a single $J$ value. 
For the construction based on 
Eq.~(\ref{eq:sub-1}), the $SU(2)$ spins in the range 
$|J_1-J_2| \leq J \leq J_1+J_2$ provide a complete set of $(2J_1+1)(2J_2+1)$ linearly independent states and 
the octahedral irrep states on the right side of
Eq.~(\ref{eq:Rmat}) can be expanded as a linear combination of the octahedral irrep states 
subduced from a single spin as follows,  
\begin{equation}
  \big| \Lambda_{\Lambda_1\otimes\Lambda_2}, r \big\rangle  = \sum_{J^{\prime}=|J_1-J_2|}^{J_1+J_2}
  {\cal A}_{ \Lambda_{\Lambda_1\otimes\Lambda_2}, ^n\!\Lambda[J^{\prime}] } 
  \big| ^n\!\Lambda, r, \left[J^{\prime}\right] \big\rangle
  \label{eq:sum_J'}
\end{equation}  
where ${\cal A}_{\Lambda_{\Lambda_1\otimes\Lambda_2}, ^n\!\Lambda[J^{\prime}],}$ is a matrix in the 
$\Lambda_{\Lambda_1\otimes\Lambda_2}$ and $^n\!\Lambda[J^{\prime}]$ indices.  

Substituting Eqs.~(\ref{eq:Clebsch_O}) and (\ref{eq:sum_J'}) into Eq.~(\ref{eq:sub-1}) and using Eq.~(\ref{eq:SJM}) gives
\begin{eqnarray}
  \sum_{^n\!\Lambda, r } \big| ^n\!\Lambda, r, \left[J\right]\big\rangle {\cal S}^{J, M}_{^n\!\Lambda, r}
  &=&  \sum_{ J^{\prime}, \Lambda_{\Lambda_1\otimes\Lambda_2} , r }
  {\cal A}_{ \Lambda_{\Lambda_1\otimes\Lambda_2}, ^n\!\Lambda[J^{\prime}] }
  \nonumber \\ 
  &&\!\!\!\!\! \times \big| ^n\!\Lambda, r, \left[J^{\prime}\right] \big\rangle
  {\cal R}^{J, M}_{\Lambda_{\Lambda_1\otimes\Lambda_2}, r}
  %  \, C \begin{pmatrix} \Lambda_1 & \Lambda_2 & \Lambda \\  r_1 & r_2 & r \end{pmatrix} \times
  % \nonumber \\
  % S^{J_1, m_1}_{\Lambda_1, r_1} S^{J_2, m_2}_{\Lambda_2, r_2}
  %\,  \left\langle J_1 m_1 J_2 m_2 \Big| J M \right\rangle. 
  \label{eq:expansion_2}
\end{eqnarray}
Because the octahedral states $\big| ^n\!\Lambda, r, [J] \big\rangle$ that are subduced from a single spin form an orthonormal set, 
their coefficients must be the same on both sides of Eq.~(\ref{eq:expansion_2}).  Thus, the subduction
matrix for total spin, $J$, obeys 
\begin{eqnarray}
  {\cal S}^{J,M}_{^n\!\Lambda, r} 
  = \sum_{\Lambda_{\Lambda_1\otimes\Lambda_2}} &&\!\!\!\!\!\! {\cal A}_{ \Lambda_{\Lambda_1\otimes\Lambda_2},
    ^n\!\Lambda [J]} 
  {\cal R}^{J, M}_{\Lambda_{\Lambda_1\otimes\Lambda_2}, r}
  % \nonumber \\
  %  & &S^{J_1, m_1}_{\Lambda_1, r_1} S^{J_2, m_2}_{\Lambda_2, r_2}
  % \big\langle J_1 m_1 J_2 m_2 \big| J M \big\rangle.
  \label{eq:sub-coef}
\end{eqnarray}
As will become evident, matrix ${\cal A}$ is orthogonal: ${\cal A} {\cal A}^T = 1$.  When the above equation is multiplied by 
$|[J,M]\rangle$ on both sides, and then summed over $M$, we get
\begin{eqnarray}
  \sum_M {\cal S}^{J,M}_{^n\!\Lambda, r} \left| \big[ J,M \right] \big\rangle &=& 
  \sum_{\Lambda_{\Lambda_1\otimes\Lambda_2}} 
  {\cal A}_{ \Lambda_{\Lambda_1\otimes\Lambda_2}, ^n\!\Lambda [J]} 
  \nonumber \\
  &\times& \sum_M {\cal R}^{J,M}_{\Lambda_{\Lambda_1\otimes\Lambda_2}, r}  \big| \left[ J,M \right] \big\rangle.
\end{eqnarray}
Using Eq.~(\ref{eq:SJM}),  this becomes 
\begin{eqnarray}
  \big| ^n\!\Lambda, r, \left[J\right]\big\rangle &=&
  \sum_{\Lambda_{\Lambda_1\otimes\Lambda_2}} {\cal A}_{ \Lambda_{\Lambda_1\otimes\Lambda_2},
    ^n\!\Lambda [J]} 
  \big| \Lambda_{\Lambda_1\otimes\Lambda_2}, r \big\rangle, 
  \nonumber \\
\end{eqnarray}
where 
\begin{equation}
  \big| \Lambda_{\Lambda_1\otimes\Lambda_2}, r \big\rangle =  
  \sum_M  {\cal R}^{J,M}_{\Lambda_{\Lambda_1\otimes\Lambda_2}, r} \big| \left[ J,M \right] \big\rangle .
  \label{eq:RJM_JM}
\end{equation}

Table~\ref{table:S1/2S1S3/2} shows the subduction matrices for the elementary states of spin $\frac{1}{2}$, $1$ and $\frac{3}{2}$.  They are unit matrices in the basis used. 
Starting with the known subduction matrices for spins $J_1=1$ and $J_2=\frac{3}{2}$, the 
subduction matrix for spin $\frac{5}{2}$ can be obtained as follows.    Evaluate 
Eq.~(\ref{eq:expansion_1}) to   obtain matrices ${\cal R}^{\frac{5}{2}, M}_{H^{\prime}_{T_1\otimes H}, r}$,
${\cal R}^{\frac{3}{2}, M}_{H_{T_1\otimes H}, r}$ and ${\cal R}^{\frac{5}{2}, M}_{G_{2\, T_1\otimes H}, r}$.    
Matrix ${\cal R}^{\frac{5}{2}, M}_{H^{\prime}_{T_1\otimes H}, r}$
is equal within an overall constant factor ${\cal A}_{H^{\prime}[\frac{5}{2}], H_{T_1\otimes H}}$ to 
subduction matrix  ${\cal S}^{\frac{5}{2}, M}_{H, r}$. Similarly, ${\cal R}^{\frac{3}{2}, M}_{H_{T_1\otimes H}, r}$
and ${\cal R}^{\frac{5}{2}, M}_{G_{2\, T_1\otimes H}, r}$ are equal within overall factors to 
${\cal S}^{\frac{3}{2}, M}_{H, r}$ and ${\cal S}^{\frac{5}{2}, M}_{G_2, r}$  . 
The overall factors lead to ${\cal S}$ matrices normalized as in 
Eq.~(\ref{eq:SJM_norm}).  The subduction 
matrices that are obtained for spin $\frac{5}{2}$ are given in Table~\ref{table:S5/2}.
The overall factors determine matrix ${\cal A}_{^n\!\Lambda[J], \Lambda_{\Lambda_1\otimes\Lambda_2}}$,
which is  a unit matrix in this example, i.e., Eq.~(\ref{eq:sum_J'}) takes the form,
\begin{eqnarray}
  \begin{pmatrix} 
    \left|H_{T_1 \otimes H} , r\right \rangle  \\
    \left|H^{\prime}_{T_1 \otimes H} , r\right \rangle \\
    \left| G_{2\, T_1 \otimes H} , r\right \rangle 
  \end{pmatrix}
  =                    
  \begin{pmatrix}  1&0&0\\  
    0&1&0\\
    0&0&1
  \end{pmatrix}
  \begin{pmatrix}   
    \left|H , r, \left[ \frac{3}{2}\right] \right \rangle \\
    \left|  H, r, \left[ \frac{5}{2}\right] \right\rangle \\
    \left|  G_2, r,  \left[ \frac{5}{2}\right] \right\rangle
  \end{pmatrix}   .
  \nonumber \\
  \label{eq:G2_H_5/2} 
\end{eqnarray} 
This block-matrix equation holds when the row indices, $r$, are the same on both sides.  
Note that orthonormality of the states on the right side implies that
the states on the left side are orthonormal. 

For higher spins, the Clebsch-Gordan coefficients of the octahedral group do not provide a 
block-diagonal result for the subduction, i.e., matrix ${\cal A}$ takes a nontrivial form. This is demonstrated for
the construction of spin $\frac{7}{2}$ based on $J_1=1$ and $J_2 = \frac{5}{2}$, which involves use 
of the previous subduction, $ \left[ \frac{5}{2} \right] \rightarrow H \oplus G_2$.  Three orthogonal
occurrences of irrep $H$ are produced in the $T_1\otimes \left[\frac{5}{2}\right] = T_1\otimes \big(H \oplus G_2\big)$ direct products and 
two orthogonal occurrences of $G_2$ are produced.  Taking into account the fact that 
two different $H$ irreps are produced by each
$T_1 \otimes H$ product, i.e., $T_1 \otimes H  \rightarrow G_1 \oplus H \oplus H^{\prime} \oplus G_2$,  the three $H$ states
that occur are $H_{T_1 \otimes H}$, \, $ H^{\prime}_{T_1 \otimes H}\, $ and $H_{T_1 \otimes G_2}$. 
The two different $G_2$ states that occur are $G_{2\, T_1 \otimes H}$ and  $G_{2\, T_1 \otimes G_2}$.

The procedure is similar to the spin $\frac{5}{2}$ case.  First one evaluates 
Eq.~(\ref{eq:expansion_1}) to   obtain 
matrices ${\cal R}^{\frac{7}{2}, M}_{G_{1\, T_1\otimes H}, r}$, ${\cal R}^{\frac{7}{2}, M}_{H_{T_1\otimes H}, r}$, ${\cal R}^{\frac{5}{2}, M}_{H^{\prime}_{T_1\otimes H}, r}$, ${\cal R}^{\frac{3}{2}, M}_{H_{T_1\otimes H}, r}$,  
${\cal R}^{\frac{7}{2}, M}_{G_{2\, T_1\otimes H}, r}$ and ${\cal R}^{\frac{5}{2}, M}_{G_{2\, T_1\otimes G_2}, r}$.
Each ${\cal R}$ matrix corresponds to a single 
${\cal S}$ matrix of the set ${\cal S}^{\frac{7}{2}, M}_{G_1, r}$
${\cal S}^{\frac{7}{2}, M}_{H, r}$, ${\cal S}^{\frac{5}{2}, M}_{H, r}$, ${\cal S}^{\frac{3}{2}, M}_{H, r}$,  
${\cal S}^{\frac{7}{2}, M}_{G_2, r}$ and ${\cal S}^{\frac{5}{2}, M}_{G_2, r}$.  In this case the subduction 
matrices for spin $\frac{7}{2}$ can be obtained simply by imposing the normalization condition
$\sum_M  {\cal S}^{J,M}_{\Lambda, r} {\cal S}^{J,M}_{\Lambda, r} = 1$.
That gives the subduction matrices of Table~\ref{table:S7/2}.

The octahedral states that result from the Clebsch-Gordan formula are linear combinations of the states
subduced from a single $J$ value.   This is expressed by 
the block-matrix equation, 
\begin{eqnarray}
  \begin{pmatrix}  \left|G_{1~T_1 \otimes H} , r\right \rangle \\
    \left|H_{T_1 \otimes H} , r\right \rangle  \\
    \left|H^{\prime}_{T_1 \otimes H} , r\right \rangle \\
    \left|H_{T_1 \otimes G_2} , r\right \rangle \\
    \left| G_{2\, T_1 \otimes H} , r\right \rangle \\
    \left|G_{2\, T_1 \otimes G_2} , r\right \rangle
  \end{pmatrix}
  =     {\cal A}^{T_1\otimes \frac{5}{2}} \times                
  \begin{pmatrix}  \left|G_1, r, \left[\frac{7}{2}\right] \right\rangle \\
    \left|H , r, \left[ \frac{7}{2}\right] \right \rangle \\
    \left|  H, r  \left[ \frac{5}{2}\right] \right\rangle \\
    \left|  H, r  \left[ \frac{3}{2}\right] \right\rangle \\
    \left| G_2 , r, \left[ \frac{7}{2}\right] \right \rangle \\
    \left|  G_2, r  \left[ \frac{5}{2}\right] \right\rangle
  \end{pmatrix}   .
  \label{eq:G1_H_G2_7/2} 
\end{eqnarray} 
where matrix ${\cal A}^{T_1\otimes \frac{5}{2}}$ is
\begin{eqnarray}
  {\cal A}^{T_1\otimes \frac{5}{2}} = 
  \begin{pmatrix}  1 ~~& 0 & 0 & 0 & 0 & 0 \\  
    0 & \sqrt{\frac{4}{7}} &  \sqrt{\frac{3}{7}} & 0 & 0 & 0 \\
    0 &  -\sqrt{\frac{1}{7}} & \sqrt{\frac{4}{21}} & -\sqrt{\frac{2}{3}} & 0 & 0 \\
    0 &  -\sqrt{\frac{2}{7}} & \sqrt{\frac{8}{21}} & \sqrt{\frac{1}{3}} & 0 & 0 \\
    0 & 0 & 0 & 0 & \sqrt{\frac{5}{21}} & -\sqrt{\frac{16}{21}} \\
    0 & 0 & 0 & 0 & -\sqrt{\frac{16}{21}} & -\sqrt{\frac{5}{21}} 
  \end{pmatrix} .
  \nonumber \\
\end{eqnarray}

Note that row indices, $r$, are the same on both sides of Eq.~(\ref{eq:G1_H_G2_7/2}) so that 
it connects the 18 states on the left side (considering all allowed values of the row indices)  to 18 states
on the right side.                                                                     
The states on the right side are orthonormal as in Eq.~(\ref{eq:oct_orthog_2}) and the rows of the matrix in Eq.~(\ref{eq:G1_H_G2_7/2}) are orthogonal to one another.  It follows that the states on the left 
side also form an orthonormal set by construction.  Matrix equation~(\ref{eq:G1_H_G2_7/2}) takes the form of Eq.~(\ref{eq:sum_J'}) and it provides a nontrivial example of matrix ${\cal A}$, demonstrating the relation between irreducible 
states resulting from direct products of octahedral states and the irreducible states that are  
subduced from a single spin.  It is straightforward to solve for the irreducible states subduced from a single spin by applying ${\cal A}^{-1} = {\cal A}^T$ to both sides of Eq.~(\ref{eq:G1_H_G2_7/2}).
That demonstrates that the irreducible octahedral states subduced from a single $J$ are linear
combinations of the irreducible octahedral states resulting from the Clebsch-Gordan formula.   

The same reasoning can be applied to the subduction of spin $\frac{9}{2}$ based on $J_1 = 1$ and $J_2 = \frac{7}{2}$. Table \ref{table:9/2occurrences} shows the linearly independent states for this case.  A total of four orthogonal occurrences of $H$, two of $G_1$ and two of $G_2$ arise from $T_1
\otimes \left[\frac{7}{2}\right] = T_1\otimes( G_1\oplus H \oplus G_2)$.  
Two of  the $H$ irreps reproduce the previous results for subduction of  spins $\frac{5}{2}$ and $\frac{7}{2}$.
The other two $H$ irreps are yet-to-be-determined subductions of spin $\frac{9}{2}$.  The two $G_1$ irreps provide subductions of spins $\frac{9}{2}$ and $\frac{7}{2}$.  The two 
$G_2$ irreps provide subductions of spins $\frac{7}{2}$ and $\frac{5}{2}$. 
The new element for spin $\frac{9}{2}$ is that two occurrences of irrep $H$ are unknown.
They are determined as follows.  First matrices ${\cal R}^{J,M}_{\Lambda_{\Lambda_1\otimes \Lambda_2}, r}$
are calculated.  They are related to the subduction matrices by a matrix ${\cal A}^{T_1\otimes \frac{7}{2}}$, which also relates the states formed from the ${\cal R}$ and ${\cal S}$ matrices as in Eqs.~(\ref{eq:SJM_JM}) and (\ref{eq:RJM_JM}).  That relation can be expressed as follows,
%\begin{comment}
\begin{eqnarray}
  \begin{pmatrix}  \left|G_{1~T_1 \otimes G_1} , r\right \rangle \\
    \left|G_{1~T_1 \otimes H} , r\right \rangle \\
    \left|H_{1~T_1 \otimes G_1} , r\right \rangle \\
    \left|H_{T_1 \otimes H} , r\right \rangle  \\
    \left|H^{\prime}_{T_1 \otimes H} , r\right \rangle \\
    \left|H_{T_1 \otimes G_2} , r\right \rangle \\
    \left| G_{2\, T_1 \otimes H} , r\right \rangle \\
    \left|G_{2\, T_1 \otimes G_2} , r\right \rangle
  \end{pmatrix}
  =    {\cal A}^{T_1\otimes \frac{7}{2}} \times                
  \begin{pmatrix}  \left|G_1, r, \left[\frac{9}{2}\right] \right\rangle \\
    \left|G_1, r, \left[\frac{7}{2}\right] \right\rangle \\
    \left| \,^1H , r, \left[ \frac{9}{2}\right] \right \rangle \\
    \left| \,^2H , r, \left[ \frac{9}{2}\right] \right \rangle \\
    \left|  H, r  \left[ \frac{7}{2}\right] \right\rangle \\
    \left|  H, r  \left[ \frac{5}{2}\right] \right\rangle \\
    \left| G_2 , r, \left[ \frac{7}{2}\right] \right \rangle \\
    \left|  G_2, r  \left[ \frac{5}{2}\right] \right\rangle
  \end{pmatrix}   .
  \label{eq:G1_H_H_9/2} 
  \nonumber \\
\end{eqnarray}  
where matrix ${\cal A}^{T_1\otimes \frac{7}{2}}$ is 
\begin{eqnarray}
  \begin{pmatrix}  \sqrt{\tfrac{20}{27}} & -\sqrt{\frac{7}{27}} & 0 & 0 & 0 & 0 & 0 &0 \\  
    -\sqrt{\frac{7}{27}} & -\sqrt{\frac{20}{27}}& 0 & 0 & 0 & 0 & 0 &0 \\
    0 & 0 & a_1 & b_1 & \sqrt{\frac{10}{27}} & \sqrt{\frac{3}{8}} & 0 & 0 \\
    0 &  0 & a_2 & b_2 & \sqrt{\frac{1}{945}} & -\sqrt{\frac{3}{7}} & 0 & 0 \\
    0 &  0 & a_3 & b_3 & -\sqrt{\frac{12}{35}} & \sqrt{\frac{3}{28}} &  0 & 0 \\
    0 & 0 & a_4 & b_4 & -\sqrt{\frac{2}{7}} & \sqrt{\frac{5}{56}} & 0 & 0  \\
    0  & 0 & 0 & 0 & 0 & 0 & \sqrt{\frac{4}{7}} & -\sqrt{\frac{3}{7}} \\
    0  & 0 & 0 & 0 & 0 & 0 & \sqrt{\frac{3}{7}} & \sqrt{\frac{4}{7}} 
  \end{pmatrix} .
  \nonumber \\
\end{eqnarray}
% \end{comment}
Constants $a_n$ and $b_n$ express the unknown parts of the matrix that
connect the four $H$ irreps on the left side to linear 
combinations involving  
$\big| ^1H , r, \left[ \frac{9}{2}\right] \big \rangle$
and $\big| ^2H , r, \left[ \frac{9}{2}\right] \big \rangle$.   

The upper-left block of the matrix equation can be solved for 
\begin{eqnarray}
  \Big|G_1, r, \left[\tfrac{9}{2}\right] \Big\rangle = \sqrt{\tfrac{20}{27}}  \Big|G_{1\, T_1 \otimes G_1} , r\Big \rangle 
  -   \sqrt{\tfrac{7}{27}}    \Big|G_{1\, T_1 \otimes H} , r\Big \rangle.
  \nonumber \\
  \label{eq:S_G1_9/2}
\end{eqnarray}
Subduction matrix ${\cal S}^{\frac{9}{2}, M}_{G_1, r}$ is determined by a linear combination of
${\cal R}$ matrices with the same coefficients as in  Eq.~(\ref{eq:S_G1_9/2}), i.e.,
\begin{equation}
  {\cal S}^{\tfrac{9}{2}, M}_{G_1, r} = \sqrt{\tfrac{20}{27}} {\cal R}^{\tfrac{9}{2}, M}_{G_{1\, T_1\otimes G_1}, r}
  -   \sqrt{\tfrac{7}{27}}   {\cal R}^{\tfrac{9}{2}, M}_{G_{1\, T_1\otimes H}, r}.
\end{equation}

The middle block gives four equations for the $H$ irreps. They can be reduced to three by making linear combinations
of  pairs of equations to eliminate the $\left| H , r, \left[ \frac{5}{2}\right] \right \rangle$ terms, and then further 
reduced to two equations by making linear combinations of pairs of equations to eliminate the $\left| H , r, \left[ \frac{7}{2}\right] \right \rangle$ terms.
The resulting equations provide candidates for the $^1H$ and $^2H$ irrep states subduced from spin $\frac{9}{2}$ as follows,
\begin{multline}
  a_1^{''}  \Big| \,^1H , r, \left[ \tfrac{9}{2}\right] \Big \rangle  +
  b_1^{''}  \Big| \,^2H , r, \left[ \tfrac{9}{2}\right] \Big \rangle  
  = \nonumber \\\sqrt{\tfrac{5}{96}} \Big|H_{T_1 \otimes G_1} , r\Big \rangle +
 \sqrt{\tfrac{7}{60}} \Big|H_{T_1 \otimes H} , r\Big \rangle +
  \sqrt{\tfrac{21}{320}} \Big|H^{\prime}_{T_1 \otimes H} , r\Big \rangle 
\end{multline}
\begin{multline}
  a_2^{''}  \Big| \,^1H , r, \left[ \tfrac{9}{2}\right] \Big \rangle +
  b_2^{''}  \Big| \,^2H , r, \left[ \tfrac{9}{2}\right] \Big \rangle  = \nonumber \\
  \tfrac{5}{24} \Big|H_{T_1 \otimes G_1} , r\Big \rangle +
\sqrt{\tfrac{7}{72}} \Big|H_{T_1 \otimes H} , r\Big \rangle -
  \sqrt{\tfrac{21}{320}} \Big|H_{T_1 \otimes G_2} , r\Big \rangle ,
  \nonumber \\
\end{multline}
where $a_n^{''}$ and $b_n^{''}$ are combinations of the unknown constants $a_n$ and $b_n$.
There remain four unknown constants here and three equations that constrain them in order that
states $\left| ^1H , r, \left[ \frac{9}{2}\right] \right \rangle $ and $\left| ^2H , r, \left[ \frac{9}{2}\right] \right \rangle $ 
are orthonormal.  A one-parameter family of solutions exists.   It is sufficient for our purpose 
to obtain a single solution by choosing $b_1^{''} = 0$.  From the first equation we find      
\begin{eqnarray}
  \Big| \,^1H , r, \left[ \tfrac{9}{2}\right] \Big \rangle  =&& \!\!\!\!\!  \sqrt{\tfrac{2}{9}} \Big|H_{T_1 \otimes G_1} , r\Big \rangle +
  \sqrt{\tfrac{112}{225}} \Big|H_{T_1 \otimes H} , r\Big \rangle +
  \nonumber \\
  && \sqrt{\tfrac{7}{25}} \Big|H^{\prime}_{T_1 \otimes H} , r\Big \rangle ,
  \label{eq:1H_9/2}
\end{eqnarray}
where constant $a_1^{''}$ was determined by normalizing the state.          
The  second equation is then used to obtain a state that is orthonormal to the first one, which determines constants $a_2^{''}$ and $b_2^{''}$, and yields
\begin{eqnarray}
  \Big| \,^2H , r, \left[ \tfrac{9}{2}\right] \Big\rangle =&&\!\!\!\! \sqrt{\tfrac{7}{216}}
  \Big|H_{T_1 \otimes G_1} , r\Big \rangle +
  \sqrt{\tfrac{49}{675}} \Big|H_{T_1 \otimes H} , r\Big \rangle 
  - 
  \nonumber \\
  && \sqrt{\tfrac{27}{100}} \Big|H^{\prime}_{T_1 \otimes H} , r\Big \rangle 
  + \sqrt{\tfrac{5}{8}} \Big|H_{T_1 \otimes G_2} , r\Big \rangle .
  \nonumber 
  \label{eq:2H_9/2}
\end{eqnarray}
The coefficients appearing in Eq.~(\ref{eq:1H_9/2}) give the values of $a_1, a_2, a_3$ ( $a_4= 0) $
and the coefficients appearing in Eq.(\ref{eq:2H_9/2}) give the values of $b_1, b_2, b_3$ and $b_4$.
They complete the determination of matrix ${\cal A}$ and can be used to show that it is orthogonal.    
Subduction matrices are determined as linear combinations of ${\cal R}$ matrices with the 
same coefficients that appear above for the $\,^1H$ and $\,^2H$ states,
\begin{eqnarray}
  {\cal S}^{\frac{9}{2}, M}_{\,^1\! H, r} &=&  \sqrt{\tfrac{2}{9}} {\cal R}^{\frac{9}{2}, M}_{H_{T_1 \otimes G_1} , r}+ 
  \sqrt{\tfrac{112}{225}} {\cal R}^{\frac{9}{2}, M}_{H_{T_1 \otimes H} , r} +
  \nonumber \\
  & &
  \sqrt{\tfrac{7}{25}} {\cal R}^{\frac{9}{2}, M}_{H^{\prime}_{T_1 \otimes H} , r} 
  \nonumber \\
  {\cal S}^{\tfrac{9}{2}, M}_{\,^2\! H, r}  &=&\sqrt{\tfrac{7}{216}}\ {\cal R}^{\frac{9}{2}, M}_{H_{T_1 \otimes G_1} , r} +
  \sqrt{\tfrac{49}{675}} {\cal R}^{\frac{9}{2}, M}_{H_{T_1 \otimes H} , r} -
  \nonumber \\
  & &      \sqrt{\tfrac{27}{100}} {\cal R}^{\frac{9}{2}, M}_{H^{\prime}_{T_1 \otimes H} , r} +
  \sqrt{\tfrac{5}{8}} {\cal R}^{\frac{9}{2}, M}_{H^{\prime}_{T_1 \otimes G_2} , r}  .
\end{eqnarray}
The resulting subduction matrices for spin $\frac{9}{2}$ are given in Table~\ref{table:S9/2}.

%% file: subduce_tables.tex
%%%%%%%%%%%  1/2, 1 and 3/2 table  %%%%%%%%%%%%%
\begin{table}[!h] 
  \caption{ Elementary subduction matrices ${\cal S}^{\frac{1}{2},m}_{G_1, r}$, ${\cal S}^{1, m}_{T_1, r}$ and ${\cal S}^{\frac{3}{2},m}_{H,r} $. \label{table:S1/2S1S3/2}}

\begin{tabular}{ccc}
  \begin{tabular}{l|cc}
  \multicolumn{3}{c}{$J=\tfrac{1}{2}\to G_1$}\\
   $\begin{smallmatrix} & m \\ r & \end{smallmatrix}$
   & $+\tfrac{1}{2}$ & $-\tfrac{1}{2}$  \\
   \hline
   $1$ & $1$ & $0$ \\
   $2$ & $0$ & $1$
  \end{tabular}
  
  &
	\quad \begin{tabular}{l|ccc}
	 \multicolumn{4}{c}{$J=1 \to T_1$}\\
  	 $\begin{smallmatrix} & m \\ r & \end{smallmatrix}$& $+1$ & $0$ & $-1$ \\
	  \hline
	 $1$ & $1$ & $0$ & $0$ \\ 
     $2$ & $0$ & $1$ & $0$ \\
     $3$ & $0$ & $0$ & $1$ \\
  \end{tabular}	
  
  & 
  	\quad \begin{tabular}{l|cccc}
	\multicolumn{5}{c}{$J=\tfrac{3}{2} \to H$}\\
  	 $\begin{smallmatrix} & m \\ r & \end{smallmatrix}$& $+\tfrac{3}{2}$ & $+\tfrac{1}{2}$ & $-\tfrac{1}{2}$ & $-\tfrac{3}{2}$ \\
	  \hline
	 $1$ & $1$ & $0$ & $0$ & $0$ \\ 
     $2$ & $0$ & $1$ & $0$ & $0$\\
     $3$ & $0$ & $0$ & $1$ & $0$\\
     $4$ & $0$ & $0$ & $0$ & $1$\\
  \end{tabular}	  
  \end{tabular} 
\end{table}

%%%%%%%%%%%%%   5/2  table   %%%%%%%%%%%%%%%%%%%%%%%%%%%%%%%

\begin{table}[!h]
  \caption{ Subduction matrices ${\cal S}^{\frac{5}{2},m}_{H, r}$ and ${\cal S}^{\frac{5}{2}, m}_{G_2, r}$ 
    \label{table:S5/2} }
    
 \begin{tabular}{l|cccccc}
  \multicolumn{7}{c}{$J=\tfrac{5}{2}\to H$}\\
   $\begin{smallmatrix} & m \\ r & \end{smallmatrix}$
   & $+\tfrac{5}{2}$ & $+\tfrac{3}{2}$ & $+\tfrac{1}{2}$ &  $-\tfrac{1}{2}$ & $-\tfrac{3}{2}$ &  $-\tfrac{5}{2}$ \\
   \hline
   $1$ & $0$ & $+\sqrt{\tfrac{1}{6}}$ & $0$ & $0$ & $0$ & $+\sqrt{\tfrac{5}{6}}$  \\
   $2$ & $0$ & $0$ & $-1$ & $0$ & $0$ & $0$ \\
   $3$ & $0$ & $0$ & $0$ & $+1$ & $0$ & $0$ \\
   $4$ & $-\sqrt{\tfrac{5}{6}}$  & $0$ & $0$ & $0$ & $-\sqrt{\tfrac{1}{6}}$ & $0$
  \end{tabular}
 % \\
 %   \vspace{.5cm}
  \begin{tabular}{l|cccccc}
  \multicolumn{7}{c}{$J=\tfrac{5}{2}\to G_2$}\\
   $\begin{smallmatrix} & m \\ r & \end{smallmatrix}$
   & $+\tfrac{5}{2}$ & $+\tfrac{3}{2}$ & $+\tfrac{1}{2}$ &  $-\tfrac{1}{2}$ & $-\tfrac{3}{2}$ &  $-\tfrac{5}{2}$ \\
   \hline
   $1$ & $+\sqrt{\tfrac{1}{6}}$ & $0$ & $0$ & $0$ & $-\sqrt{\tfrac{5}{6}}$ & $0$ \\
   $2$ & $0$ & $-\sqrt{\tfrac{5}{6}}$ & $0$ & $0$ & $0$ & $+\sqrt{\tfrac{1}{6}}$ \\
  \end{tabular}
\end{table}

%%%%%%%%%%%%%%  7/2  table %%%%%%%%%%%%%%%%%%%%%%%%

\begin{table}[!h]
  \caption{Subduction matrices ${\cal S}^{\frac{7}{2},m}_{G_1, r}$,  ${\cal S}^{\frac{7}{2},m}_{H, r}$ and ${\cal S}^{\frac{7}{2}, m}_{G_2, r}$.  \label{table:S7/2}  }
  
   \begin{tabular}{l|cccccccc}
  \multicolumn{9}{c}{$J=\tfrac{7}{2}\to G_1$}\\
   $\begin{smallmatrix} & m \\ r & \end{smallmatrix}$
   & $+\tfrac{7}{2}$ & $+\tfrac{5}{2}$ & $+\tfrac{3}{2}$ & $+\tfrac{1}{2}$ &  $-\tfrac{1}{2}$ & $-\tfrac{3}{2}$ &  $-\tfrac{5}{2}$ & $-\tfrac{7}{2}$\\
   \hline
   $1$ & $0$ &  $0$ & $0$ & $+\sqrt{\tfrac{7}{12}}$ & $0$& $0$& $0$ & $+\sqrt{\tfrac{5}{12}}$  \\
   $2$ & $-\sqrt{\tfrac{5}{12}}$  & $0$ & $0$ & $0$ & $-\sqrt{\tfrac{7}{12}}$ & $0$& $0$& $0$ 
  \end{tabular}  
  \\
  \vspace{.5cm}
  
   \begin{tabular}{l|cccccccc}
  \multicolumn{9}{c}{$J=\tfrac{7}{2}\to H$}\\
   $\begin{smallmatrix} & m\\ r & \end{smallmatrix}$
   & $+\tfrac{7}{2}$ & $+\tfrac{5}{2}$ & $+\tfrac{3}{2}$ & $+\tfrac{1}{2}$ &  $-\tfrac{1}{2}$ & $-\tfrac{3}{2}$ &  $-\tfrac{5}{2}$ & $-\tfrac{7}{2}$\\
   \hline
   $1$ & $0$ &  $0$ & $+\sqrt{\tfrac{3}{4}}$ & $0$ & $0$& $0$ & $+\sqrt{\tfrac{1}{4}}$  & $0$ \\
   $2$ & $0$ &  $0$ & $0$ & $-\sqrt{\tfrac{5}{12}}$ & $0$& $0$& $0$ & $+\sqrt{\tfrac{7}{12}}$  \\
   $3$ & $+\sqrt{\tfrac{7}{12}}$ &  $0$ & $0$ & $0$ & $-\sqrt{\tfrac{5}{12}}$ &  $0$ & $0$ & $0$ \\
   $4$ & $0$ &  $+\sqrt{\tfrac{1}{4}}$ & $0$  & $0$& $0$  & $+\sqrt{\tfrac{3}{4}}$ & $0$& $0$  \\   
  \end{tabular}
\\
  \vspace{.5cm}
  
     \begin{tabular}{l|cccccccc}
  \multicolumn{9}{c}{$J=\tfrac{7}{2}\to G_2$}\\
   $\begin{smallmatrix} & m \\ r & \end{smallmatrix}$
   & $+\tfrac{7}{2}$ & $+\tfrac{5}{2}$ & $+\tfrac{3}{2}$ & $+\tfrac{1}{2}$ &  $-\tfrac{1}{2}$ & $-\tfrac{3}{2}$ &  $-\tfrac{5}{2}$ & $-\tfrac{7}{2}$\\
   \hline
   $1$ & $0$ &  $+\sqrt{\tfrac{3}{4}}$ & $0$ & $0$ & $0$& $-\sqrt{\tfrac{1}{4}}$ & $0$ & $0$  \\
   $2$ & $0$ & $0$  & $+\sqrt{\tfrac{1}{4}}$ & $0$ & $0$ & $0$ &  $-\sqrt{\tfrac{3}{4}}$ & $0$ 
  \end{tabular}
    
\end{table}
%
%
%%%%%%%%%%%%%%    9/2  tables  %%%%%%%%%%%%%%%%%

\begin{table}[!h] 
 \caption{ Subduction matrices ${\cal S}^{\frac{9}{2},m}_{G_1, r}$,  ${\cal S}^{\frac{9}{2},m}_{^1H, r}$ and ${\cal S}^{\frac{9}{2}, m}_{^2H, r}$.  \label{table:S9/2}}

   \begin{tabular}{l|cccccccccc}
  \multicolumn{11}{c}{$J=\tfrac{9}{2}\to G_1$}\\
   $\begin{smallmatrix} & m \\ r & \end{smallmatrix}$
   & $+\tfrac{9}{2}$  & $+\tfrac{7}{2}$ & $+\tfrac{5}{2}$ & $+\tfrac{3}{2}$ & $+\tfrac{1}{2}$ &  $-\tfrac{1}{2}$ & $-\tfrac{3}{2}$ &  $-\tfrac{5}{2}$ & $-\tfrac{7}{2}$ & $-\tfrac{9}{2}$ \\
   \hline
   $1$ & $-\sqrt{\tfrac{3}{8}}$ &  $0$ & $0$ & $0$ & $-\sqrt{\tfrac{7}{12}}$ &  $0$ & $0$ & $0$ & $-\sqrt{\tfrac{1}{24}}$ & $0$ \\
   $2$ & $0$ & $-\sqrt{\tfrac{1}{24}}$ & $0$ & $0$ & $0$ & $-\sqrt{\tfrac{7}{12}}$ &  $0$ & $0$ & $0$ & $-\sqrt{\tfrac{3}{8}}$ \\
  \end{tabular}
  \\
  \vspace{.5cm}
 \begin{tabular}{l|cccccccccc}
  \multicolumn{11}{c}{$J=\tfrac{9}{2}\to \,^1\!H$}\\
   $\begin{smallmatrix} & m \\ r & \end{smallmatrix}$
   & $+\tfrac{9}{2}$  & $+\tfrac{7}{2}$ & $+\tfrac{5}{2}$ & $+\tfrac{3}{2}$ & $+\tfrac{1}{2}$ &  $-\tfrac{1}{2}$ & $-\tfrac{3}{2}$ &  $-\tfrac{5}{2}$ & $-\tfrac{7}{2}$ & $-\tfrac{9}{2}$ \\
   \hline
   $1$ &   $0$ & $0$ & $0$ & $+\sqrt{\tfrac{7}{10}}$ &  $0$ & $0$ & $0$ & $+\sqrt{\tfrac{3}{10}}$ & $0$ & $0 $\\
   $2$ &   $-\sqrt{\tfrac{5}{8}}$ & $0$ & $0$ & $0$ &  $+\sqrt{\tfrac{7}{20}}$ &  $0$ & $0$ & $0$ & $+\sqrt{\tfrac{1}{40}}$ & $0$\\
   $3$ & $0$ & $-\sqrt{\tfrac{1}{40}}$ & $0$& $0$& $0$ & $-\sqrt{\tfrac{7}{20}}$ & $0$& $0$& $0$ & $+\sqrt{\tfrac{5}{8}}$ \\
    $4$ &   $0$ & $0$  & $-\sqrt{\tfrac{3}{10}}$ &  $0$ & $0$ & $0$ & $-\sqrt{\tfrac{7}{10}}$ & $0$ & $0 $ & $0$ \\
     \end{tabular}  
      \\
  \vspace{.5cm}
 \begin{tabular}{l|cccccccccc}
  \multicolumn{11}{c}{$J=\tfrac{9}{2}\to \,^2\!H$}\\
   $\begin{smallmatrix} & m \\ r & \end{smallmatrix}$
   & $+\tfrac{9}{2}$  & $+\tfrac{7}{2}$ & $+\tfrac{5}{2}$ & $+\tfrac{3}{2}$ & $+\tfrac{1}{2}$ &  $-\tfrac{1}{2}$ & $-\tfrac{3}{2}$ &  $-\tfrac{5}{2}$ & $-\tfrac{7}{2}$ & $-\tfrac{9}{2}$ \\
   \hline
   $1$ &   $0$ & $0$ & $0$ & $+\sqrt{\tfrac{3}{10}}$ &  $0$ & $0$ & $0$ & $-\sqrt{\tfrac{7}{10}}$ & $0$ & $0 $\\
   $2$ &   $0$ & $0$ & $0$ & $0$ &  $-\sqrt{\tfrac{1}{15}}$ &  $0$ & $0$ & $0$ & $+\sqrt{\tfrac{14}{15}}$ & $0$\\
   $3$ & $0$ & $-\sqrt{\tfrac{14}{15}}$ & $0$& $0$& $0$ & $+\sqrt{\tfrac{1}{15}}$ & $0$& $0$& $0$ & $0$ \\
    $4$ &   $0$ & $0$  & $+\sqrt{\tfrac{7}{10}}$ &  $0$ & $0$ & $0$ & $-\sqrt{\tfrac{3}{10}}$ & $0$ & $0 $ & $0$ \\
     \end{tabular}

\end{table}